\documentclass[11pt]{article}

\usepackage[a4paper,margin=30mm]{geometry}
\usepackage[round,authoryear]{natbib}
\usepackage{microtype}
\usepackage{pp-manuscript}
\usepackage[hidelinks]{hyperref}
\usepackage[capitalise,nameinlink,noabbrev]{cleveref}

\newcommand{\papertitle}{Who Are a Firm's Economic Peers? Mapping a Latent
Information Field with Language Models}

\newcommand{\paperauthors}{Marcus Gawronsky, Chun-Sung Huang}

\newcommand{\paperaffiliations}{Department of Finance and Tax,
  University of Cape
Town}

\newcommand{\paperdate}{September 2026}

\newcommand{\paperkeywords}{Spatial asset pricing; Language models; Economic peers;
Optimal transport; Wasserstein barycentric reconstruction; Residual covariance}
\newcommand{\paperjel}{G12; G11; C21; C58}

\ppDeclareValue{post-evaluation-date-start}{2023-01-03}
\ppDeclareValue{post-evaluation-date-end}{2026-04-06}
\ppDeclareValue{post-eval2023-equal-support-rho}{0.843530}
\ppDeclareValue{post-eval2023-equal-support-rho-ci-lower}{0.827323}
\ppDeclareValue{post-eval2023-equal-support-rho-ci-upper}{0.856492}
\ppDeclareValue{post-eval2023-equal-support-loglik}{64843.300633}
\ppDeclareValue{post-eval2023-equal-support-trading-days}{250}
\ppDeclareValue{post-eval2023-equal-support-lambda}{5.390997}
\ppDeclareValue{post-eval2023-equal-support-lambda-ci-lower}{4.791168}
\ppDeclareValue{post-eval2023-equal-support-lambda-ci-upper}{5.968267}
\ppDeclareValue{post-eval2023-w-co-mentions-rho}{0.676182}
\ppDeclareValue{post-eval2023-w-co-mentions-rho-ci-lower}{0.643087}
\ppDeclareValue{post-eval2023-w-co-mentions-rho-ci-upper}{0.705994}
\ppDeclareValue{post-eval2023-w-co-mentions-loglik}{65202.747160}
\ppDeclareValue{post-eval2023-w-co-mentions-trading-days}{250}
\ppDeclareValue{post-eval2023-w-co-mentions-lambda}{2.088150}
\ppDeclareValue{post-eval2023-w-co-mentions-lambda-ci-lower}{1.801800}
\ppDeclareValue{post-eval2023-w-co-mentions-lambda-ci-upper}{2.401294}
\ppDeclareValue{post-eval2023-w-flat-rho}{0.844901}
\ppDeclareValue{post-eval2023-w-flat-rho-ci-lower}{0.826540}
\ppDeclareValue{post-eval2023-w-flat-rho-ci-upper}{0.860441}
\ppDeclareValue{post-eval2023-w-flat-loglik}{65710.680010}
\ppDeclareValue{post-eval2023-w-flat-trading-days}{250}
\ppDeclareValue{post-eval2023-w-flat-lambda}{5.447476}
\ppDeclareValue{post-eval2023-w-flat-lambda-ci-lower}{4.765022}
\ppDeclareValue{post-eval2023-w-flat-lambda-ci-upper}{6.165452}
\ppDeclareValue{post-eval2023-w-h-rho}{0.820919}
\ppDeclareValue{post-eval2023-w-h-rho-ci-lower}{0.796881}
\ppDeclareValue{post-eval2023-w-h-rho-ci-upper}{0.837668}
\ppDeclareValue{post-eval2023-w-h-loglik}{64282.817509}
\ppDeclareValue{post-eval2023-w-h-trading-days}{250}
\ppDeclareValue{post-eval2023-w-h-lambda}{4.584051}
\ppDeclareValue{post-eval2023-w-h-lambda-ci-lower}{3.923213}
\ppDeclareValue{post-eval2023-w-h-lambda-ci-upper}{5.160226}
\ppDeclareValue{post-eval2024-equal-support-rho}{0.824979}
\ppDeclareValue{post-eval2024-equal-support-rho-ci-lower}{0.800962}
\ppDeclareValue{post-eval2024-equal-support-rho-ci-upper}{0.844081}
\ppDeclareValue{post-eval2024-equal-support-loglik}{63315.614667}
\ppDeclareValue{post-eval2024-equal-support-trading-days}{252}
\ppDeclareValue{post-eval2024-equal-support-lambda}{4.713595}
\ppDeclareValue{post-eval2024-equal-support-lambda-ci-lower}{4.024161}
\ppDeclareValue{post-eval2024-equal-support-lambda-ci-upper}{5.413569}
\ppDeclareValue{post-eval2024-w-co-mentions-rho}{0.617399}
\ppDeclareValue{post-eval2024-w-co-mentions-rho-ci-lower}{0.578746}
\ppDeclareValue{post-eval2024-w-co-mentions-rho-ci-upper}{0.649489}
\ppDeclareValue{post-eval2024-w-co-mentions-loglik}{63603.283163}
\ppDeclareValue{post-eval2024-w-co-mentions-trading-days}{252}
\ppDeclareValue{post-eval2024-w-co-mentions-lambda}{1.613692}
\ppDeclareValue{post-eval2024-w-co-mentions-lambda-ci-lower}{1.373867}
\ppDeclareValue{post-eval2024-w-co-mentions-lambda-ci-upper}{1.852978}
\ppDeclareValue{post-eval2024-w-flat-rho}{0.819993}
\ppDeclareValue{post-eval2024-w-flat-rho-ci-lower}{0.792792}
\ppDeclareValue{post-eval2024-w-flat-rho-ci-upper}{0.841637}
\ppDeclareValue{post-eval2024-w-flat-loglik}{64134.200955}
\ppDeclareValue{post-eval2024-w-flat-trading-days}{252}
\ppDeclareValue{post-eval2024-w-flat-lambda}{4.555345}
\ppDeclareValue{post-eval2024-w-flat-lambda-ci-lower}{3.826075}
\ppDeclareValue{post-eval2024-w-flat-lambda-ci-upper}{5.314625}
\ppDeclareValue{post-eval2024-w-h-rho}{0.784373}
\ppDeclareValue{post-eval2024-w-h-rho-ci-lower}{0.757685}
\ppDeclareValue{post-eval2024-w-h-rho-ci-upper}{0.804731}
\ppDeclareValue{post-eval2024-w-h-loglik}{62661.154942}
\ppDeclareValue{post-eval2024-w-h-trading-days}{252}
\ppDeclareValue{post-eval2024-w-h-lambda}{3.637640}
\ppDeclareValue{post-eval2024-w-h-lambda-ci-lower}{3.126864}
\ppDeclareValue{post-eval2024-w-h-lambda-ci-upper}{4.121136}
\ppDeclareValue{post-eval2025-equal-support-rho}{0.852154}
\ppDeclareValue{post-eval2025-equal-support-rho-ci-lower}{0.783938}
\ppDeclareValue{post-eval2025-equal-support-rho-ci-upper}{0.890506}
\ppDeclareValue{post-eval2025-equal-support-loglik}{60104.412884}
\ppDeclareValue{post-eval2025-equal-support-trading-days}{249}
\ppDeclareValue{post-eval2025-equal-support-lambda}{5.763777}
\ppDeclareValue{post-eval2025-equal-support-lambda-ci-lower}{3.628310}
\ppDeclareValue{post-eval2025-equal-support-lambda-ci-upper}{8.132956}
\ppDeclareValue{post-eval2025-w-co-mentions-rho}{0.694275}
\ppDeclareValue{post-eval2025-w-co-mentions-rho-ci-lower}{0.581560}
\ppDeclareValue{post-eval2025-w-co-mentions-rho-ci-upper}{0.779272}
\ppDeclareValue{post-eval2025-w-co-mentions-loglik}{60508.066638}
\ppDeclareValue{post-eval2025-w-co-mentions-trading-days}{249}
\ppDeclareValue{post-eval2025-w-co-mentions-lambda}{2.270911}
\ppDeclareValue{post-eval2025-w-co-mentions-lambda-ci-lower}{1.389830}
\ppDeclareValue{post-eval2025-w-co-mentions-lambda-ci-upper}{3.530464}
\ppDeclareValue{post-eval2025-w-flat-rho}{0.852420}
\ppDeclareValue{post-eval2025-w-flat-rho-ci-lower}{0.780818}
\ppDeclareValue{post-eval2025-w-flat-rho-ci-upper}{0.896331}
\ppDeclareValue{post-eval2025-w-flat-loglik}{61027.771072}
\ppDeclareValue{post-eval2025-w-flat-trading-days}{249}
\ppDeclareValue{post-eval2025-w-flat-lambda}{5.775972}
\ppDeclareValue{post-eval2025-w-flat-lambda-ci-lower}{3.562411}
\ppDeclareValue{post-eval2025-w-flat-lambda-ci-upper}{8.646102}
\ppDeclareValue{post-eval2025-w-h-rho}{0.832306}
\ppDeclareValue{post-eval2025-w-h-rho-ci-lower}{0.750979}
\ppDeclareValue{post-eval2025-w-h-rho-ci-upper}{0.875397}
\ppDeclareValue{post-eval2025-w-h-loglik}{59505.603266}
\ppDeclareValue{post-eval2025-w-h-trading-days}{249}
\ppDeclareValue{post-eval2025-w-h-lambda}{4.963233}
\ppDeclareValue{post-eval2025-w-h-lambda-ci-lower}{3.015732}
\ppDeclareValue{post-eval2025-w-h-lambda-ci-upper}{7.025498}
\ppDeclareValue{post-eval2026-equal-support-rho}{0.822976}
\ppDeclareValue{post-eval2026-equal-support-rho-ci-lower}{0.795165}
\ppDeclareValue{post-eval2026-equal-support-rho-ci-upper}{0.849881}
\ppDeclareValue{post-eval2026-equal-support-loglik}{14597.506858}
\ppDeclareValue{post-eval2026-equal-support-trading-days}{64}
\ppDeclareValue{post-eval2026-equal-support-lambda}{4.648961}
\ppDeclareValue{post-eval2026-equal-support-lambda-ci-lower}{3.881990}
\ppDeclareValue{post-eval2026-equal-support-lambda-ci-upper}{5.661363}
\ppDeclareValue{post-eval2026-w-co-mentions-rho}{0.671319}
\ppDeclareValue{post-eval2026-w-co-mentions-rho-ci-lower}{0.610624}
\ppDeclareValue{post-eval2026-w-co-mentions-rho-ci-upper}{0.741044}
\ppDeclareValue{post-eval2026-w-co-mentions-loglik}{14859.870562}
\ppDeclareValue{post-eval2026-w-co-mentions-trading-days}{64}
\ppDeclareValue{post-eval2026-w-co-mentions-lambda}{2.042467}
\ppDeclareValue{post-eval2026-w-co-mentions-lambda-ci-lower}{1.568209}
\ppDeclareValue{post-eval2026-w-co-mentions-lambda-ci-upper}{2.861659}
\ppDeclareValue{post-eval2026-w-flat-rho}{0.856100}
\ppDeclareValue{post-eval2026-w-flat-rho-ci-lower}{0.821592}
\ppDeclareValue{post-eval2026-w-flat-rho-ci-upper}{0.889807}
\ppDeclareValue{post-eval2026-w-flat-loglik}{14943.097803}
\ppDeclareValue{post-eval2026-w-flat-trading-days}{64}
\ppDeclareValue{post-eval2026-w-flat-lambda}{5.949280}
\ppDeclareValue{post-eval2026-w-flat-lambda-ci-lower}{4.605129}
\ppDeclareValue{post-eval2026-w-flat-lambda-ci-upper}{8.075004}
\ppDeclareValue{post-eval2026-w-h-rho}{0.764726}
\ppDeclareValue{post-eval2026-w-h-rho-ci-lower}{0.729271}
\ppDeclareValue{post-eval2026-w-h-rho-ci-upper}{0.795506}
\ppDeclareValue{post-eval2026-w-h-loglik}{14414.304308}
\ppDeclareValue{post-eval2026-w-h-trading-days}{64}
\ppDeclareValue{post-eval2026-w-h-lambda}{3.250354}
\ppDeclareValue{post-eval2026-w-h-lambda-ci-lower}{2.693726}
\ppDeclareValue{post-eval2026-w-h-lambda-ci-upper}{3.890110}
\ppDeclareValue{post-pooled-2023-2026-equal-support-rho}{0.840574}
\ppDeclareValue{post-pooled-2023-2026-equal-support-rho-ci-lower}{0.813871}
\ppDeclareValue{post-pooled-2023-2026-equal-support-rho-ci-upper}{0.864629}
\ppDeclareValue{post-pooled-2023-2026-equal-support-loglik}{202124.561952}
\ppDeclareValue{post-pooled-2023-2026-equal-support-trading-days}{815}
\ppDeclareValue{post-pooled-2023-2026-equal-support-lambda}{5.272507}
\ppDeclareValue{post-pooled-2023-2026-equal-support-lambda-ci-lower}{4.372628}
\ppDeclareValue{post-pooled-2023-2026-equal-support-lambda-ci-upper}{6.387118}
\ppDeclareValue{post-pooled-2023-2026-w-co-mentions-rho}{0.667318}
\ppDeclareValue{post-pooled-2023-2026-w-co-mentions-rho-ci-lower}{0.618456}
\ppDeclareValue{post-pooled-2023-2026-w-co-mentions-rho-ci-upper}{0.716434}
\ppDeclareValue{post-pooled-2023-2026-w-co-mentions-loglik}{203493.833042}
\ppDeclareValue{post-pooled-2023-2026-w-co-mentions-trading-days}{815}
\ppDeclareValue{post-pooled-2023-2026-w-co-mentions-lambda}{2.005877}
\ppDeclareValue{post-pooled-2023-2026-w-co-mentions-lambda-ci-lower}{1.620932}
\ppDeclareValue{post-pooled-2023-2026-w-co-mentions-lambda-ci-upper}{2.526519}
\ppDeclareValue{post-pooled-2023-2026-w-flat-rho}{0.842849}
\ppDeclareValue{post-pooled-2023-2026-w-flat-rho-ci-lower}{0.814506}
\ppDeclareValue{post-pooled-2023-2026-w-flat-rho-ci-upper}{0.868627}
\ppDeclareValue{post-pooled-2023-2026-w-flat-loglik}{205146.310284}
\ppDeclareValue{post-pooled-2023-2026-w-flat-trading-days}{815}
\ppDeclareValue{post-pooled-2023-2026-w-flat-lambda}{5.363295}
\ppDeclareValue{post-pooled-2023-2026-w-flat-lambda-ci-lower}{4.391019}
\ppDeclareValue{post-pooled-2023-2026-w-flat-lambda-ci-upper}{6.611919}
\ppDeclareValue{post-pooled-2023-2026-w-h-rho}{0.812335}
\ppDeclareValue{post-pooled-2023-2026-w-h-rho-ci-lower}{0.779518}
\ppDeclareValue{post-pooled-2023-2026-w-h-rho-ci-upper}{0.842240}
\ppDeclareValue{post-pooled-2023-2026-w-h-loglik}{200099.699399}
\ppDeclareValue{post-pooled-2023-2026-w-h-trading-days}{815}
\ppDeclareValue{post-pooled-2023-2026-w-h-lambda}{4.328634}
\ppDeclareValue{post-pooled-2023-2026-w-h-lambda-ci-lower}{3.535517}
\ppDeclareValue{post-pooled-2023-2026-w-h-lambda-ci-upper}{5.338730}

\ppDeclareValue{transport-calibration-date-start}{2018-03-19}
\ppDeclareValue{transport-calibration-date-end}{2022-12-30}
\ppDeclareValue{transport-calibration-trading-days}{1207}
\ppDeclareValue{transport-evaluation-date-start}{2023-01-03}
\ppDeclareValue{transport-evaluation-date-end}{2026-04-06}
\ppDeclareValue{transport-evaluation-trading-days}{815}
\ppDeclareValue{transport-n-assets}{100}
\ppDeclareValue{transport-w1-rho}{0.683673}
\ppDeclareValue{transport-w1-rho-ci-lower}{0.654985}
\ppDeclareValue{transport-w1-rho-ci-upper}{0.708792}
\ppDeclareValue{transport-w1-lambda}{2.161287}
\ppDeclareValue{transport-w1-lambda-ci-lower}{1.898427}
\ppDeclareValue{transport-w1-lambda-ci-upper}{2.433966}
\ppDeclareValue{transport-w1-loglik}{209488.342184}
\ppDeclareValue{transport-w1-loglik-per-firm-day}{2.570409}
\ppDeclareValue{transport-w1-unadjusted-rho}{0.840563}
\ppDeclareValue{transport-w1-factor-conditioning-rho-change}{-0.156890}
\ppDeclareValue{transport-w2-rho}{0.686596}
\ppDeclareValue{transport-w2-rho-ci-lower}{0.658084}
\ppDeclareValue{transport-w2-rho-ci-upper}{0.711631}
\ppDeclareValue{transport-w2-lambda}{2.190765}
\ppDeclareValue{transport-w2-lambda-ci-lower}{1.924696}
\ppDeclareValue{transport-w2-lambda-ci-upper}{2.467774}
\ppDeclareValue{transport-w2-loglik}{209476.031412}
\ppDeclareValue{transport-w2-loglik-per-firm-day}{2.570258}
\ppDeclareValue{transport-w2-unadjusted-rho}{0.842849}
\ppDeclareValue{transport-w2-factor-conditioning-rho-change}{-0.156253}
\ppDeclareValue{transport-n-factors}{6}
\ppDeclareValue{transport-delta-rho}{0.002922}
\ppDeclareValue{transport-delta-rho-ci-lower}{0.002266}
\ppDeclareValue{transport-delta-rho-ci-upper}{0.003582}
\ppDeclareValue{transport-delta-loglik}{-12.310771}
\ppDeclareValue{transport-delta-loglik-ci-lower}{-22.563834}
\ppDeclareValue{transport-delta-loglik-ci-upper}{-2.779265}
\ppDeclareValue{transport-w1-predictive-score}{2.526065}
\ppDeclareValue{transport-w2-predictive-score}{2.525957}
\ppDeclareValue{transport-delta-predictive-score}{-0.000109}
\ppDeclareValue{transport-delta-predictive-score-ci-lower}{-0.000315}
\ppDeclareValue{transport-delta-predictive-score-ci-upper}{0.000105}
\ppDeclareValue{transport-equivalence-margin}{0.000495}
\ppDeclareValue{transport-equivalence-ci-lower}{-0.000282}
\ppDeclareValue{transport-equivalence-ci-upper}{0.000066}
\ppDeclareValue{transport-equivalence-status}{equivalent}
\ppDeclareValue{transport-weight-correlation}{0.997944}
\ppDeclareValue{transport-active-support-jaccard}{0.938057}
\ppDeclareValue{transport-signal-correlation}{0.999775}
\ppDeclareValue{transport-relative-signal-difference-sd}{0.022043}
\ppDeclareValue{transport-factor-residualized-signal-correlation}{0.999212}
\ppDeclareValue{transport-factor-residualized-relative-signal-difference-sd}{0.040903}
\ppDeclareValue{transport-minimum-row-correlation}{0.993498}
\ppDeclareValue{transport-weakly-separated}{true}

\ppDeclareValue{transport-joint-w1-rho}{0.683673}
\ppDeclareValue{transport-joint-w2-rho}{0.000000}
\ppDeclareValue{transport-joint-rho-total}{0.683673}
\ppDeclareValue{transport-joint-theta-w1}{1.000000}
\ppDeclareValue{transport-joint-loglik}{209488.342184}
\ppDeclareValue{transport-joint-loglik-per-firm-day}{2.570409}
\ppDeclareValue{transport-joint-loglik-gain-w1}{0.000000}
\ppDeclareValue{transport-joint-loglik-gain-w2}{12.310771}
\ppDeclareValue{transport-joint-channel-correlation}{-0.928860}
\ppDeclareValue{transport-joint-channel-correlation-status}{estimated}
\ppDeclareValue{transport-joint-w1-only-boundary-share}{0.961000}
\ppDeclareValue{transport-joint-w2-only-boundary-share}{0.000000}
\ppDeclareValue{transport-joint-rho-w1-ci-lower}{0.611772}
\ppDeclareValue{transport-joint-rho-w1-ci-upper}{0.708781}
\ppDeclareValue{transport-joint-rho-w2-ci-lower}{0.000000}
\ppDeclareValue{transport-joint-rho-w2-ci-upper}{0.058233}
\ppDeclareValue{transport-joint-rho-total-ci-lower}{0.654266}
\ppDeclareValue{transport-joint-rho-total-ci-upper}{0.709059}
\ppDeclareValue{transport-joint-theta-w1-ci-lower}{0.915324}
\ppDeclareValue{transport-joint-theta-w1-ci-upper}{1.000000}
\ppDeclareValue{transport-joint-qlr-w2-given-w1-statistic}{0.000000}
\ppDeclareValue{transport-joint-qlr-w2-given-w1-pvalue}{1.000000}
\ppDeclareValue{transport-joint-qlr-w2-given-w1-exceedances}{2000}
\ppDeclareValue{transport-joint-qlr-w2-given-w1-decision}{do\_not\_reject}
\ppDeclareValue{transport-joint-qlr-w1-given-w2-statistic}{24.621543}
\ppDeclareValue{transport-joint-qlr-w1-given-w2-pvalue}{0.000500}
\ppDeclareValue{transport-joint-qlr-w1-given-w2-exceedances}{0}
\ppDeclareValue{transport-joint-qlr-w1-given-w2-decision}{reject}
\ppDeclareValue{transport-joint-incremental-conclusion}{w1\_adds\_beyond\_w2\_but\_not\_reverse}
\ppDeclareValue{transport-joint-follow-up-status}{results-aware diagnostic added after the standalone W1/W2 comparison}

\ppDeclareValue{transport-factor-ff3-n-factors}{3}
\ppDeclareValue{transport-factor-ff3-w1-rho}{0.699439}
\ppDeclareValue{transport-factor-ff3-w1-rho-ci-lower}{0.672530}
\ppDeclareValue{transport-factor-ff3-w1-rho-ci-upper}{0.727102}
\ppDeclareValue{transport-factor-ff3-w1-loglik}{208275.409195}
\ppDeclareValue{transport-factor-ff3-w1-loglik-per-firm-day}{2.555526}
\ppDeclareValue{transport-factor-ff3-w1-predictive-score}{2.531221}
\ppDeclareValue{transport-factor-ff3-w2-rho}{0.702358}
\ppDeclareValue{transport-factor-ff3-w2-rho-ci-lower}{0.675461}
\ppDeclareValue{transport-factor-ff3-w2-rho-ci-upper}{0.730013}
\ppDeclareValue{transport-factor-ff3-w2-loglik}{208265.732868}
\ppDeclareValue{transport-factor-ff3-w2-loglik-per-firm-day}{2.555408}
\ppDeclareValue{transport-factor-ff3-w2-predictive-score}{2.531151}
\ppDeclareValue{transport-factor-ff3-delta-rho}{0.002919}
\ppDeclareValue{transport-factor-ff3-delta-predictive-score}{-0.000070}
\ppDeclareValue{transport-factor-ff3-mom-n-factors}{4}
\ppDeclareValue{transport-factor-ff3-mom-w1-rho}{0.689639}
\ppDeclareValue{transport-factor-ff3-mom-w1-rho-ci-lower}{0.661626}
\ppDeclareValue{transport-factor-ff3-mom-w1-rho-ci-upper}{0.715005}
\ppDeclareValue{transport-factor-ff3-mom-w1-loglik}{208917.109905}
\ppDeclareValue{transport-factor-ff3-mom-w1-loglik-per-firm-day}{2.563400}
\ppDeclareValue{transport-factor-ff3-mom-w1-predictive-score}{2.529959}
\ppDeclareValue{transport-factor-ff3-mom-w2-rho}{0.692483}
\ppDeclareValue{transport-factor-ff3-mom-w2-rho-ci-lower}{0.664338}
\ppDeclareValue{transport-factor-ff3-mom-w2-rho-ci-upper}{0.717859}
\ppDeclareValue{transport-factor-ff3-mom-w2-loglik}{208904.793693}
\ppDeclareValue{transport-factor-ff3-mom-w2-loglik-per-firm-day}{2.563249}
\ppDeclareValue{transport-factor-ff3-mom-w2-predictive-score}{2.529888}
\ppDeclareValue{transport-factor-ff3-mom-delta-rho}{0.002844}
\ppDeclareValue{transport-factor-ff3-mom-delta-predictive-score}{-0.000071}
\ppDeclareValue{transport-factor-ff5-n-factors}{5}
\ppDeclareValue{transport-factor-ff5-w1-rho}{0.693998}
\ppDeclareValue{transport-factor-ff5-w1-rho-ci-lower}{0.665763}
\ppDeclareValue{transport-factor-ff5-w1-rho-ci-upper}{0.721858}
\ppDeclareValue{transport-factor-ff5-w1-loglik}{208862.422026}
\ppDeclareValue{transport-factor-ff5-w1-loglik-per-firm-day}{2.562729}
\ppDeclareValue{transport-factor-ff5-w1-predictive-score}{2.527230}
\ppDeclareValue{transport-factor-ff5-w2-rho}{0.697010}
\ppDeclareValue{transport-factor-ff5-w2-rho-ci-lower}{0.668700}
\ppDeclareValue{transport-factor-ff5-w2-rho-ci-upper}{0.725111}
\ppDeclareValue{transport-factor-ff5-w2-loglik}{208852.921614}
\ppDeclareValue{transport-factor-ff5-w2-loglik-per-firm-day}{2.562613}
\ppDeclareValue{transport-factor-ff5-w2-predictive-score}{2.527099}
\ppDeclareValue{transport-factor-ff5-delta-rho}{0.003012}
\ppDeclareValue{transport-factor-ff5-delta-predictive-score}{-0.000131}
\ppDeclareValue{transport-factor-ff5-mom-n-factors}{6}
\ppDeclareValue{transport-factor-ff5-mom-w1-rho}{0.683673}
\ppDeclareValue{transport-factor-ff5-mom-w1-rho-ci-lower}{0.654985}
\ppDeclareValue{transport-factor-ff5-mom-w1-rho-ci-upper}{0.708792}
\ppDeclareValue{transport-factor-ff5-mom-w1-loglik}{209488.342184}
\ppDeclareValue{transport-factor-ff5-mom-w1-loglik-per-firm-day}{2.570409}
\ppDeclareValue{transport-factor-ff5-mom-w1-predictive-score}{2.526065}
\ppDeclareValue{transport-factor-ff5-mom-w2-rho}{0.686596}
\ppDeclareValue{transport-factor-ff5-mom-w2-rho-ci-lower}{0.658084}
\ppDeclareValue{transport-factor-ff5-mom-w2-rho-ci-upper}{0.711631}
\ppDeclareValue{transport-factor-ff5-mom-w2-loglik}{209476.031412}
\ppDeclareValue{transport-factor-ff5-mom-w2-loglik-per-firm-day}{2.570258}
\ppDeclareValue{transport-factor-ff5-mom-w2-predictive-score}{2.525957}
\ppDeclareValue{transport-factor-ff5-mom-delta-rho}{0.002922}
\ppDeclareValue{transport-factor-ff5-mom-delta-predictive-score}{-0.000109}

\ppDeclareValue{spatial-value-factor-ff3-n-factors}{3}
\ppDeclareValue{spatial-value-factor-ff3-parameter-count}{402}
\ppDeclareValue{spatial-value-factor-ff3-loglik}{208265.732868}
\ppDeclareValue{spatial-value-factor-ff3-aic}{-415727.465736}
\ppDeclareValue{spatial-value-factor-ff3-hqc-dates}{-415001.788803}
\ppDeclareValue{spatial-value-factor-ff3-hqc-firm-days}{-414581.329868}
\ppDeclareValue{spatial-value-factor-ff3-bic-dates}{-413836.784115}
\ppDeclareValue{spatial-value-factor-ff3-bic-firm-days}{-411985.505700}
\ppDeclareValue{spatial-value-factor-ff3-mom-n-factors}{4}
\ppDeclareValue{spatial-value-factor-ff3-mom-parameter-count}{502}
\ppDeclareValue{spatial-value-factor-ff3-mom-loglik}{208904.793693}
\ppDeclareValue{spatial-value-factor-ff3-mom-aic}{-416805.587387}
\ppDeclareValue{spatial-value-factor-ff3-mom-hqc-dates}{-415899.393803}
\ppDeclareValue{spatial-value-factor-ff3-mom-hqc-firm-days}{-415374.343093}
\ppDeclareValue{spatial-value-factor-ff3-mom-bic-dates}{-414444.586954}
\ppDeclareValue{spatial-value-factor-ff3-mom-bic-firm-days}{-412132.791520}
\ppDeclareValue{spatial-value-factor-ff5-n-factors}{5}
\ppDeclareValue{spatial-value-factor-ff5-parameter-count}{602}
\ppDeclareValue{spatial-value-factor-ff5-loglik}{208852.921614}
\ppDeclareValue{spatial-value-factor-ff5-aic}{-416501.843228}
\ppDeclareValue{spatial-value-factor-ff5-hqc-dates}{-415415.132994}
\ppDeclareValue{spatial-value-factor-ff5-hqc-firm-days}{-414785.490510}
\ppDeclareValue{spatial-value-factor-ff5-bic-dates}{-413670.523984}
\ppDeclareValue{spatial-value-factor-ff5-bic-firm-days}{-410898.211532}
\ppDeclareValue{spatial-value-factor-ff5-mom-n-factors}{6}
\ppDeclareValue{spatial-value-factor-ff5-mom-parameter-count}{702}
\ppDeclareValue{spatial-value-factor-ff5-mom-loglik}{209476.031412}
\ppDeclareValue{spatial-value-factor-ff5-mom-aic}{-417548.062825}
\ppDeclareValue{spatial-value-factor-ff5-mom-hqc-dates}{-416280.835941}
\ppDeclareValue{spatial-value-factor-ff5-mom-hqc-firm-days}{-415546.601682}
\ppDeclareValue{spatial-value-factor-ff5-mom-bic-dates}{-414246.424769}
\ppDeclareValue{spatial-value-factor-ff5-mom-bic-firm-days}{-411013.595299}
\ppDeclareValue{spatial-value-spherical-none-n-factors}{6}
\ppDeclareValue{spatial-value-spherical-none-parameter-count}{701}
\ppDeclareValue{spatial-value-spherical-none-loglik}{205748.307526}
\ppDeclareValue{spatial-value-spherical-none-aic}{-410094.615052}
\ppDeclareValue{spatial-value-spherical-none-hqc-dates}{-408829.193334}
\ppDeclareValue{spatial-value-spherical-none-hqc-firm-days}{-408096.004993}
\ppDeclareValue{spatial-value-spherical-none-bic-dates}{-406797.680185}
\ppDeclareValue{spatial-value-spherical-none-bic-firm-days}{-403569.455884}
\ppDeclareValue{spatial-value-spherical-none-rho}{0.000000}
\ppDeclareValue{spatial-value-spherical-none-rho-ci-lower}{0.000000}
\ppDeclareValue{spatial-value-spherical-none-rho-ci-upper}{0.000000}
\ppDeclareValue{spatial-value-spherical-none-held-out-score}{2.468881}
\ppDeclareValue{spatial-value-spherical-w1-n-factors}{6}
\ppDeclareValue{spatial-value-spherical-w1-parameter-count}{702}
\ppDeclareValue{spatial-value-spherical-w1-loglik}{209488.342184}
\ppDeclareValue{spatial-value-spherical-w1-aic}{-417572.684367}
\ppDeclareValue{spatial-value-spherical-w1-hqc-dates}{-416305.457483}
\ppDeclareValue{spatial-value-spherical-w1-hqc-firm-days}{-415571.223224}
\ppDeclareValue{spatial-value-spherical-w1-bic-dates}{-414271.046312}
\ppDeclareValue{spatial-value-spherical-w1-bic-firm-days}{-411038.216841}
\ppDeclareValue{spatial-value-spherical-w1-rho}{0.683673}
\ppDeclareValue{spatial-value-spherical-w1-rho-ci-lower}{0.654985}
\ppDeclareValue{spatial-value-spherical-w1-rho-ci-upper}{0.708792}
\ppDeclareValue{spatial-value-spherical-w1-held-out-score}{2.526065}
\ppDeclareValue{spatial-value-spherical-w2-n-factors}{6}
\ppDeclareValue{spatial-value-spherical-w2-parameter-count}{702}
\ppDeclareValue{spatial-value-spherical-w2-loglik}{209476.031412}
\ppDeclareValue{spatial-value-spherical-w2-aic}{-417548.062825}
\ppDeclareValue{spatial-value-spherical-w2-hqc-dates}{-416280.835941}
\ppDeclareValue{spatial-value-spherical-w2-hqc-firm-days}{-415546.601682}
\ppDeclareValue{spatial-value-spherical-w2-bic-dates}{-414246.424769}
\ppDeclareValue{spatial-value-spherical-w2-bic-firm-days}{-411013.595299}
\ppDeclareValue{spatial-value-spherical-w2-rho}{0.686596}
\ppDeclareValue{spatial-value-spherical-w2-rho-ci-lower}{0.658084}
\ppDeclareValue{spatial-value-spherical-w2-rho-ci-upper}{0.711631}
\ppDeclareValue{spatial-value-spherical-w2-held-out-score}{2.525957}
\ppDeclareValue{spatial-value-diagonal-none-n-factors}{6}
\ppDeclareValue{spatial-value-diagonal-none-parameter-count}{800}
\ppDeclareValue{spatial-value-diagonal-none-loglik}{212593.999475}
\ppDeclareValue{spatial-value-diagonal-none-aic}{-423587.998950}
\ppDeclareValue{spatial-value-diagonal-none-hqc-dates}{-422143.865748}
\ppDeclareValue{spatial-value-diagonal-none-hqc-firm-days}{-421307.131550}
\ppDeclareValue{spatial-value-diagonal-none-bic-dates}{-419825.448459}
\ppDeclareValue{spatial-value-diagonal-none-bic-firm-days}{-416141.312310}
\ppDeclareValue{spatial-value-diagonal-none-rho}{0.000000}
\ppDeclareValue{spatial-value-diagonal-none-rho-ci-lower}{0.000000}
\ppDeclareValue{spatial-value-diagonal-none-rho-ci-upper}{0.000000}
\ppDeclareValue{spatial-value-diagonal-none-held-out-score}{2.486224}
\ppDeclareValue{spatial-value-diagonal-w1-n-factors}{6}
\ppDeclareValue{spatial-value-diagonal-w1-parameter-count}{801}
\ppDeclareValue{spatial-value-diagonal-w1-loglik}{216943.536756}
\ppDeclareValue{spatial-value-diagonal-w1-aic}{-432285.073513}
\ppDeclareValue{spatial-value-diagonal-w1-hqc-dates}{-430839.135145}
\ppDeclareValue{spatial-value-diagonal-w1-hqc-firm-days}{-430001.355029}
\ppDeclareValue{spatial-value-diagonal-w1-bic-dates}{-428517.819834}
\ppDeclareValue{spatial-value-diagonal-w1-bic-firm-days}{-424829.078515}
\ppDeclareValue{spatial-value-diagonal-w1-rho}{0.706081}
\ppDeclareValue{spatial-value-diagonal-w1-rho-ci-lower}{0.677855}
\ppDeclareValue{spatial-value-diagonal-w1-rho-ci-upper}{0.729855}
\ppDeclareValue{spatial-value-diagonal-w1-held-out-score}{2.552188}
\ppDeclareValue{spatial-value-diagonal-w2-n-factors}{6}
\ppDeclareValue{spatial-value-diagonal-w2-parameter-count}{801}
\ppDeclareValue{spatial-value-diagonal-w2-loglik}{216916.030007}
\ppDeclareValue{spatial-value-diagonal-w2-aic}{-432230.060014}
\ppDeclareValue{spatial-value-diagonal-w2-hqc-dates}{-430784.121645}
\ppDeclareValue{spatial-value-diagonal-w2-hqc-firm-days}{-429946.341530}
\ppDeclareValue{spatial-value-diagonal-w2-bic-dates}{-428462.806335}
\ppDeclareValue{spatial-value-diagonal-w2-bic-firm-days}{-424774.065016}
\ppDeclareValue{spatial-value-diagonal-w2-rho}{0.707906}
\ppDeclareValue{spatial-value-diagonal-w2-rho-ci-lower}{0.679852}
\ppDeclareValue{spatial-value-diagonal-w2-rho-ci-upper}{0.731353}
\ppDeclareValue{spatial-value-diagonal-w2-held-out-score}{2.551743}
\ppDeclareValue{spatial-value-spherical-w1-versus-none-loglik-gain}{3740.034658}
\ppDeclareValue{spatial-value-spherical-w1-versus-none-qlr-statistic}{7480.069315}
\ppDeclareValue{spatial-value-spherical-w1-versus-none-qlr-pvalue}{0.000500}
\ppDeclareValue{spatial-value-spherical-w1-versus-none-held-out-score-gain}{0.057184}
\ppDeclareValue{spatial-value-spherical-w1-versus-none-held-out-score-gain-ci-lower}{0.049738}
\ppDeclareValue{spatial-value-spherical-w1-versus-none-held-out-score-gain-ci-upper}{0.065565}
\ppDeclareValue{spatial-value-spherical-w1-versus-none-qlr-decision}{reject\_rho\_zero}
\ppDeclareValue{spatial-value-spherical-w1-versus-none-held-out-decision}{improves}
\ppDeclareValue{spatial-value-spherical-w2-versus-none-loglik-gain}{3727.723886}
\ppDeclareValue{spatial-value-spherical-w2-versus-none-qlr-statistic}{7455.447773}
\ppDeclareValue{spatial-value-spherical-w2-versus-none-qlr-pvalue}{0.000500}
\ppDeclareValue{spatial-value-spherical-w2-versus-none-held-out-score-gain}{0.057076}
\ppDeclareValue{spatial-value-spherical-w2-versus-none-held-out-score-gain-ci-lower}{0.049624}
\ppDeclareValue{spatial-value-spherical-w2-versus-none-held-out-score-gain-ci-upper}{0.065259}
\ppDeclareValue{spatial-value-spherical-w2-versus-none-qlr-decision}{reject\_rho\_zero}
\ppDeclareValue{spatial-value-spherical-w2-versus-none-held-out-decision}{improves}
\ppDeclareValue{spatial-value-diagonal-w1-versus-none-loglik-gain}{4349.537282}
\ppDeclareValue{spatial-value-diagonal-w1-versus-none-qlr-statistic}{8699.074563}
\ppDeclareValue{spatial-value-diagonal-w1-versus-none-qlr-pvalue}{0.000500}
\ppDeclareValue{spatial-value-diagonal-w1-versus-none-held-out-score-gain}{0.065964}
\ppDeclareValue{spatial-value-diagonal-w1-versus-none-held-out-score-gain-ci-lower}{0.055795}
\ppDeclareValue{spatial-value-diagonal-w1-versus-none-held-out-score-gain-ci-upper}{0.075885}
\ppDeclareValue{spatial-value-diagonal-w1-versus-none-qlr-decision}{reject\_rho\_zero}
\ppDeclareValue{spatial-value-diagonal-w1-versus-none-held-out-decision}{improves}
\ppDeclareValue{spatial-value-diagonal-w2-versus-none-loglik-gain}{4322.030532}
\ppDeclareValue{spatial-value-diagonal-w2-versus-none-qlr-statistic}{8644.061064}
\ppDeclareValue{spatial-value-diagonal-w2-versus-none-qlr-pvalue}{0.000500}
\ppDeclareValue{spatial-value-diagonal-w2-versus-none-held-out-score-gain}{0.065519}
\ppDeclareValue{spatial-value-diagonal-w2-versus-none-held-out-score-gain-ci-lower}{0.055425}
\ppDeclareValue{spatial-value-diagonal-w2-versus-none-held-out-score-gain-ci-upper}{0.075410}
\ppDeclareValue{spatial-value-diagonal-w2-versus-none-qlr-decision}{reject\_rho\_zero}
\ppDeclareValue{spatial-value-diagonal-w2-versus-none-held-out-decision}{improves}
\ppDeclareValue{spatial-value-identity-spherical-w1-coefficient-error}{0.000000000000}
\ppDeclareValue{spatial-value-identity-spherical-w1-fitted-mean-error}{0.000000000000}
\ppDeclareValue{spatial-value-identity-spherical-w2-coefficient-error}{0.000000000000}
\ppDeclareValue{spatial-value-identity-spherical-w2-fitted-mean-error}{0.000000000000}
\ppDeclareValue{spatial-value-identity-diagonal-w1-coefficient-error}{0.000000000000}
\ppDeclareValue{spatial-value-identity-diagonal-w1-fitted-mean-error}{0.000000000000}
\ppDeclareValue{spatial-value-identity-diagonal-w2-coefficient-error}{0.000000000000}
\ppDeclareValue{spatial-value-identity-diagonal-w2-fitted-mean-error}{0.000000000000}

\ppDeclareValue{post-pooled-2023-2026-factor-two-field-rho-b}{0.423956}
\ppDeclareValue{post-pooled-2023-2026-factor-two-field-rho-n}{0.237126}
\ppDeclareValue{post-pooled-2023-2026-factor-two-field-rho-total}{0.661082}
\ppDeclareValue{post-pooled-2023-2026-factor-two-field-theta}{0.641306}
\ppDeclareValue{post-pooled-2023-2026-factor-two-field-lambda-b}{1.250908}
\ppDeclareValue{post-pooled-2023-2026-factor-two-field-lambda-n}{0.699656}
\ppDeclareValue{post-pooled-2023-2026-factor-two-field-loglik}{209882.716226}
\ppDeclareValue{post-pooled-2023-2026-factor-two-field-trading-days}{815}
\ppDeclareValue{post-pooled-2023-2026-factor-two-field-loglik-gain-w-flat}{406.684814}
\ppDeclareValue{post-pooled-2023-2026-factor-two-field-loglik-gain-co-mentions}{642.439072}
\ppDeclareValue{post-pooled-2023-2026-factor-two-field-channel-correlation}{-0.510364}
\ppDeclareValue{post-pooled-2023-2026-factor-two-field-boundary-share-b}{0.000000}
\ppDeclareValue{post-pooled-2023-2026-factor-two-field-boundary-share-n}{0.000000}
\ppDeclareValue{post-pooled-2023-2026-factor-two-field-rho-b-ci-lower}{0.392873}
\ppDeclareValue{post-pooled-2023-2026-factor-two-field-rho-b-ci-upper}{0.451404}
\ppDeclareValue{post-pooled-2023-2026-factor-two-field-rho-n-ci-lower}{0.212209}
\ppDeclareValue{post-pooled-2023-2026-factor-two-field-rho-n-ci-upper}{0.260382}
\ppDeclareValue{post-pooled-2023-2026-factor-two-field-rho-total-ci-lower}{0.631682}
\ppDeclareValue{post-pooled-2023-2026-factor-two-field-rho-total-ci-upper}{0.685179}
\ppDeclareValue{post-pooled-2023-2026-factor-two-field-lambda-b-ci-lower}{1.078046}
\ppDeclareValue{post-pooled-2023-2026-factor-two-field-lambda-b-ci-upper}{1.409495}
\ppDeclareValue{post-pooled-2023-2026-factor-two-field-lambda-n-ci-lower}{0.596522}
\ppDeclareValue{post-pooled-2023-2026-factor-two-field-lambda-n-ci-upper}{0.797123}
\ppDeclareValue{post-pooled-2023-2026-factor-two-field-theta-ci-lower}{0.607152}
\ppDeclareValue{post-pooled-2023-2026-factor-two-field-theta-ci-upper}{0.676793}
\ppDeclareValue{post-pooled-2023-2026-factor-two-field-qlr-news-given-barycentric-statistic}{813.369628}
\ppDeclareValue{post-pooled-2023-2026-factor-two-field-qlr-news-given-barycentric-pvalue}{0.000500}
\ppDeclareValue{post-pooled-2023-2026-factor-two-field-qlr-news-given-barycentric-exceedances}{0}
\ppDeclareValue{post-pooled-2023-2026-factor-two-field-qlr-barycentric-given-news-statistic}{1284.878144}
\ppDeclareValue{post-pooled-2023-2026-factor-two-field-qlr-barycentric-given-news-pvalue}{0.000500}
\ppDeclareValue{post-pooled-2023-2026-factor-two-field-qlr-barycentric-given-news-exceedances}{0}
\ppDeclareValue{post-pooled-2023-2026-factor-two-field-held-out-score-factor-only}{2.468881}
\ppDeclareValue{post-pooled-2023-2026-factor-two-field-held-out-score-barycentric}{2.525957}
\ppDeclareValue{post-pooled-2023-2026-factor-two-field-held-out-score-news}{2.519320}
\ppDeclareValue{post-pooled-2023-2026-factor-two-field-held-out-score-joint}{2.530229}
\ppDeclareValue{post-pooled-2023-2026-factor-two-field-held-out-joint-minus-factor-only-estimate}{0.061348}
\ppDeclareValue{post-pooled-2023-2026-factor-two-field-held-out-joint-minus-factor-only-ci-lower}{0.052994}
\ppDeclareValue{post-pooled-2023-2026-factor-two-field-held-out-joint-minus-factor-only-ci-upper}{0.070661}
\ppDeclareValue{post-pooled-2023-2026-factor-two-field-held-out-joint-minus-barycentric-estimate}{0.004272}
\ppDeclareValue{post-pooled-2023-2026-factor-two-field-held-out-joint-minus-barycentric-ci-lower}{0.002419}
\ppDeclareValue{post-pooled-2023-2026-factor-two-field-held-out-joint-minus-barycentric-ci-upper}{0.005857}
\ppDeclareValue{post-pooled-2023-2026-factor-two-field-held-out-joint-minus-news-estimate}{0.010909}
\ppDeclareValue{post-pooled-2023-2026-factor-two-field-held-out-joint-minus-news-ci-lower}{0.009318}
\ppDeclareValue{post-pooled-2023-2026-factor-two-field-held-out-joint-minus-news-ci-upper}{0.012724}

\ppDeclareValue{post-eval2023-two-field-rho-b}{0.604862}
\ppDeclareValue{post-eval2023-two-field-rho-n}{0.228782}
\ppDeclareValue{post-eval2023-two-field-rho-total}{0.833645}
\ppDeclareValue{post-eval2023-two-field-theta}{0.725564}
\ppDeclareValue{post-eval2023-two-field-lambda-b}{3.635963}
\ppDeclareValue{post-eval2023-two-field-lambda-n}{1.375261}
\ppDeclareValue{post-eval2023-two-field-loglik}{65832.546347}
\ppDeclareValue{post-eval2023-two-field-trading-days}{250}
\ppDeclareValue{post-eval2023-two-field-loglik-gain-w-flat}{121.866337}
\ppDeclareValue{post-eval2023-two-field-loglik-gain-co-mentions}{629.799187}
\ppDeclareValue{post-eval2023-two-field-boundary-w-flat-rho}{0.844900}
\ppDeclareValue{post-eval2023-two-field-boundary-w-flat-rho-ci-lower}{0.826540}
\ppDeclareValue{post-eval2023-two-field-boundary-w-flat-rho-ci-upper}{0.860442}
\ppDeclareValue{post-eval2023-two-field-boundary-w-flat-lambda}{5.447471}
\ppDeclareValue{post-eval2023-two-field-boundary-w-flat-lambda-ci-lower}{4.765025}
\ppDeclareValue{post-eval2023-two-field-boundary-w-flat-lambda-ci-upper}{6.165457}
\ppDeclareValue{post-eval2023-two-field-boundary-co-mentions-rho}{0.676181}
\ppDeclareValue{post-eval2023-two-field-boundary-co-mentions-rho-ci-lower}{0.643087}
\ppDeclareValue{post-eval2023-two-field-boundary-co-mentions-rho-ci-upper}{0.705994}
\ppDeclareValue{post-eval2023-two-field-boundary-co-mentions-lambda}{2.088150}
\ppDeclareValue{post-eval2023-two-field-boundary-co-mentions-lambda-ci-lower}{1.801800}
\ppDeclareValue{post-eval2023-two-field-boundary-co-mentions-lambda-ci-upper}{2.401295}
\ppDeclareValue{post-eval2023-two-field-channel-correlation}{-0.915201}
\ppDeclareValue{post-eval2023-two-field-boundary-share-b}{0.000000}
\ppDeclareValue{post-eval2023-two-field-boundary-share-n}{0.000000}
\ppDeclareValue{post-eval2023-two-field-rho-b-ci-lower}{0.571949}
\ppDeclareValue{post-eval2023-two-field-rho-b-ci-upper}{0.637920}
\ppDeclareValue{post-eval2023-two-field-rho-n-ci-lower}{0.188580}
\ppDeclareValue{post-eval2023-two-field-rho-n-ci-upper}{0.269696}
\ppDeclareValue{post-eval2023-two-field-rho-total-ci-lower}{0.814761}
\ppDeclareValue{post-eval2023-two-field-rho-total-ci-upper}{0.847909}
\ppDeclareValue{post-eval2023-two-field-theta-ci-lower}{0.680434}
\ppDeclareValue{post-eval2023-two-field-theta-ci-upper}{0.771172}
\ppDeclareValue{post-eval2023-two-field-lambda-b-ci-lower}{3.248447}
\ppDeclareValue{post-eval2023-two-field-lambda-b-ci-upper}{3.986797}
\ppDeclareValue{post-eval2023-two-field-lambda-n-ci-lower}{1.054338}
\ppDeclareValue{post-eval2023-two-field-lambda-n-ci-upper}{1.749240}
\ppDeclareValue{post-eval2024-two-field-rho-b}{0.623536}
\ppDeclareValue{post-eval2024-two-field-rho-n}{0.183779}
\ppDeclareValue{post-eval2024-two-field-rho-total}{0.807315}
\ppDeclareValue{post-eval2024-two-field-theta}{0.772358}
\ppDeclareValue{post-eval2024-two-field-lambda-b}{3.236030}
\ppDeclareValue{post-eval2024-two-field-lambda-n}{0.953776}
\ppDeclareValue{post-eval2024-two-field-loglik}{64213.679591}
\ppDeclareValue{post-eval2024-two-field-trading-days}{252}
\ppDeclareValue{post-eval2024-two-field-loglik-gain-w-flat}{79.478636}
\ppDeclareValue{post-eval2024-two-field-loglik-gain-co-mentions}{610.396429}
\ppDeclareValue{post-eval2024-two-field-boundary-w-flat-rho}{0.819993}
\ppDeclareValue{post-eval2024-two-field-boundary-w-flat-rho-ci-lower}{0.792792}
\ppDeclareValue{post-eval2024-two-field-boundary-w-flat-rho-ci-upper}{0.841637}
\ppDeclareValue{post-eval2024-two-field-boundary-w-flat-lambda}{4.555343}
\ppDeclareValue{post-eval2024-two-field-boundary-w-flat-lambda-ci-lower}{3.826073}
\ppDeclareValue{post-eval2024-two-field-boundary-w-flat-lambda-ci-upper}{5.314623}
\ppDeclareValue{post-eval2024-two-field-boundary-co-mentions-rho}{0.617399}
\ppDeclareValue{post-eval2024-two-field-boundary-co-mentions-rho-ci-lower}{0.578746}
\ppDeclareValue{post-eval2024-two-field-boundary-co-mentions-rho-ci-upper}{0.649489}
\ppDeclareValue{post-eval2024-two-field-boundary-co-mentions-lambda}{1.613692}
\ppDeclareValue{post-eval2024-two-field-boundary-co-mentions-lambda-ci-lower}{1.373867}
\ppDeclareValue{post-eval2024-two-field-boundary-co-mentions-lambda-ci-upper}{1.852979}
\ppDeclareValue{post-eval2024-two-field-channel-correlation}{-0.817928}
\ppDeclareValue{post-eval2024-two-field-boundary-share-b}{0.000000}
\ppDeclareValue{post-eval2024-two-field-boundary-share-n}{0.000000}
\ppDeclareValue{post-eval2024-two-field-rho-b-ci-lower}{0.578760}
\ppDeclareValue{post-eval2024-two-field-rho-b-ci-upper}{0.665889}
\ppDeclareValue{post-eval2024-two-field-rho-n-ci-lower}{0.148998}
\ppDeclareValue{post-eval2024-two-field-rho-n-ci-upper}{0.219273}
\ppDeclareValue{post-eval2024-two-field-rho-total-ci-lower}{0.779498}
\ppDeclareValue{post-eval2024-two-field-rho-total-ci-upper}{0.829711}
\ppDeclareValue{post-eval2024-two-field-theta-ci-lower}{0.727396}
\ppDeclareValue{post-eval2024-two-field-theta-ci-upper}{0.813811}
\ppDeclareValue{post-eval2024-two-field-lambda-b-ci-lower}{2.703272}
\ppDeclareValue{post-eval2024-two-field-lambda-b-ci-upper}{3.846129}
\ppDeclareValue{post-eval2024-two-field-lambda-n-ci-lower}{0.740281}
\ppDeclareValue{post-eval2024-two-field-lambda-n-ci-upper}{1.182419}
\ppDeclareValue{post-eval2025-two-field-rho-b}{0.602242}
\ppDeclareValue{post-eval2025-two-field-rho-n}{0.239218}
\ppDeclareValue{post-eval2025-two-field-rho-total}{0.841460}
\ppDeclareValue{post-eval2025-two-field-theta}{0.715711}
\ppDeclareValue{post-eval2025-two-field-lambda-b}{3.798671}
\ppDeclareValue{post-eval2025-two-field-lambda-n}{1.508880}
\ppDeclareValue{post-eval2025-two-field-loglik}{61163.629257}
\ppDeclareValue{post-eval2025-two-field-trading-days}{249}
\ppDeclareValue{post-eval2025-two-field-loglik-gain-w-flat}{135.858186}
\ppDeclareValue{post-eval2025-two-field-loglik-gain-co-mentions}{655.562619}
\ppDeclareValue{post-eval2025-two-field-boundary-w-flat-rho}{0.852420}
\ppDeclareValue{post-eval2025-two-field-boundary-w-flat-rho-ci-lower}{0.780817}
\ppDeclareValue{post-eval2025-two-field-boundary-w-flat-rho-ci-upper}{0.896331}
\ppDeclareValue{post-eval2025-two-field-boundary-w-flat-lambda}{5.775974}
\ppDeclareValue{post-eval2025-two-field-boundary-w-flat-lambda-ci-lower}{3.562408}
\ppDeclareValue{post-eval2025-two-field-boundary-w-flat-lambda-ci-upper}{8.646091}
\ppDeclareValue{post-eval2025-two-field-boundary-co-mentions-rho}{0.694275}
\ppDeclareValue{post-eval2025-two-field-boundary-co-mentions-rho-ci-lower}{0.581560}
\ppDeclareValue{post-eval2025-two-field-boundary-co-mentions-rho-ci-upper}{0.779272}
\ppDeclareValue{post-eval2025-two-field-boundary-co-mentions-lambda}{2.270910}
\ppDeclareValue{post-eval2025-two-field-boundary-co-mentions-lambda-ci-lower}{1.389830}
\ppDeclareValue{post-eval2025-two-field-boundary-co-mentions-lambda-ci-upper}{3.530463}
\ppDeclareValue{post-eval2025-two-field-channel-correlation}{-0.036037}
\ppDeclareValue{post-eval2025-two-field-boundary-share-b}{0.000000}
\ppDeclareValue{post-eval2025-two-field-boundary-share-n}{0.000000}
\ppDeclareValue{post-eval2025-two-field-rho-b-ci-lower}{0.538205}
\ppDeclareValue{post-eval2025-two-field-rho-b-ci-upper}{0.632936}
\ppDeclareValue{post-eval2025-two-field-rho-n-ci-lower}{0.193820}
\ppDeclareValue{post-eval2025-two-field-rho-n-ci-upper}{0.290353}
\ppDeclareValue{post-eval2025-two-field-rho-total-ci-lower}{0.767075}
\ppDeclareValue{post-eval2025-two-field-rho-total-ci-upper}{0.886807}
\ppDeclareValue{post-eval2025-two-field-theta-ci-lower}{0.667738}
\ppDeclareValue{post-eval2025-two-field-theta-ci-upper}{0.758715}
\ppDeclareValue{post-eval2025-two-field-lambda-b-ci-lower}{2.315867}
\ppDeclareValue{post-eval2025-two-field-lambda-b-ci-upper}{5.308345}
\ppDeclareValue{post-eval2025-two-field-lambda-n-ci-lower}{0.894012}
\ppDeclareValue{post-eval2025-two-field-lambda-n-ci-upper}{2.521254}
\ppDeclareValue{post-eval2026-two-field-rho-b}{0.552125}
\ppDeclareValue{post-eval2026-two-field-rho-n}{0.289550}
\ppDeclareValue{post-eval2026-two-field-rho-total}{0.841675}
\ppDeclareValue{post-eval2026-two-field-theta}{0.655984}
\ppDeclareValue{post-eval2026-two-field-lambda-b}{3.487284}
\ppDeclareValue{post-eval2026-two-field-lambda-n}{1.828828}
\ppDeclareValue{post-eval2026-two-field-loglik}{14997.948925}
\ppDeclareValue{post-eval2026-two-field-trading-days}{64}
\ppDeclareValue{post-eval2026-two-field-loglik-gain-w-flat}{54.851122}
\ppDeclareValue{post-eval2026-two-field-loglik-gain-co-mentions}{138.078363}
\ppDeclareValue{post-eval2026-two-field-boundary-w-flat-rho}{0.856100}
\ppDeclareValue{post-eval2026-two-field-boundary-w-flat-rho-ci-lower}{0.821592}
\ppDeclareValue{post-eval2026-two-field-boundary-w-flat-rho-ci-upper}{0.889807}
\ppDeclareValue{post-eval2026-two-field-boundary-w-flat-lambda}{5.949289}
\ppDeclareValue{post-eval2026-two-field-boundary-w-flat-lambda-ci-lower}{4.605129}
\ppDeclareValue{post-eval2026-two-field-boundary-w-flat-lambda-ci-upper}{8.075009}
\ppDeclareValue{post-eval2026-two-field-boundary-co-mentions-rho}{0.671319}
\ppDeclareValue{post-eval2026-two-field-boundary-co-mentions-rho-ci-lower}{0.610624}
\ppDeclareValue{post-eval2026-two-field-boundary-co-mentions-rho-ci-upper}{0.741044}
\ppDeclareValue{post-eval2026-two-field-boundary-co-mentions-lambda}{2.042467}
\ppDeclareValue{post-eval2026-two-field-boundary-co-mentions-lambda-ci-lower}{1.568209}
\ppDeclareValue{post-eval2026-two-field-boundary-co-mentions-lambda-ci-upper}{2.861658}
\ppDeclareValue{post-eval2026-two-field-channel-correlation}{-0.910684}
\ppDeclareValue{post-eval2026-two-field-boundary-share-b}{0.000000}
\ppDeclareValue{post-eval2026-two-field-boundary-share-n}{0.000000}
\ppDeclareValue{post-eval2026-two-field-rho-b-ci-lower}{0.496550}
\ppDeclareValue{post-eval2026-two-field-rho-b-ci-upper}{0.586609}
\ppDeclareValue{post-eval2026-two-field-rho-n-ci-lower}{0.228948}
\ppDeclareValue{post-eval2026-two-field-rho-n-ci-upper}{0.372759}
\ppDeclareValue{post-eval2026-two-field-rho-total-ci-lower}{0.805999}
\ppDeclareValue{post-eval2026-two-field-rho-total-ci-upper}{0.877613}
\ppDeclareValue{post-eval2026-two-field-theta-ci-lower}{0.572157}
\ppDeclareValue{post-eval2026-two-field-theta-ci-upper}{0.718064}
\ppDeclareValue{post-eval2026-two-field-lambda-b-ci-lower}{2.878554}
\ppDeclareValue{post-eval2026-two-field-lambda-b-ci-upper}{4.250881}
\ppDeclareValue{post-eval2026-two-field-lambda-n-ci-lower}{1.202949}
\ppDeclareValue{post-eval2026-two-field-lambda-n-ci-upper}{2.978620}
\ppDeclareValue{post-pooled-2023-2026-two-field-rho-b}{0.604735}
\ppDeclareValue{post-pooled-2023-2026-two-field-rho-n}{0.226190}
\ppDeclareValue{post-pooled-2023-2026-two-field-rho-total}{0.830925}
\ppDeclareValue{post-pooled-2023-2026-two-field-theta}{0.727785}
\ppDeclareValue{post-pooled-2023-2026-two-field-lambda-b}{3.576718}
\ppDeclareValue{post-pooled-2023-2026-two-field-lambda-n}{1.337805}
\ppDeclareValue{post-pooled-2023-2026-two-field-loglik}{205542.529007}
\ppDeclareValue{post-pooled-2023-2026-two-field-trading-days}{815}
\ppDeclareValue{post-pooled-2023-2026-two-field-loglik-gain-w-flat}{396.218723}
\ppDeclareValue{post-pooled-2023-2026-two-field-loglik-gain-co-mentions}{2048.695965}
\ppDeclareValue{post-pooled-2023-2026-two-field-boundary-w-flat-rho}{0.842849}
\ppDeclareValue{post-pooled-2023-2026-two-field-boundary-w-flat-rho-ci-lower}{0.814506}
\ppDeclareValue{post-pooled-2023-2026-two-field-boundary-w-flat-rho-ci-upper}{0.868627}
\ppDeclareValue{post-pooled-2023-2026-two-field-boundary-w-flat-lambda}{5.363290}
\ppDeclareValue{post-pooled-2023-2026-two-field-boundary-w-flat-lambda-ci-lower}{4.391023}
\ppDeclareValue{post-pooled-2023-2026-two-field-boundary-w-flat-lambda-ci-upper}{6.611916}
\ppDeclareValue{post-pooled-2023-2026-two-field-boundary-co-mentions-rho}{0.667318}
\ppDeclareValue{post-pooled-2023-2026-two-field-boundary-co-mentions-rho-ci-lower}{0.618456}
\ppDeclareValue{post-pooled-2023-2026-two-field-boundary-co-mentions-rho-ci-upper}{0.716434}
\ppDeclareValue{post-pooled-2023-2026-two-field-boundary-co-mentions-lambda}{2.005877}
\ppDeclareValue{post-pooled-2023-2026-two-field-boundary-co-mentions-lambda-ci-lower}{1.620932}
\ppDeclareValue{post-pooled-2023-2026-two-field-boundary-co-mentions-lambda-ci-upper}{2.526519}
\ppDeclareValue{post-pooled-2023-2026-two-field-channel-correlation}{-0.433322}
\ppDeclareValue{post-pooled-2023-2026-two-field-boundary-share-b}{0.000000}
\ppDeclareValue{post-pooled-2023-2026-two-field-boundary-share-n}{0.000000}
\ppDeclareValue{post-pooled-2023-2026-two-field-rho-b-ci-lower}{0.578058}
\ppDeclareValue{post-pooled-2023-2026-two-field-rho-b-ci-upper}{0.629461}
\ppDeclareValue{post-pooled-2023-2026-two-field-rho-n-ci-lower}{0.198952}
\ppDeclareValue{post-pooled-2023-2026-two-field-rho-n-ci-upper}{0.255212}
\ppDeclareValue{post-pooled-2023-2026-two-field-rho-total-ci-lower}{0.801978}
\ppDeclareValue{post-pooled-2023-2026-two-field-rho-total-ci-upper}{0.857547}
\ppDeclareValue{post-pooled-2023-2026-two-field-theta-ci-lower}{0.698431}
\ppDeclareValue{post-pooled-2023-2026-two-field-theta-ci-upper}{0.756188}
\ppDeclareValue{post-pooled-2023-2026-two-field-lambda-b-ci-lower}{2.952040}
\ppDeclareValue{post-pooled-2023-2026-two-field-lambda-b-ci-upper}{4.324050}
\ppDeclareValue{post-pooled-2023-2026-two-field-lambda-n-ci-lower}{1.034137}
\ppDeclareValue{post-pooled-2023-2026-two-field-lambda-n-ci-upper}{1.748887}
\ppDeclareValue{post-pooled-2023-2026-two-field-qlr-news-given-distributional-statistic}{792.437445}
\ppDeclareValue{post-pooled-2023-2026-two-field-qlr-news-given-distributional-pvalue}{0.000500}
\ppDeclareValue{post-pooled-2023-2026-two-field-qlr-distributional-given-news-statistic}{4097.391930}
\ppDeclareValue{post-pooled-2023-2026-two-field-qlr-distributional-given-news-pvalue}{0.000500}
\ppDeclareValue{two-field-theta-points}{401}
\ppDeclareValue{two-field-rho-points}{1000}

\ppDeclareValue{overlap-jaccard-mean}{0.318561}
\ppDeclareValue{overlap-jaccard-median}{0.316452}
\ppDeclareValue{overlap-directed-share}{0.399593}
\ppDeclareValue{overlap-rank-matched-share}{0.498202}
\ppDeclareValue{overlap-support-null-share}{0.279272}
\ppDeclareValue{overlap-support-excess-pct}{12.032120}
\ppDeclareValue{overlap-support-null-pvalue}{0.001}
\ppDeclareValue{overlap-w-flat-support-size}{46.090000}
\ppDeclareValue{overlap-co-mentions-support-size}{27.660000}
\ppDeclareValue{overlap-support-draws}{1000}
\ppDeclareValue{overlap-weight-union-pearson}{0.543315}
\ppDeclareValue{overlap-weight-union-cosine}{0.640227}
\ppDeclareValue{overlap-weight-common-spearman}{0.442218}
\ppDeclareValue{overlap-weight-common-count}{18.240000}
\ppDeclareValue{overlap-weight-offdiagonal-pearson}{0.601468}
\ppDeclareValue{overlap-induced-pearson}{0.897022}
\ppDeclareValue{overlap-induced-r-squared}{0.804648}
\ppDeclareValue{overlap-induced-null-pearson}{0.707059}
\ppDeclareValue{overlap-induced-null-pvalue}{0.001}
\ppDeclareValue{overlap-induced-asset-min}{0.578802}
\ppDeclareValue{overlap-induced-asset-max}{0.984955}
\ppDeclareValue{n-tickers-priced}{100}
\ppDeclareValue{network-row-sum-residual}{0.00000000}
\ppDeclareValue{network-w2-minimum-weight}{0.000047}
\ppDeclareValue{network-dobrushin}{0.983580}
\ppDeclareValue{network-stationary-minimum}{0.004925}
\ppDeclareValue{network-stationary-residual}{5.191837e-13}
\ppDeclareValue{network-w-flat-resolvent-error}{0.627802}
\ppDeclareValue{network-w-flat-resolvent-bound}{1.838128}
\ppDeclareValue{network-w-flat-lean-neumann-bound}{\ppmissing}
\ppDeclareValue{network-w-flat-complement-norm}{1.276683}
\ppDeclareValue{network-w-flat-in-neumann-region}{false}
\ppDeclareValue{network-w-flat-g-common-mode-mean}{6.363295}
\ppDeclareValue{network-w-flat-g-common-mode-error}{1.154632e-14}
\ppDeclareValue{network-w-flat-h-common-mode-mean}{1.000000}
\ppDeclareValue{network-w-flat-h-common-mode-error}{1.776357e-15}
\ppDeclareValue{network-w-flat-own-exposure-mean}{0.170130}
\ppDeclareValue{network-w-flat-own-exposure-minimum}{0.163128}
\ppDeclareValue{network-w-flat-own-exposure-maximum}{0.182568}
\ppDeclareValue{network-w-flat-other-exposure-mean}{0.829870}
\ppDeclareValue{network-w-flat-other-exposure-minimum}{0.817432}
\ppDeclareValue{network-w-flat-other-exposure-maximum}{0.836872}

\ppDeclareValue{network-factor-w2-rho}{0.686596}
\ppDeclareValue{network-factor-w2-g-common-mode-mean}{3.190765}
\ppDeclareValue{network-factor-w2-g-common-mode-error}{3.552714e-15}
\ppDeclareValue{network-factor-w2-h-common-mode-mean}{1.000000}
\ppDeclareValue{network-factor-w2-h-common-mode-error}{1.110223e-15}
\ppDeclareValue{network-factor-w2-own-exposure-mean}{0.325230}
\ppDeclareValue{network-factor-w2-own-exposure-minimum}{0.318524}
\ppDeclareValue{network-factor-w2-own-exposure-maximum}{0.339805}
\ppDeclareValue{network-factor-w2-other-exposure-mean}{0.674770}
\ppDeclareValue{network-factor-w2-other-exposure-minimum}{0.660195}
\ppDeclareValue{network-factor-w2-other-exposure-maximum}{0.681476}

\ppDeclareValue{economic-rbf-selected-multiplier}{0.250000}
\ppDeclareValue{economic-rbf-median-bandwidth}{1.100056}
\ppDeclareValue{economic-factor-only-rho}{0.000000}
\ppDeclareValue{economic-factor-only-loglik}{205748.307526}
\ppDeclareValue{economic-factor-only-held-out-score}{2.468881}
\ppDeclareValue{economic-factor-only-held-out-gain}{0.000000}
\ppDeclareValue{economic-factor-only-held-out-gain-ci-lower}{0.000000}
\ppDeclareValue{economic-factor-only-held-out-gain-ci-upper}{0.000000}
\ppDeclareValue{economic-ledoit-wolf-rho}{\ppmissing}
\ppDeclareValue{economic-ledoit-wolf-loglik}{\ppmissing}
\ppDeclareValue{economic-ledoit-wolf-held-out-score}{2.558644}
\ppDeclareValue{economic-ledoit-wolf-held-out-gain}{0.089763}
\ppDeclareValue{economic-ledoit-wolf-held-out-gain-ci-lower}{-0.002771}
\ppDeclareValue{economic-ledoit-wolf-held-out-gain-ci-upper}{0.090337}
\ppDeclareValue{economic-w1-rho}{0.683673}
\ppDeclareValue{economic-w1-loglik}{209488.342184}
\ppDeclareValue{economic-w1-held-out-score}{2.526065}
\ppDeclareValue{economic-w1-held-out-gain}{0.057184}
\ppDeclareValue{economic-w1-held-out-gain-ci-lower}{0.049738}
\ppDeclareValue{economic-w1-held-out-gain-ci-upper}{0.065565}
\ppDeclareValue{economic-w2-rho}{0.686596}
\ppDeclareValue{economic-w2-loglik}{209476.031412}
\ppDeclareValue{economic-w2-held-out-score}{2.525957}
\ppDeclareValue{economic-w2-held-out-gain}{0.057076}
\ppDeclareValue{economic-w2-held-out-gain-ci-lower}{0.049624}
\ppDeclareValue{economic-w2-held-out-gain-ci-upper}{0.065259}
\ppDeclareValue{economic-rbf-median-rho}{0.467334}
\ppDeclareValue{economic-rbf-median-loglik}{206227.559029}
\ppDeclareValue{economic-rbf-median-held-out-score}{2.479192}
\ppDeclareValue{economic-rbf-median-held-out-gain}{0.010311}
\ppDeclareValue{economic-rbf-median-held-out-gain-ci-lower}{0.008201}
\ppDeclareValue{economic-rbf-median-held-out-gain-ci-upper}{0.012943}
\ppDeclareValue{economic-rbf-tuned-rho}{0.571598}
\ppDeclareValue{economic-rbf-tuned-loglik}{206604.044245}
\ppDeclareValue{economic-rbf-tuned-held-out-score}{2.484870}
\ppDeclareValue{economic-rbf-tuned-held-out-gain}{0.015989}
\ppDeclareValue{economic-rbf-tuned-held-out-gain-ci-lower}{0.013520}
\ppDeclareValue{economic-rbf-tuned-held-out-gain-ci-upper}{0.019079}
\ppDeclareValue{economic-centroid-rho}{0.469656}
\ppDeclareValue{economic-centroid-loglik}{209343.097980}
\ppDeclareValue{economic-centroid-held-out-score}{2.522691}
\ppDeclareValue{economic-centroid-held-out-gain}{0.053810}
\ppDeclareValue{economic-centroid-held-out-gain-ci-lower}{0.045954}
\ppDeclareValue{economic-centroid-held-out-gain-ci-upper}{0.062459}
\ppDeclareValue{economic-equal-support-rho}{0.618968}
\ppDeclareValue{economic-equal-support-loglik}{207216.296666}
\ppDeclareValue{economic-equal-support-held-out-score}{2.494105}
\ppDeclareValue{economic-equal-support-held-out-gain}{0.025224}
\ppDeclareValue{economic-equal-support-held-out-gain-ci-lower}{0.021758}
\ppDeclareValue{economic-equal-support-held-out-gain-ci-upper}{0.029367}
\ppDeclareValue{economic-co-mentions-rho}{0.465631}
\ppDeclareValue{economic-co-mentions-loglik}{209240.277154}
\ppDeclareValue{economic-co-mentions-held-out-score}{2.519320}
\ppDeclareValue{economic-co-mentions-held-out-gain}{0.050439}
\ppDeclareValue{economic-co-mentions-held-out-gain-ci-lower}{0.042751}
\ppDeclareValue{economic-co-mentions-held-out-gain-ci-upper}{0.058584}
\ppDeclareValue{economic-industry-sector-rho}{0.229354}
\ppDeclareValue{economic-industry-sector-loglik}{207873.272413}
\ppDeclareValue{economic-industry-sector-held-out-score}{2.501014}
\ppDeclareValue{economic-industry-sector-held-out-gain}{0.032133}
\ppDeclareValue{economic-industry-sector-held-out-gain-ci-lower}{0.025932}
\ppDeclareValue{economic-industry-sector-held-out-gain-ci-upper}{0.038965}
\ppDeclareValue{economic-centroid-w2-gain-share}{0.942781}
\ppDeclareValue{economic-w2-minus-centroid-estimate}{0.003266}
\ppDeclareValue{economic-w2-minus-centroid-ci-lower}{0.001354}
\ppDeclareValue{economic-w2-minus-centroid-ci-upper}{0.005442}
\ppDeclareValue{economic-w2-minus-centroid-positive-share}{0.999000}
\ppDeclareValue{economic-ledoit-wolf-minus-w2-estimate}{0.032687}
\ppDeclareValue{economic-ledoit-wolf-minus-w2-ci-lower}{-0.057862}
\ppDeclareValue{economic-ledoit-wolf-minus-w2-ci-upper}{0.031576}
\ppDeclareValue{economic-ledoit-wolf-minus-w2-positive-share}{0.364500}

\ppDeclareValue{economic-peer-amd-top-candidate}{NVDA}
\ppDeclareValue{economic-peer-amd-top-weight}{0.122489}
\ppDeclareValue{economic-peer-amd-top-distance-rank}{1}
\ppDeclareValue{economic-peer-amd-top-same-sector}{true}
\ppDeclareValue{economic-peer-amd-top-same-industry}{true}
\ppDeclareValue{economic-peer-amd-top-co-mention}{true}
\ppDeclareValue{economic-peer-amd-effective-source-count}{16.712344}
\ppDeclareValue{economic-peer-amd-incoming-mass}{0.946093}
\ppDeclareValue{point-in-time-w2-held-out-gain}{0.028203}
\ppDeclareValue{point-in-time-w2-minus-rbf-estimate}{0.018199}
\ppDeclareValue{point-in-time-w2-minus-rbf-ci-lower}{0.015156}
\ppDeclareValue{point-in-time-w2-minus-rbf-ci-upper}{0.021388}

\ppDeclareValue{paper5-balanced-cloud-size}{128}

\ppDeclareValue{transport-switch-d4096-switched-atoms}{62612}
\ppDeclareValue{transport-switch-d4096-max-difference}{-0.003530}
\ppDeclareValue{transport-switch-d4096-p95-difference}{-0.002238}
\ppDeclareValue{transport-switch-d4096-sd-difference}{-0.001118}
\ppDeclareValue{transport-switch-d4096-tail-count-difference}{-0.0762}
\ppDeclareValue{transport-switch-d4096-switched-difference}{0.000702}
\ppDeclareValue{transport-switch-d4096-max-reduction-share}{0.214747}
\ppDeclareValue{transport-switch-d4096-max-increase-share}{0.041414}
\ppDeclareValue{transport-switch-d4096-max-reduction-given-lower}{-0.018117}
\ppDeclareValue{transport-switch-d4096-max-increase-given-higher}{0.008712}
\ppDeclareValue{transport-switch-d4096-tail-reduction-share}{0.083636}
\ppDeclareValue{transport-switch-d4096-tail-trade-pattern}{true}
\ppDeclareValue{transport-switch-d64-switched-atoms}{111245}
\ppDeclareValue{transport-switch-d64-max-difference}{-0.019251}
\ppDeclareValue{transport-switch-d64-p95-difference}{-0.010815}
\ppDeclareValue{transport-switch-d64-sd-difference}{-0.005274}
\ppDeclareValue{transport-switch-d64-tail-count-difference}{-0.0622}
\ppDeclareValue{transport-switch-d64-switched-difference}{0.001979}
\ppDeclareValue{transport-switch-d64-max-reduction-share}{0.443838}
\ppDeclareValue{transport-switch-d64-max-increase-share}{0.027475}
\ppDeclareValue{transport-switch-d64-max-reduction-given-lower}{-0.044293}
\ppDeclareValue{transport-switch-d64-max-increase-given-higher}{0.014858}
\ppDeclareValue{transport-switch-d64-tail-reduction-share}{0.060000}
\ppDeclareValue{transport-switch-d64-tail-trade-pattern}{true}

\ppDeclareValue{annual-raw-article-floor}{47}

\ppDeclareValue{bridge-reference-52-short-rho}{0.771197}
\ppDeclareValue{bridge-reference-52-short-rho-ci-lower}{0.734924}
\ppDeclareValue{bridge-reference-52-short-rho-ci-upper}{0.805830}
\ppDeclareValue{bridge-reference-52-short-lambda}{3.370576}
\ppDeclareValue{bridge-reference-52-short-lambda-ci-lower}{2.772504}
\ppDeclareValue{bridge-reference-52-short-lambda-ci-upper}{4.150118}
\ppDeclareValue{bridge-reference-52-short-trading-days}{816}
\ppDeclareValue{bridge-reference-52-short-date-start}{2023-01-03}
\ppDeclareValue{bridge-reference-52-short-date-end}{2026-04-06}
\ppDeclareValue{bridge-reference-52-long-rho}{0.775863}
\ppDeclareValue{bridge-reference-52-long-rho-ci-lower}{0.742978}
\ppDeclareValue{bridge-reference-52-long-rho-ci-upper}{0.806505}
\ppDeclareValue{bridge-reference-52-long-lambda}{3.461561}
\ppDeclareValue{bridge-reference-52-long-lambda-ci-lower}{2.890711}
\ppDeclareValue{bridge-reference-52-long-lambda-ci-upper}{4.168088}
\ppDeclareValue{bridge-reference-52-long-trading-days}{885}
\ppDeclareValue{bridge-reference-52-long-date-start}{2023-01-03}
\ppDeclareValue{bridge-reference-52-long-date-end}{2026-07-15}
\ppDeclareValue{bridge-current-100-short-rho}{0.842849}
\ppDeclareValue{bridge-current-100-short-rho-ci-lower}{0.814506}
\ppDeclareValue{bridge-current-100-short-rho-ci-upper}{0.868627}
\ppDeclareValue{bridge-current-100-short-lambda}{5.363295}
\ppDeclareValue{bridge-current-100-short-lambda-ci-lower}{4.391019}
\ppDeclareValue{bridge-current-100-short-lambda-ci-upper}{6.611919}
\ppDeclareValue{bridge-current-100-short-trading-days}{815}
\ppDeclareValue{bridge-current-100-short-date-start}{2023-01-03}
\ppDeclareValue{bridge-current-100-short-date-end}{2026-04-06}
\ppDeclareValue{bridge-horizon-rho-difference}{0.004666}
\ppDeclareValue{bridge-horizon-lambda-difference}{0.090985}
\ppDeclareValue{bridge-universe-rho-difference}{0.071652}
\ppDeclareValue{bridge-universe-lambda-difference}{1.992719}
\ppDeclareValue{bridge-common-return-dates}{816}
\ppDeclareValue{bridge-common-return-max-difference}{0.00000000}

\ppDeclareValue{lean-verified-thm:wflat-sar-gate}{checked}
\ppDeclareValue{lean-disclosure-thm:wflat-sar-gate}{Machine-checked for the exact finite operator contract in the pinned Lean 4 / mathlib v4.35.0 workspace. This verifies SAR compatibility, not the economic validity of the text-derived weights.}
\ppDeclareValue{lean-verified-def:perron-limit}{checked}
\ppDeclareValue{lean-disclosure-def:perron-limit}{Formalized as the exact quantitative premise consumed by the convergence bridge. The declaration does not claim that a parquet operator satisfies the premise and does not formalize general Perron--Frobenius theory.}
\ppDeclareValue{lean-verified-thm:perron-from-geometric}{checked}
\ppDeclareValue{lean-disclosure-thm:perron-from-geometric}{Machine-checked under standard axioms only. This verifies that a quantitative contraction certificate is sufficient for the Perron limit; the empirical float64 certificate remains a separate numerical witness.}
\ppDeclareValue{lean-verified-prop:finite-rho-bound}{checked}
\ppDeclareValue{lean-disclosure-prop:finite-rho-bound}{Machine-checked under standard axioms only in the maximum absolute row-sum matrix norm. It is a finite-rho algebraic bound, not a statistical confidence interval.}
\ppDeclareValue{lean-verified-thm:replication-centrality}{checked}
\ppDeclareValue{lean-disclosure-thm:replication-centrality}{Machine-checked in the pinned Lean 4 / mathlib v4.35.0 workspace, conditional on the assumed Perron--Frobenius limit hypothesis.}
\ppDeclareValue{lean-verified-thm:stationarity}{checked}
\ppDeclareValue{lean-disclosure-thm:stationarity}{Machine-checked in the pinned Lean 4 / mathlib v4.35.0 workspace, conditional on the assumed Perron--Frobenius limit hypothesis.}
\ppDeclareValue{lean-verified-thm:rankone-collapse}{checked}
\ppDeclareValue{lean-disclosure-thm:rankone-collapse}{Machine-checked in the pinned Lean 4 / mathlib v4.35.0 workspace, conditional on the single assumed Perron--Frobenius power-limit hypothesis. The resolvent limit it previously also assumed is now derived from that hypothesis rather than assumed alongside it. The limiting common-mode variance may vanish, so the formal conclusion permits rank zero as well as rank one.}
\ppDeclareValue{lean-verified-thm:resolvent-from-perron}{checked}
\ppDeclareValue{lean-disclosure-thm:resolvent-from-perron}{Machine-checked in the pinned Lean 4 / mathlib v4.35.0 workspace under standard axioms only. The derivation is elementary linear algebra via an absorbing-idempotent resolvent split; it introduces no Perron--Frobenius content of its own, no Abel/Tauberian summation, and no spectral theory.}
\ppDeclareValue{lean-verified-thm:adjustment-equilibrium}{checked}
\ppDeclareValue{lean-disclosure-thm:adjustment-equilibrium}{Machine-checked in the pinned Lean 4 / mathlib v4.35.0 workspace by a completing-square identity rather than a differentiability argument, so the equivalence holds in any real inner-product exposure space.}
\ppDeclareValue{lean-verified-thm:spatial-closure}{checked}
\ppDeclareValue{lean-disclosure-thm:spatial-closure}{Machine-checked in the pinned Lean 4 / mathlib v4.35.0 workspace. The spatial equation is derived from the stated adjustment problem; it is not imposed as a reduced form, and the interaction weights themselves are a maintained input.}
\ppDeclareValue{lean-verified-cor:mixture-nesting}{checked}
\ppDeclareValue{lean-disclosure-cor:mixture-nesting}{Machine-checked in the pinned Lean 4 / mathlib v4.35.0 workspace. The selected declaration proves nonnegativity and unit row sums under convex mixing. The zero-diagonal clause follows analytically from zero-diagonal endpoints and is not part of this selected formal statement. No economic validity of either field is certified.}
\ppDeclareValue{lean-verified-thm:two-field-closure}{checked}
\ppDeclareValue{lean-disclosure-thm:two-field-closure}{Machine-checked in the pinned Lean 4 / mathlib v4.35.0 workspace by reduction to the one-field closure at the intensity-weighted mixture. The two-channel equation is derived from the stated adjustment problem; both fields remain maintained inputs, and the result does not establish that they are separately identified in any sample.}
\ppDeclareValue{lean-verified-prop:channel-inversion}{checked}
\ppDeclareValue{lean-disclosure-prop:channel-inversion}{Machine-checked in the pinned Lean 4 / mathlib v4.35.0 workspace. The inversion is an identity of the quadratic model; reading a fitted coefficient through it presumes the return-bridge restriction stated in the identification section.}
\ppDeclareValue{lean-verified-thm:sar-reduced-form}{checked}
\ppDeclareValue{lean-disclosure-thm:sar-reduced-form}{Machine-checked in the pinned Lean 4 / mathlib v4.35.0 workspace for Hilbert-valued exposures under the displayed norm gate.}
\ppDeclareValue{lean-verified-prop:scalar-projection}{checked}
\ppDeclareValue{lean-disclosure-prop:scalar-projection}{Machine-checked in the pinned Lean 4 / mathlib v4.35.0 workspace for scalar projections of the systematic exposure equation. Its return-QMLE interpretation remains conditional on the maintained innovation specification and is not causal.}
\ppDeclareValue{lean-verified-cor:zero-feedback-nesting}{checked}
\ppDeclareValue{lean-disclosure-cor:zero-feedback-nesting}{Machine-checked unconditionally in the pinned Lean 4 / mathlib v4.35.0 workspace: at zero feedback, peer-adjusted exposure equals stand-alone exposure exactly. This verifies the exposure identity only; it does not import the related paper's transmission or covariance assumptions.}
\ppDeclareValue{lean-verified-thm:pairwise-nesting}{checked}
\ppDeclareValue{lean-disclosure-thm:pairwise-nesting}{Machine-checked as an exact equality in the pinned Lean 4 / mathlib v4.35.0 workspace, with no optimal-coupling existence theorem required. It identifies the pairwise quadratic Wasserstein term as the two-firm case of the multi-firm dispersion functional.}
\ppDeclareValue{lean-verified-thm:spatial-attenuation}{checked}
\ppDeclareValue{lean-disclosure-thm:spatial-attenuation}{Machine-checked in the pinned Lean 4 / mathlib v4.35.0 workspace for a nonnegative row-stochastic matrix centered at strictly positive stationary probability weights. No reversibility, symmetry, or spectral premise is used; the upper factor of one is exact and the lower factor is a universal worst case.}

\newcommand{\ppLeanStatusList}{%
  \begin{itemize}
    \item \Cref{thm:wflat-sar-gate}: \ppvalue{lean-disclosure-thm:wflat-sar-gate}
    \item \Cref{def:perron-limit}: \ppvalue{lean-disclosure-def:perron-limit}
    \item \Cref{thm:perron-from-geometric}: \ppvalue{lean-disclosure-thm:perron-from-geometric}
    \item \Cref{prop:finite-rho-bound}: \ppvalue{lean-disclosure-prop:finite-rho-bound}
    \item \Cref{thm:replication-centrality}: \ppvalue{lean-disclosure-thm:replication-centrality}
    \item \Cref{thm:stationarity}: \ppvalue{lean-disclosure-thm:stationarity}
    \item \Cref{thm:rankone-collapse}: \ppvalue{lean-disclosure-thm:rankone-collapse}
    \item \Cref{thm:resolvent-from-perron}: \ppvalue{lean-disclosure-thm:resolvent-from-perron}
    \item \Cref{thm:adjustment-equilibrium}: \ppvalue{lean-disclosure-thm:adjustment-equilibrium}
    \item \Cref{thm:spatial-closure}: \ppvalue{lean-disclosure-thm:spatial-closure}
    \item \Cref{cor:mixture-nesting}: \ppvalue{lean-disclosure-cor:mixture-nesting}
    \item \Cref{thm:two-field-closure}: \ppvalue{lean-disclosure-thm:two-field-closure}
    \item \Cref{prop:channel-inversion}: \ppvalue{lean-disclosure-prop:channel-inversion}
    \item \Cref{thm:sar-reduced-form}: \ppvalue{lean-disclosure-thm:sar-reduced-form}
    \item \Cref{prop:scalar-projection}: \ppvalue{lean-disclosure-prop:scalar-projection}
    \item \Cref{cor:zero-feedback-nesting}: \ppvalue{lean-disclosure-cor:zero-feedback-nesting}
    \item \Cref{thm:pairwise-nesting}: \ppvalue{lean-disclosure-thm:pairwise-nesting}
    \item \Cref{thm:spatial-attenuation}: \ppvalue{lean-disclosure-thm:spatial-attenuation}
  \end{itemize}%
}

\title{\papertitle}
\author{\paperauthors\\\paperaffiliations}
\date{\paperdate}

\begin{document}

\maketitle

\begin{abstract}
  Identifying a firm's economic peers matters for benchmarking, risk measurement,
and understanding return dependence, yet a diversified firm may have no single
natural peer.
We use language-model representations of corporate news to identify the
combination of firms that jointly represents each firm's information
environment.
Each firm is represented by a distribution of article embeddings, and
target-anchored Wasserstein barycentric reconstruction produces a directed map
of economic-peer weights.
We use this map as the interaction structure in a spatial factor model and ask
whether return dependence follows peer relations measured from news without
using returns.
We construct peer weights from 2018--2022 news for 100 large-cap US-traded
firms and evaluate their financial relevance in subsequent returns.
Relative to a matched model with the Fama--French five factors and momentum,
the information field improves held-out joint-density fit over 2023--2026,
conditional on realized factor returns.
Because unrestricted firm coefficients leave conventional factor betas
unchanged, the improvement operates through residual covariance: the field
helps organize which firms' unexpected returns move together after common
factor exposures are accounted for.
Joint reconstruction contributes substantially more than pairwise Wasserstein
proximity or equal weighting of the selected peers, while centroid
reconstruction retains most of the gain.
Explicit news co-mentions provide complementary information.
Language models thus make the interaction structure in spatial asset pricing
an object of measurement: economically interpretable peer relations can be
constructed from text before their financial relevance is assessed in returns.

\end{abstract}


\medskip

\noindent\textit{JEL classification:} \paperjel

\section*{Introduction}
\label{sec:introduction}

A firm's economic peers matter for benchmarking, risk measurement, and the
estimation of return dependence, yet the economically relevant peer of a
diversified firm need not be another individual firm.
Industry membership, supply-chain links, geographic proximity, and shared news
coverage each provide economically meaningful definitions of relatedness.
In spatial finance, such relations determine the interaction structure through
which firms' returns are connected
\citep{fernandez_spatial_2011,kou_asset_2018,ge_news-implied_2023}.
Our starting point is that the relevant peer can instead be a weighted
combination of firms, each representing a different part of the target's
economic activities.

A diversified firm illustrates the distinction.
One candidate may resemble its technology business, another its customer base,
and another its production or supply-chain environment.
No individual firm need closely resemble the whole.
Pairwise similarity ranks each candidate separately; joint representation asks
which combination best represents the target and how much each member
contributes.
A firm can therefore be an important peer even when it is not especially close
to the target on average.
Its relevance depends on what it contributes alongside the other members of the
basket.

Language models make this compositional peer structure measurable from firms'
information environments.
A text embedding model maps each article to a fixed-dimensional vector, with
semantically related texts placed nearby in the resulting representation space.
This geometry, commonly used for semantic retrieval, lets a collection of
articles be treated as a distribution of economic information.
A firm's articles therefore form a distribution of positions, which we call its
information footprint.
Two firms can have the same average position while occupying different
footprints, much as retail chains with the same average store location can
serve different regions.
We use Wasserstein geometry to align each candidate peer's articles with those
of the target, then choose the nonnegative weights, summing to one and
excluding the target, that best reconstruct the target from the aligned peer
footprints.
This target-anchored barycentric reconstruction produces a customized
information-peer basket for every firm.
Across firms, these baskets form a directed map $W^\flat$ of a latent
information field: firm $j$ can contribute strongly to firm $i$'s basket while
$i$ contributes little to $j$'s.

The broader contribution is to make the interaction structure in a spatial
finance model an object of measurement.
Spatial asset-pricing models require a matrix $W$ describing which firms
interact and with what weight.
Rather than specifying this structure from geography, industry
classifications, or the returns subsequently used for evaluation, we measure it
from an earlier information environment and hold it fixed.
This distinction matters because economically relevant information about firms,
such as their products, customers, suppliers, technologies, and risks, can
accumulate independently of the history of return co-movement from which
cross-firm dependence would otherwise be inferred.
Subsequent returns therefore evaluate the financial relevance of the measured
field rather than determine its construction.

We study \ppvalue{n-tickers-priced} continuously covered large-cap firms
traded in the United States from a Nasdaq-100-based universe.
We construct the field from 2018--2022 news, hold it fixed, and evaluate it on
\ppvalue{transport-evaluation-trading-days} trading days in 2023--2026, with
2026 covering a partial year.
For the held-out exercise, all return-model parameters are fitted on
2018--2022 returns before the later observations are scored.
A spatial factor model connects the field to returns, and a quadratic
exposure-adjustment problem, in which a firm balances its stand-alone exposure
against alignment with its information peers, gives the spatial coefficient an
economic interpretation.
With unrestricted firm coefficients, reduced-form factor betas remain exactly
equal to firm-by-firm least-squares estimates.
Standard factors therefore continue to describe exposure to common systematic
risks, and the field's contribution lies in residual covariance.

The frozen field improves held-out density scores relative to a matched model
with the Fama--French five factors and momentum, conditional on realized factor
returns, with paired intervals above zero.
Joint representation is the source of the gain: pairwise Wasserstein
proximity captures less than a third of it, and equal weighting of the selected
peers captures less than half.
Centroid reconstruction retains most of the gain, while the full distribution
adds a smaller, separately measurable increment.
Explicit news co-mentions capture a complementary dimension of relatedness, and
each field improves held-out joint density conditional on the other.
A conventional residual-covariance estimator has a higher held-out point
score, although the paired difference is statistically uncertain; the
contribution is an interpretable peer structure whose financial relevance can
be evaluated out of period.
The reconstruction advantage is stable across transport costs and, in sample,
across alternative embedding models.
The primary embedding model was released after the news window, so the
representation measures earlier news retrospectively.
We assess sensitivity to this retrospective measurement with an encoder
whose reported training cutoff precedes the article window.

\Cref{sec:literature} reviews related work.
\Cref{sec:interaction-operator} constructs the field,
\Cref{sec:spatial-closure} develops the exposure-adjustment model, and
\Cref{sec:identification-estimation} connects it to returns.
\Cref{sec:empirical-design} describes the design,
\Cref{sec:results} reports the evidence, and
\Cref{sec:limitations,sec:conclusion} discuss implications and conclude.

\section{Related Literature}\label{sec:literature}

Text-based finance establishes that firms' information environments contain
financially relevant variation.
Attention measures, textual factors, embedding-based return and
stochastic-discount-factor models, and learned graphs for factor exposures
use that information in different ways
\citep{ben_rephael_information_consumption_2019,cong_textual_factors_2024,
wang2025newsnet,son2022graph,uddin2024network}.
Our question concerns the relations among firms that can be measured from
these representations.
We construct a basket of information peers for each target, then evaluate
whether the resulting cross-sectional structure organizes return dependence.
The distinction is between extracting information about an individual firm
and measuring how several firms jointly represent its information environment.

News networks provide a direct informational definition of a peer.
Economic linkages inferred from news predict returns, and sentence-level
co-mentions trace contagion and aggregate risk
\citep{scherbina_economic_2013,schwenkler_network_2020}.
\citet{ge_news-implied_2023} place a co-mention matrix inside a spatial factor
model, and \citet{ge_network_guided_covariance_2026} study whether auxiliary
network information improves covariance estimation.
These approaches begin with an observed or count-weighted graph.
Our reconstruction uses the content of firms' coverage to select combinations
of peers, including firms that need not appear together in an article.
A persistent co-mention field therefore provides both a natural comparison
and a complementary source of information that can enter the model jointly.

Spatial finance supplies the framework for evaluating these peer structures.
Spatial asset-pricing models describe how local interaction modifies
factor-based pricing, conditional on an interaction matrix $W$
\citep{fernandez_spatial_2011,kou_asset_2018,ge_news-implied_2023}.
In the spatial CAPM and spatial APT of \citet{kou_asset_2018}, assets occupy
point-valued locations and inverse geographic distance supplies the weights.
The spatial multiplier accumulates direct and higher-order interactions.
Classical spatial-autoregressive inference also conditions on a known $W$;
estimating weights from the outcomes entering the spatial lag creates
mechanical reflection \citep{anselin_1988,lesage_pace_2009,kelejian_prucha_2010}.
We measure peer weights from news without fitting them to the return outcomes
used in evaluation and hold the resulting field fixed during estimation.

Linear-quadratic network games explain how individual objectives generate
aggregate propagation on a given graph.
Costly local complementarity yields best responses summarized by a network
resolvent \citep{ballester_networks_2006}.
Our adjustment model applies this logic to factor exposures.
Firms balance departures from stand-alone exposure against misalignment with
peer exposures, and the spatial coefficient records their relative weights.
The measured information field supplies the peer baskets to which this
adjustment applies.

The construction connects these finance questions to Wasserstein geometry.
A distribution of articles retains information about a firm's internal
composition; if every distribution collapses to a point mass, Wasserstein
distance reduces to the underlying point distance.
Pairwise transport provides distances and article-level correspondence.
Target-anchored reconstruction uses those correspondences to select the
combination of firms that jointly represents each target.
It borrows the geometry of the Wasserstein barycenter
\citep{agueh_carlier_barycenters_2011}, while fixing the target and estimating
peer weights instead of optimizing over a free center distribution.
The centroid and pairwise-distance comparators separate the roles of
within-firm distributions and joint reconstruction in the return evidence.
Recent work uses distribution-valued firm characteristics for pairwise
covariance bounds \citep{gawronsky_continuous_2026} and portfolio allocation
\citep{gawronsky_distance_mpt_2026}; the object here is the multi-firm
information field.

\section{Measuring the Information Field}
\label{sec:interaction-operator}

Measuring the information field requires a peer basket for every firm.
The basket represents the target through a combination of other firms'
information footprints.
We collect its weights in a row of an interaction matrix $W$.
Nonnegative weights summing to one make each basket a peer average, and a zero
self-weight requires the target to be represented by other firms.

\begin{definition}[Row-stochastic, zero-diagonal interaction
  matrix]\label{def:interaction-matrix}
  An interaction matrix is any $W \in \R^{n \times n}$ with $W_{ij} \geq 0$ for
  all $i$ and $j$, $W_{ii} = 0$ for all $i$, and $\sum_j W_{ij} = 1$ for all
  $i$.
  A firm never interacts with itself, weights every other firm nonnegatively,
  and spends exactly one unit of interaction weight on the remaining firms.
\end{definition}

Each row can average a firm-level quantity in its original units, much as
benchmark weights make a benchmark return comparable with a portfolio return.
The definition specifies the form of a peer basket.
To select its members and weights, we use language-model representations of
news and find the combination that best represents the target's information
footprint within the observed universe.

A text embedding model maps each article to a fixed-dimensional vector in a
representation space, with semantically related articles lying close together.
A firm's articles therefore form an empirical characteristic law
$\widehat C_i$ in the embedding space, whose population counterpart is $C_i$.
A diversified firm occupies many positions at once, so the law records spread,
multimodality, and internal composition that one averaged embedding would
suppress.
Quadratic Wasserstein distance compares two such laws by aligning their
article clouds so as to minimize average squared displacement.
That alignment supplies article-level correspondence, but a distance alone
does not say how several candidate firms jointly represent one target.

Consider a three-firm universe with target $A$ and candidate peers $B$ and $C$.
After separately aligning the article positions of $B$ and $C$ with those of
$A$, suppose the best common convex approximation assigns coefficient $0.60$ to
the aligned positions of $B$ and $0.40$ to those of $C$ at every matched index.
In the order $(A,B,C)$, the resulting target row is
$W^\flat_{A\cdot}=(0,0.60,0.40)$.
The 60--40 weights are coordinates on aligned peer positions, not probabilities
of drawing whole articles from $B$ or $C$.

\begin{definition}[Target-anchored Wasserstein barycentric reconstruction]
  \label{def:w-flat}
  Write the equally sized empirical clouds as
  $\widehat C_i=M^{-1}\sum_{m=1}^M\delta_{x_{im}}$.
  Let $\mathcal A_M$ be the finite set of permutations of the $M$ support
  indices.
  For every ordered pair $i\ne j$, first fix one optimal balanced assignment
  under the implementation's predetermined deterministic selection rule,
  \begin{equation}\label{eq:w2-assignment}
    \pi_{ij}\in\operatorname*{argmin}_{\pi\in\mathcal A_M}
    \frac{1}{M}\sum_{m=1}^M
    \lVert x_{im}-x_{j,\pi(m)}\rVert_2^2,
    \qquad y_{ijm}=x_{j,\pi_{ij}(m)}.
  \end{equation}
  Conditional on these target-specific assignments, define
  \begin{equation}\label{eq:w2-barycentre}
    w_i\in\operatorname*{argmin}_{w\in\Delta_{-i}}
    \frac{1}{M}\sum_{m=1}^M
    \left\lVert x_{im}-\sum_{j\ne i}w_j y_{ijm}\right\rVert_2^2,
    \qquad
    \Delta_{-i}=\{w:w_j\geq0,\ \sum_{j\ne i}w_j=1\}.
  \end{equation}
  The output is the barycentric interaction field
  $W^\flat_{ij}=(w_i)_j$ for $j\ne i$ and $W^\flat_{ii}=0$.
\end{definition}

Equation~\eqref{eq:w2-assignment} fixes a transport alignment for each ordered
target--candidate pair, and Equation~\eqref{eq:w2-barycentre} selects the
target's simplex coordinates by minimizing reconstruction loss across all
aligned candidates jointly.
Replacing squared with unsquared norms in both stages gives the linear-cost
field $W^\flat_1$, used only to test sensitivity to the transport order.
Because the second objective is joint, a row of $W^\flat$ is not a scalar
transformation of pairwise distances.
Every other eligible firm enters the candidate set, self-links are excluded,
and the assignments are target specific.
The squared reconstruction loss, simplex normalization, equal cloud size, and
the rule that selects among tied optimal assignments are maintained design
choices.

The target remains fixed throughout this reconstruction.
An unrestricted Wasserstein barycenter holds input laws and weights fixed and
optimizes over a new center distribution $Q$, for example through
$\inf_Q\sum_{j\ne i}a_j W_2^2(C_j,Q)$
\citep{agueh_carlier_barycenters_2011}.
Here the center is the target itself, and the optimization chooses the peer
coordinates $w_i$ that best reconstruct it.

Economically, a row of $W^\flat$ is a customized information-peer basket for
one firm.
The weights are distributional spanning weights: they identify the combination
of other firms whose aligned information footprints jointly come closest to the
target's, and the objective value records how well that convex combination
represents it.
Like benchmark weights, they are nonnegative and sum to one; economically, they
measure target-specific information-peer relevance.
Because the target determines both the alignments and the reconstruction
problem, $W^\flat$ is directed even though pairwise Wasserstein distance is
symmetric.
Firm $j$ can contribute strongly to reconstructing firm $i$ while $i$
contributes little to reconstructing $j$.
The collection of these target-specific relations is the measured map of the
latent information field.

\Cref{fig:w2-operator-construction} shows how article-level correspondence
becomes firm-level reconstruction relevance.

\begin{figure}[htbp]
  \centering
  \import{images/}{w2_operator_construction\ppfiguresuffix.pgf}
  \caption{Target-anchored reconstruction of one row of $W^\flat$.
    For target firm $i$, separate target-to-peer transport assignments establish
    article-level correspondence.
    Holding those assignments fixed, simplex coordinates combine the aligned
    peer clouds to reconstruct the target.
  The fitted coordinates form a nonnegative unit-sum row with zero self-weight.}
  \label{fig:w2-operator-construction}
  \figalttext{Four panels read left to right. Target article points are
    separately matched to three peer clouds; three decreasing simplex weights
    combine aligned peer points into a reconstruction of the target. The same
    weights become an interaction row with zero self-weight and unit row sum.
  The diagram illustrates the construction for one target.}
\end{figure}

Read from left to right, transport establishes correspondence, simplex
reconstruction selects joint coordinates, and the fitted row becomes the
target's peer average.
The simplex constraint and leave-one-out policy imply, row by row,

\[
  W^\flat_{ij}\geq0,\qquad W^\flat_{ii}=0,\qquad
  \sum_{j} W^\flat_{ij}=1.
\]

The construction therefore satisfies \Cref{def:interaction-matrix} with no
separate normalization of the rows and no kernel bandwidth.

\begin{proposition}[SAR compatibility]\label{thm:wflat-sar-gate}
  Every operator satisfying \Cref{def:w-flat} is entrywise nonnegative,
  row-stochastic, and zero-diagonal. In the induced $\ell^\infty$ operator
  norm, $\lVert W^\flat\rVert\leq1$; hence $|\rho|<1$ implies
  $\lVert\rho W^\flat\rVert<1$ and is sufficient for the SAR resolvent.
\end{proposition}

The $\ell^\infty$ norm is the maximum absolute row sum, so the proposition
follows from the display above: the constructed field enters the spatial
multiplier without further normalization.
\Cref{sec:results-anatomy} displays actual rows of the estimated field.

\subsection{Comparators}\label{sec:interaction-comparators}

The comparators identify which aspects of the peer baskets carry information
about returns.
They distinguish the contribution of fitted weights, within-firm
distributions, joint reconstruction, and the source used to define a link.

The equal-active-support comparator holds fixed the active peer set selected by
$W^\flat$ and changes only the relative weights within that set.
For each row, a peer is active when $W^\flat_{ij}>10^{-8}$, a numerical zero threshold.
Writing $A_i=\{j\ne i:W^\flat_{ij}>10^{-8}\}$, the
equal-active-support matrix is

\[
  W^{\mathrm{eq}}_{ij}
  =\frac{\mathbf{1}\{j\in A_i\}}{|A_i|}.
\]

It has exactly the same support as $W^\flat$ but weights every active peer
equally, so it measures what the fitted spanning weights add to peer
selection.

The centroid comparator holds fixed the target-anchored reconstruction problem
and changes only the representation of a firm.
It collapses each firm's articles to their mean embedding
$\bar x_i=M^{-1}\sum_{m=1}^M x_{im}$ and then solves the same
leave-one-out simplex problem in that averaged space, choosing for each target
the nonnegative weights summing to one that minimize
$\lVert\bar x_i-\sum_{j\ne i}w_{ij}\bar x_j\rVert_2^2$.
It therefore retains target-specific joint reconstruction while discarding the
within-firm dispersion that the distributional representation preserves, and it
separates the two components of the barycentric construction.

The radial basis function (RBF) comparator holds fixed the empirical laws,
their pairwise quadratic
Wasserstein distances, and the candidate universe, and replaces joint target
reconstruction with a separate proximity weight for each target--candidate
pair.
For $D_{ij}=W_2(\widehat C_i,\widehat C_j)$ and bandwidth $h>0$, it sets

\[
  W^h_{ij}
  =\frac{\mathbf{1}\{j\ne i\}\exp(-D_{ij}^2/h)}
  {\sum_{k\ne i}\exp(-D_{ik}^2/h)}.
\]

The primary bandwidth is the median off-diagonal value of $D_{ij}^2$.
The comparison measures whether peer dependence follows pairwise proximity or
the joint reconstruction of a fixed target.

The persistent news co-mention comparator holds fixed the firm universe, the
2018--2022 construction window, and the row-stochastic role, and changes the
information object and link rule.
The matrix $W^{\mathrm{news}}$ links firms that are the only two tagged names in
an article and co-occur in at least two calendar years during that period;
total co-mention counts are symmetrized and row normalized.
It contrasts language-model geometry with explicit shared coverage.
Because it satisfies \Cref{def:interaction-matrix}, it can also enter the
adjustment problem as a second channel beside $W^\flat$, and
\Cref{sec:results} uses it in both roles.

The industry/sector comparator assigns equal weight to all other firms in the
target's industry. If that set is empty, it falls back to the target's sector;
if the sector is also a singleton, it assigns equal weight to the remaining
priced firms. This rule is applied once to the frozen universe and then held
fixed through the return comparison.

\section{Peer Adjustment and Spatial Closure}\label{sec:spatial-closure}

\subsection{Stand-alone and peer-adjusted exposures}\label{sec:model}

The measured field describes which firms jointly represent each target's
information environment.
An exposure-adjustment model gives those peer baskets a role in firms'
responses to common risks.
Begin with the familiar factor model, in which an exposure is a vector of $K$
factor loadings in $\R^K$.
A firm's \emph{stand-alone exposure} $\xi_i$ is the loading vector implied by
its own information before any peer adjustment.
Peer adjustment maps $\xi_i$ into the \emph{peer-adjusted exposure} $B_i$, and
a factor bridge links $B_i$ to returns.

The quadratic criterion below provides a reduced-form representation of the
costs of operational, financing, or portfolio reconfiguration.
Keeping $\xi_i$ and $B_i$ separate distinguishes the exposure implied by a
firm's own information from the exposure after alignment with its peers.

Embedding positions describe information content, and factor loadings measure
sensitivity to common shocks.
A transmission map connects these different quantities.
Draw $X_i\sim C_i$, let $U_i$ collect idiosyncratic transmission randomness, and
let a common measurable map $T$ convert the information draw into the exposure
space:

\begin{equation}\label{eq:primitive-exposure}
  \xi_i = T(X_i,U_i).
\end{equation}

The characteristic law is measured in embedding-space units and $\xi_i$ in
factor-exposure units.
The map $T$ and its inputs are latent.

The empirical implementation uses $\R^K$ throughout; the supporting dispersion
appendix records the corresponding Hilbert-space extension.
Let $F_t\in\R^K$ be a centered common factor innovation with covariance matrix
$\Gamma$, and let $e_t^i$ be a centered idiosyncratic return component
that is square-integrable, orthogonal to $F_t$, and uncorrelated across firms.
The return bridge specifies how peer-adjusted exposure enters centered excess
returns:

\begin{equation}\label{eq:hilbert-return}
  \widetilde r_t^i = B_i^{\prime}F_t+e_t^i.
\end{equation}

Here $B_i$ has loading units, $F_t$ has factor-innovation units, and their inner
product has return units.
The equilibrium below is solved pointwise for each realization of $\xi$, so $B$
is random whenever $\xi$ is, even when $W$ and the adjustment weight are fixed.

The adjustment problem below balances departure from stand-alone exposure
against misalignment with information peers.
Its equilibrium has spatial coefficient
\[
  \rho=\frac{\lambda}{1+\lambda},
\]
where $\lambda$ is the relative penalty on peer misalignment.
\Cref{fig:spatial-exposure-arc} locates this model relation within the empirical
sequence from text to peer weights and return evaluation.

\begin{figure}[htbp]
  \centering
  \import{images/}{spatial_exposure_arc\ppfiguresuffix.pgf}
  \caption{From text geometry to model-implied adjustment. Text distributions
    construct the frozen barycentric interaction field $W^\flat$. In the latent
    model, firm-specific information generates stand-alone exposure $\xi_i$,
    which adjustment transforms into peer-adjusted exposure $B_i$. The return
    bridge connects exposure to returns, and quasi-maximum-likelihood
    estimation (QMLE) on the frozen field estimates the working-model
    coefficient $\hat\rho$ and hence the adjustment index
    $\hat\lambda=\hat\rho/(1-\hat\rho)$. Solid
    arrows denote construction or estimation;
  dashed arrows denote model relations.}
  \label{fig:spatial-exposure-arc}
  \figalttext{Flow diagram with three layers. In the observed and constructed
    layer, firm text distributions produce a frozen peer field; this field and
    returns feed quasi-maximum likelihood estimation. In the latent layer, text
    generates stand-alone exposure, and the peer field adjusts it before a model
    bridge maps exposure to returns. The estimated coefficient is converted to
    an adjustment index. Solid arrows are empirical construction or estimation;
  dashed arrows are maintained model relations.}
\end{figure}

\subsection{Exposure adjustment}\label{sec:exposure-adjustment}

Peer adjustment follows an as-if reduced-form criterion that summarizes costly
operational, financing, or portfolio reconfiguration.
The criterion balances fidelity to the firm's stand-alone exposure against
alignment with peer-weighted exposures, and it accommodates several peer
definitions at once.
Joint information reconstruction and explicit news links describe different
forms of peer relevance, so we state the criterion for the barycentric field
$W^\flat$ and the news-link field $W^{\mathrm{news}}$, each satisfying
\Cref{def:interaction-matrix}; each coefficient carries its field's
superscript.
Setting $\lambda^{\mathrm{news}}=0$ leaves a single field; the single-field
results hold for
any $W$ satisfying \Cref{def:interaction-matrix}, including every comparator.
The main text writes exposures as vectors in $\mathbb R^K$; the same arguments
extend to a real separable Hilbert space as noted in the appendix.

\begin{definition}[Peer adjustment problem]\label{def:adjustment-problem}
  \label{def:two-field-adjustment}
  Let $\lambda^\flat,\lambda^{\mathrm{news}}\geq0$.
  Holding the profile $b_{-i}$ of peer exposures fixed, firm $i$ chooses
  $a\in\mathbb R^K$ to minimize
  \begin{equation}\label{eq:two-field-objective}
    \frac{1}{2}\norm{a-\xi_i}^2
    +\frac{\lambda^\flat}{2}\sum_j W^\flat_{ij}\norm{a-b_j}^2
    +\frac{\lambda^{\mathrm{news}}}{2}\sum_j W^{\mathrm{news}}_{ij}\norm{a-b_j}^2 .
  \end{equation}
  A profile $b$ is an equilibrium when every $b_i$ minimizes
  \eqref{eq:two-field-objective} at $b_{-i}$.
\end{definition}

All terms are measured in squared exposure units.
The first penalizes departure from the stand-alone exposure $\xi_i$, and the
others penalize disagreement with each peer profile; $\lambda^\flat$ and
$\lambda^{\mathrm{news}}$
are dimensionless penalty weights relative to the stand-alone term.
The quadratic form and the fixed fields are maintained inputs to the model.
Write
\[
  \begin{gathered}
    \rho^\flat=\frac{\lambda^\flat}{1+\lambda^\flat+\lambda^{\mathrm{news}}},
    \qquad
    \rho^{\mathrm{news}}
    =\frac{\lambda^{\mathrm{news}}}{1+\lambda^\flat+\lambda^{\mathrm{news}}},
    \\
    s=\rho^\flat+\rho^{\mathrm{news}},
    \qquad
    R=\rho^\flat W^\flat+\rho^{\mathrm{news}} W^{\mathrm{news}}.
  \end{gathered}
\]

\begin{proposition}[Peer-adjustment equilibrium]\label{prop:peer-adjustment}
  \label{thm:adjustment-equilibrium}\label{thm:spatial-closure}
  \label{thm:two-field-closure}\label{cor:mixture-nesting}
  \label{thm:sar-reduced-form}\label{prop:channel-inversion}
  \label{cor:zero-feedback-nesting}
  Let $\lambda^\flat,\lambda^{\mathrm{news}}\geq0$ and let
  $W^\flat,W^{\mathrm{news}}$ satisfy
  \Cref{def:interaction-matrix}.
  (i) A profile $b$ is an equilibrium if and only if, for every $i$,
  \begin{equation}\label{eq:adjustment-stationarity}
    b_i-\xi_i+\lambda^\flat\sum_j W^\flat_{ij}(b_i-b_j)
    +\lambda^{\mathrm{news}}\sum_j W^{\mathrm{news}}_{ij}(b_i-b_j)=0 .
  \end{equation}
  (ii) Every equilibrium satisfies the spatial autoregression
  \begin{equation}\label{eq:two-field-sar}
    B=\rho^\flat\,W^\flat B+\rho^{\mathrm{news}}\,W^{\mathrm{news}}B+(1-s)\,\xi .
  \end{equation}
  With a single field this is
  \begin{equation}\label{eq:hilbert-sar}
    B=\rho\,WB+(1-\rho)\,\xi,
    \qquad
    \rho=\frac{\lambda}{1+\lambda}.
  \end{equation}
  (iii) If $s>0$, then $R=sW(\theta)$ for the mixture
  $W(\theta)=\theta W^\flat+(1-\theta)W^{\mathrm{news}}$ with
  $\theta=\lambda^\flat/(\lambda^\flat+\lambda^{\mathrm{news}})$, and
  $W(\theta)$ satisfies
  \Cref{def:interaction-matrix}.
  In every case $\norm{R}=s<1$ in the $\ell^\infty$-induced operator norm,
  so the equilibrium is unique and
  \begin{equation}\label{eq:hilbert-reduced-form}
    B=(1-s)(I-R)^{-1}\xi .
  \end{equation}
  (iv) The intensities are recovered from the coefficients by
  \begin{equation}\label{eq:channel-inversion}
    \lambda^\flat=\frac{\rho^\flat}{1-s},
    \qquad
    \lambda^{\mathrm{news}}=\frac{\rho^{\mathrm{news}}}{1-s},
  \end{equation}
  which with a single field reads $\lambda=\rho/(1-\rho)$.
  (v) At zero feedback, $\lambda^\flat=\lambda^{\mathrm{news}}=0$,
  peer-adjusted exposure
  equals stand-alone exposure: $B=\xi$.
\end{proposition}

Because the rows of each field sum to one, completing the square writes
\eqref{eq:two-field-objective} as
$\tfrac12(1+\lambda^\flat+\lambda^{\mathrm{news}})\norm{a-m_i}^2$ plus terms
free of $a$, where
\[
  m_i=\frac{\xi_i+\lambda^\flat\sum_j W^\flat_{ij}b_j
    +\lambda^{\mathrm{news}}\sum_j
  W^{\mathrm{news}}_{ij}b_j}{1+\lambda^\flat+\lambda^{\mathrm{news}}}.
\]
The unique minimizer is $a=m_i$, which is (i) and, after division by
$1+\lambda^\flat+\lambda^{\mathrm{news}}$, (ii); no differentiability
argument is required.
The mixture is entrywise nonnegative with a zero diagonal, and its rows are
convex combinations of unit row sums.
The nonnegative rows of $R$ therefore sum to $s=(\lambda^\flat+\lambda^{\mathrm{news}})/
(1+\lambda^\flat+\lambda^{\mathrm{news}})<1$, and the Neumann series for
$(I-R)^{-1}$ gives (iii).
Part (iv) follows from $1-s=(1+\lambda^\flat+\lambda^{\mathrm{news}})^{-1}$,
and (v) from $R=0$.

Equation~\eqref{eq:hilbert-sar} expresses spatial dependence as the equilibrium
of the adjustment problem: peer-adjusted exposure is a convex balance between
the peer average and the stand-alone exposure.
The coefficient $\rho$ is therefore a model-implied index of relative peer
adjustment.
A value $\rho=1/2$ gives both equal weight, and values approaching one place
progressively greater weight on peer alignment.
With two fields, $s$ is the total peer-adjustment weight, $\rho^\flat$ and
$\rho^{\mathrm{news}}$
divide it between the channels, and $1-s$ anchors exposure to the stand-alone
value; \eqref{eq:channel-inversion} measures each channel's coefficient against
that anchor.
Stability is automatic on this nonnegative-intensity branch.
The working return likelihood of \Cref{sec:estimation} searches the wider
interval $|\rho|<1$, a distinction \Cref{sec:estimand} keeps explicit.
The multiplier aggregates direct and iterated peer adjustment, while the factor
$1-s$ preserves the scale of the stand-alone exposures.
Part (v) recovers the exposure object of the pairwise covariance restriction in
\citet{gawronsky_continuous_2026}.

Peer adjustment also compresses cross-sectional heterogeneity in a controlled
way.
Under a nonnegative row-stochastic field, adjustment cannot increase
stationary-weighted exposure dispersion and retains at least
$\{(1-\rho)/(1+\rho)\}^2$ of it (\Cref{thm:spatial-attenuation} in
\ppAppendixRef{sec:dispersion}), so stronger peer alignment narrows exposure
differences across firms without erasing them.
\Cref{sec:identification-estimation} carries the one-field and two-field
equilibria to returns.

\section{From Exposure Adjustment to Returns}
\label{sec:identification-estimation}

\subsection{Scalar projection and the return equation}
\label{sec:exposures-to-returns}

The financial test asks whether return dependence follows the measured
information field.
Connecting that test to the exposure model requires a bridge from exposure
vectors to scalar returns.
Scalar projection carries the spatial operator and coefficient to the
systematic return component; a working quasi-likelihood then specifies the
observable return model.
The exposure vectors $\xi_i$ and $B_i$ and transmission map $T$ remain latent.
The application observes text laws $\widehat C_i$ and returns $r_t$, with
$W^\flat$ constructed from the pre-evaluation news window.
Projection onto a scalar risk direction preserves the exposure equilibrium.

\begin{proposition}[Scalar projection]\label{prop:scalar-projection}
  For every $h\in\mathbb R^K$, the scalar field $B_i^{\prime}h$ satisfies the
  ordinary spatial autoregression with the same $W$ and $\rho$, with reduced
  form $(1-\rho)(I-\rho W)^{-1}$ applied to $\xi_i^{\prime}h$.
\end{proposition}
By linearity of the dot product,
\[
  B_i^{\prime}h
  =\rho\sum_j W_{ij}B_j^{\prime}h
  +(1-\rho)\xi_i^{\prime}h.
\]
Stacking preserves the same $W$ and $\rho$; applying the inverse in
\Cref{thm:sar-reduced-form} gives the stated reduced form.

Define the systematic return component and the projected stand-alone exposure
by $s_t^i=B_i^{\prime}F_t$ and $z_t^i=\xi_i^{\prime}F_t$, and stack them
across firms.
Conditional on the realized factor direction $F_t$, scalar projection gives
$s_t=\rho Ws_t+(1-\rho)z_t$.
Since centered excess returns satisfy $\widetilde r_t=s_t+e_t$, substitution
yields

\begin{equation}\label{eq:model-return-innovation}
  \widetilde r_t
  =\rho W\widetilde r_t+x_t^{\mathrm{model}},
  \qquad
  x_t^{\mathrm{model}}
  =(1-\rho)z_t+(I-\rho W)e_t.
\end{equation}

Equation~\eqref{eq:model-return-innovation} is the return equation implied by
the exposure model.
Filtering the idiosyncratic components by $I-\rho W$ and adding the projected
stand-alone exposure generally makes $x_t^{\mathrm{model}}$ nonspherical and
correlated across firms, even when the components of $e_t$ are
cross-sectionally uncorrelated.
Allowing an intercept to absorb centering and mean components gives the
empirical notation

\begin{equation}\label{eq:spatial-lag}
  r_t = \rho W r_t + x_t .
\end{equation}

\subsection{Estimand}
\label{sec:estimand}

Three coefficients remain distinct throughout the paper.
The structural $\rho$ governs peer adjustment of latent exposures in
\Cref{sec:spatial-closure} and lies on the nonnegative-intensity branch $0\leq\rho<1$ of
\Cref{prop:peer-adjustment}, where $\lambda=\rho/(1-\rho)$.
For the concentrated working log-likelihood $\ell_t^Q(\rho;W)$ of
\Cref{sec:estimation}, the population target is the pseudo-true coefficient

\[
  \rho^\star
  \in\operatorname*{argmax}_{|\rho|<1}
  \E[\ell_t^Q(\rho;W)],
\]

and $\hat\rho$ is its sample estimate for a supplied $W$.
Identifying the population target with the structural coefficient requires the
population quasi-score to have its unique zero at the structural value despite
the richer innovation in \eqref{eq:model-return-innovation}.
This condition is additional to the algebraic result supplied by scalar
projection.

The fitted $\hat\rho$ measures conditional return dependence along the
supplied information structure.
Its transformation $\hat\lambda=\hat\rho/(1-\hat\rho)$ is an adjustment index
for this working return model.
Evaluation returns do not determine their own peer weights.
Text and returns can nevertheless share common determinants, so causal
interpretation requires an additional exclusion restriction or shock design.
This estimand and interpretation govern the empirical comparisons below.

\subsection{What the spatial field adds to a factor model}
\label{sec:factor-invariance}

The primary specification conditions excess returns on common factors with
unrestricted firm coefficients,
\begin{equation}\label{eq:factor-spatial-lag}
  (I-\rho W)r_t^e
  =\alpha+A f_t+\varepsilon_t,
  \qquad \varepsilon_t\mathrel{\overset{\mathrm{work}}{\sim}}
  \mathrm{iid}(0,\sigma^2I),
\end{equation}
where $r_t^e$ subtracts the daily risk-free rate, $\alpha$ contains
firm-specific intercepts, each row of $A$ contains one firm's loadings, $f_t$
contains the factor returns, and $W$ is the supplied interaction field,
$W^\flat$ in the primary specification.
Unrestricted firm coefficients pin down what the field can contribute.

\begin{proposition}[Factor-beta invariance]\label{prop:factor-beta-invariance}
  On a balanced panel with fixed $W$, common regressors $[1,f_t]$, and
  unrestricted firm coefficients, let $C_{\mathrm{OLS}}$ collect the
  coefficients of firm-by-firm least-squares regressions of $r_t^e$ on
  $[1,f_t]$.
  For every $\rho$ with $I-\rho W$ invertible, the profiled coefficients are
  $C(\rho)=(I-\rho W)C_{\mathrm{OLS}}$, and
  \begin{equation}\label{eq:factor-reduced-form-identity}
    (I-\rho W)^{-1}C(\rho)=C_{\mathrm{OLS}}.
  \end{equation}
\end{proposition}
Profiling applies least squares to $(I-\rho W)r_t^e$ on the common design, and
least squares is linear in the dependent variable, which gives $C(\rho)$ and
hence \eqref{eq:factor-reduced-form-identity}.

Under these conditions, standard factor exposures and fitted means remain
equal to their firm-by-firm regression estimates.
The field adds information through residual covariance: which firms'
unexpected returns move together, and how strongly.
The held-out comparisons evaluate how well alternative peer definitions
organize that joint distribution.

\section{Data and Empirical Design}
\label{sec:empirical-design}

\subsection{Data and timing}
\label{sec:panel-data}

The design evaluates the financial relevance of peer relations measured from
an earlier information window.
We constructed the barycentric field and every comparator field from
information ending in 2022 and froze each matrix.
We then evaluated conditional return dependence on
\ppvalue{post-pooled-2023-2026-w-flat-trading-days} common trading days from
\ppvalue{post-evaluation-date-start} through
\ppvalue{post-evaluation-date-end}.
No evaluation return enters the text geometry or any comparator matrix.
Return data serve two roles.
The reported post-corpus QMLE estimates are fitted on 2023--2026 returns, while
the held-out density exercise fits every parameter on 2018--2022 returns and
evaluates it on 2023--2026.

We assembled firm text from Nasdaq's public per-symbol archive of syndicated
news \citep{nasdaq_news_headlines_2026}, whose ticker-indexed retrieval and
dated article bodies support one reproducible collection rule across the
universe.
Ticker assignment follows the archive and is not manually validated at the
article level.
Daily simple returns are Yahoo Finance adjusted-close ratios minus one,
retrieved through a version-fixed \texttt{yfinance} client
\citep{yahoo_finance_historical_2026,aroussi_yfinance_2026}.
The primary specification subtracts the daily risk-free rate and aligns the
excess-return panel with the daily Fama--French five factors and momentum.
We keep only dates on which every firm has a return; missing returns are not filled.

The universe is a prespecified Nasdaq-100-based frame admitted by an annual
coverage screen requiring at least \ppvalue{annual-raw-article-floor} raw
articles per firm in each year from 2018 through 2022.
Intersecting firms that have a frozen embedding cloud with firms that have a
compatible return history yields \ppvalue{n-tickers-priced} continuously and
heavily covered large-cap firms, a survivor-conditioned cross-section.
The annual floor preserves coverage in every construction year; both that
floor and the balanced cloud size below were fixed before the return
comparisons.
The return window ends on 6 April 2026, the final common date before HOLX's
take-private,
so the 2026 fold covers
\ppvalue{post-eval2026-w-flat-trading-days} dates while the 2023--2025 folds
cover complete years.
\ppAppendixRef{sec:app-sample} reports sector composition and bridges the
estimates to an earlier 52-firm specification.

Each firm's information environment is a balanced cloud of
\ppvalue{paper5-balanced-cloud-size} articles, selected from the 2018--2022
window by a deterministic URL-hash order.
The common cloud size puts every empirical law on the support required by the
balanced assignment in \eqref{eq:w2-assignment}; it makes firms comparable and,
by design, removes raw news volume from the geometry.
It also fixes the assignment problem's computational support across firms.
We encoded each article with the full-width Qwen3-Embedding-8B model, which is
trained to place semantically related texts near one another
\citep{qwen3_embedding_2025}, and row-normalized its 4,096-coordinate vector.
The encoder was released after the 2018--2022 article window, so the primary
representation is a retrospective measurement of earlier news.
The earlier-cutoff comparison in \Cref{sec:results-representation}
examines dependence on this retrospective measurement.
Each article has equal mass in the firm's empirical distribution.
The representation exercises in \Cref{sec:results-representation} vary output
width, model capacity, model family, and reported information vintage
(\ppAppendixRef{sec:p5-representation-ablation}).

\subsection{Estimation, inference, and out-of-period evaluation}
\label{sec:estimation}

We estimate conditional return dependence by quasi-maximum likelihood
(QMLE).
The primary specification is
\eqref{eq:factor-spatial-lag} with the barycentric field $W^\flat$, firm-specific
intercepts, and loadings on the market, size, value,
profitability, investment, and momentum factors.
At every candidate $\rho$, least squares profiles all firm intercepts and
loadings jointly with $\sigma^2$, and the concentrated objective retains the
spatial Jacobian $T\log|I-\rho W|$.
The search covers the stable region $|\rho|<1$.
The factor-only benchmark sets $\rho=0$ with the same intercepts, loadings,
dates, and innovation specification, so comparisons with it isolate the value
of the field.
Following \citet{ge_news-implied_2023}, a control ladder repeats the primary
specification under FF3, FF3 plus momentum, and FF5 on the same dates; FF5 plus
momentum remains primary throughout.
A robustness specification profiles firm-specific innovation variances
$D=\operatorname{diag}(\sigma_1^2,\ldots,\sigma_N^2)$, because ignored
heteroskedasticity can move a spatial point estimate as well as its standard
error \citep{kelejian_prucha_2010}.
Maximized quasi-log-likelihood (QLL) describes fit on the estimation panel.
The appendix reports the associated working AIC, Hannan--Quinn (HQC), and BIC
criteria, their definitions, and parameter counts
(\ppAppendixRef{sec:app-information-criteria}).

Two bootstrap procedures supply inference. A restricted-null stationary
bootstrap tests $H_0:\rho=0$ through the quasi-likelihood-ratio (QLR) statistic
by independently resampling standardized firm residuals on stationary-block
paths, restoring the fitted variances and factor means before refitting both
models. Interval estimates and every held-out comparison use 2,000 paired
date-block refits on one common resampling schedule, which refits the
calibration model, the evaluation coefficients, and the scored parameters on
each draw; all models and both transport orders share every draw. Both
procedures condition on the corpus, embeddings, and constructed fields, so they
propagate return sampling and nuisance-parameter uncertainty. Implementation
details are given in \ppAppendixRef{sec:app-inference}.

The main financial evaluation asks whether the field improves the joint
distribution assigned to later returns.
All parameters are fitted on 2018--2022 returns and held fixed while 2023--2026
returns are scored by the mean Gaussian quasi-log score per firm-day.
Higher scores mean that the model assigns greater density to the realized
cross-section of returns.
Evaluation-date factors are observed when scoring, so the comparison concerns
conditional joint density given realized factor returns.
Paired intervals resample calibration and evaluation dates on separate
stationary-block schedules shared by all models.

\subsection{Transport order and the comparator fields}
\label{sec:two-field-estimation}

The transport-order comparison repeats the target-anchored construction under
linear ($p=1$) and quadratic ($p=2$) displacement cost, holding fixed the firm
clouds, ground space, leave-one-out candidates, and simplex.
For $W^\flat_1$, each row minimizes the sum of Euclidean reconstruction norms
after order-one optimal assignments; for $W^\flat$, the sum of squared norms
after order-two assignments.
Tables and the comparison below label the two orders W1 and W2.
Changing the order changes how embedding-space displacement enters the
measurement of the field; the exposure-adjustment criterion that maps $\rho$
into $\lambda$ is the same for both, so the comparison measures sensitivity to
the transport geometry rather than the curvature of adjustment costs.
Before the final estimation we fixed an equivalence margin for the
W2-minus-W1 held-out score and a weak-separation rule that classifies the
fields as nearly collinear from the correlation of their induced peer-return
signals; when the separation rule binds, the design reports paired contrasts
rather than interpreting a joint coefficient allocation. Both thresholds and
their construction are stated in \ppAppendixRef{sec:app-inference}.
A Matryoshka
dimension ladder repeats both constructions at
$D\in\{4096,1024,256,64\}$ coordinates of the same model
\citep{kusupati_matryoshka_2022,qwen3_embedding_2025} to locate where any
difference between the orders enters (\ppAppendixRef{sec:p5-matryoshka}).

Each comparator field changes one element of the peer definition, holding the
firms, dates, factor controls, and likelihood fixed.

\begin{table}[H]
  \centering
  \caption{Comparator specifications and the element of the dependence
    definition each one changes. All fields use firm information dated through
    2022 and are held fixed throughout the 2023--2026 evaluation; covariance
  inputs are calibrated on 2018--2022 returns.}
  \label{tab:comparator-design}
  \small
  \begin{tabular}{p{0.3\linewidth}p{0.62\linewidth}}
    \toprule
    Specification & Relative to the barycentric field \\
    \midrule
    Equal active support & Same peers; equal instead of fitted weights \\
    RBF--Wasserstein (median or validation-tuned bandwidth) & Same
    distributions and pairwise distances; separate distance-decay weight per
    pair instead of joint reconstruction \\
    Centroid reconstruction & Same simplex reconstruction; each firm's cloud
    collapsed to its mean embedding \\
    Persistent news co-mentions & Explicit shared coverage instead of
    language-model geometry \\
    Industry/sector & Conventional classification \\
    Direct Ledoit--Wolf residual covariance & No peer field; constant-correlation
    shrinkage of the factor-residual covariance \\
    W1 barycentric & Linear instead of quadratic transport cost \\
    \bottomrule
  \end{tabular}
\end{table}

The equal-support, RBF, and co-mention fields are defined in
\Cref{sec:interaction-comparators}.
The centroid field solves the leave-one-out simplex problem on mean
embeddings, and the tuned RBF selects its bandwidth multiplier from
$\{0.25,0.5,1,2,4\}$ using only a 2018--2020 fit and 2021--2022 validation
split before refitting through 2022.
Because these fields are non-nested, their held-out gains are compared through
paired stationary-block intervals rather than QLR statistics.

As a structure-agnostic covariance benchmark, we estimate the Ledoit--Wolf
linear-shrinkage covariance of the factor residuals in the 2018--2022
calibration sample~\citep{ledoit_wolf_2004}. The estimator shrinks the sample
covariance toward a constant-correlation target that preserves each residual
variance and replaces pairwise correlations with their cross-sectional mean;
the shrinkage intensity is estimated from the calibration residuals. We hold
the factor fitted means and this covariance fixed when scoring the same
2023--2026 observations. This benchmark asks whether the information field
adds held-out density beyond a conventional covariance estimator without an
economic peer structure.

The relation between the barycentric and news-link fields is estimated jointly,
as the two-channel model of \Cref{thm:two-field-closure} prescribes, under the
same primary factor-conditioned return specification.
Scalar projection applies to \eqref{eq:two-field-sar} exactly as to the
one-field equilibrium, giving

\begin{equation}\label{eq:two-field-spatial-lag}
  (I-\rho^\flat W^\flat-\rho^{\mathrm{news}} W^{\mathrm{news}})r_t^e
  =\alpha+A f_t+\varepsilon_t .
\end{equation}

This comparison uses the firm-specific intercepts and factor loadings of the
primary model. Its matched-random overlap tests use the same frozen fields and
return panel.
The raw-return, common-intercept counterpart, which also supplies the annual
re-estimates, is reported in the appendix as a corroborating specification:
\[
  r_t=\alpha\1+\rho W r_t+\varepsilon_t,
  \qquad \varepsilon_t\mathrel{\overset{\mathrm{work}}{\sim}}
  \mathrm{iid}(0,\sigma^2I),
\]
extended to two fields as in \eqref{eq:two-field-spatial-lag}.
Both matrices enter unchanged: residualizing one on the other would create
negative entries and non-unit row sums and destroy the peer-average
interpretation of \Cref{def:interaction-matrix}, so the coefficient would no
longer map to an adjustment intensity.
Residualizing the induced signals $W^\flat r_t$ and $W^{\mathrm{news}}r_t$
instead would use the contemporaneous returns and so would not yield
predetermined fields for the spatial likelihood.
Because $I-\rho^\flat W^\flat-\rho^{\mathrm{news}}
W^{\mathrm{news}}=I-(\rho^\flat+\rho^{\mathrm{news}})W(\theta)$,
the concentrated objective is the one-field objective at the mixture, and the
restrictions $0\leq\theta\leq1$ and $0\leq\rho^\flat+\rho^{\mathrm{news}}<1$ impose
$\lambda^\flat,\lambda^{\mathrm{news}}\geq0$.
The single-field fits are the boundaries $\theta=1$ and $\theta=0$ of this
family, so the joint quasi-log-likelihood cannot fall below either.
The statistic
$\mathrm{QLR}^{\mathrm{news}}=2\{\sup_{\rho^\flat,\rho^{\mathrm{news}}\geq 0}
  \ell^Q(\rho^\flat,\rho^{\mathrm{news}})
-\sup_{\rho^{\mathrm{news}}=0}\ell^Q(\rho^\flat,\rho^{\mathrm{news}})\}$ tests
$\rho^{\mathrm{news}}=0$, $\mathrm{QLR}^\flat$ tests $\rho^\flat=0$
symmetrically, and each has its
reference distribution simulated under the restricted single-field fit because
the null lies on the boundary
(\ppAppendixRef{sec:p5-two-field-implementation}).

\section{Results}\label{sec:results}

\subsection{Information structure beyond standard factors}
\label{sec:results-factor-value}

The information field measured from news organizes residual return dependence
in the later evaluation period.
With all parameters fitted on 2018--2022 returns, the barycentric field
$W^\flat$ raises the
2023--2026 mean conditional quasi-log score by
\ppnum[4]{spatial-value-spherical-w2-versus-none-held-out-score-gain} per
firm-day relative to the factor-only model, with 95\% paired interval
$[\ppnum[4]{spatial-value-spherical-w2-versus-none-held-out-score-gain-ci-lower},
\ppnum[4]{spatial-value-spherical-w2-versus-none-held-out-score-gain-ci-upper}]$.
The positive gain means that information-peer relations improve the joint
distribution assigned to subsequent returns after conditioning on standard
factors.
The linear-cost W1 field also improves the score, with its
interval above zero.

\Cref{tab:spatial-value} compares the fields with the matched factor-only
model, holding FF5-plus-momentum controls, firm coefficients, dates, and the
innovation specification fixed.
The field adds a single spatial coefficient.
On the post-corpus estimation panel, W1 gains
\ppnum[1]{spatial-value-spherical-w1-versus-none-loglik-gain} QLL units and W2
gains \ppnum[1]{spatial-value-spherical-w2-versus-none-loglik-gain}, and the
restricted-null bootstrap rejects zero spatial feedback for both fields, at
$p=\ppnum[4]{spatial-value-spherical-w1-versus-none-qlr-pvalue}$ for W1 and
$p=\ppnum[4]{spatial-value-spherical-w2-versus-none-qlr-pvalue}$ for W2.
The information criteria also favor each spatial model
(\ppAppendixRef{sec:app-information-criteria}).

\begin{table}[H]
  \centering
  \caption{Primary spatial-value comparison under spherical innovations.
    Every row uses FF5 plus momentum and the same firm-specific coefficients.
    QLL is maximized on the post-corpus panel. Restricted-null QLR $p$-values
    use 2,000 independently resampled firm paths under the working $\rho=0$
    null; zero exceedances are reported at the add-one floor.
    Out-of-period conditional score gains are relative to the factor-only model,
    with 95\% paired stationary-bootstrap intervals; higher is better.
  Evaluation-date factors are observed when scoring.}
  \label{tab:spatial-value}
  \begin{tabular}{lrrrr}
    \toprule
    Model & $\hat\rho$ & QLL & QLR $p$ & Held-out gain \\
    \midrule
    Factor only &
    \ppnum[3]{spatial-value-spherical-none-rho} &
    \ppnum[1]{spatial-value-spherical-none-loglik} &
    --- & --- \\
    Target-anchored W1 &
    \ppnum[3]{spatial-value-spherical-w1-rho} &
    \ppnum[1]{spatial-value-spherical-w1-loglik} &
    \ppnum[4]{spatial-value-spherical-w1-versus-none-qlr-pvalue} &
    $\ppnum[4]{spatial-value-spherical-w1-versus-none-held-out-score-gain}$
    $[\ppnum[4]{spatial-value-spherical-w1-versus-none-held-out-score-gain-ci-lower},
    \ppnum[4]{spatial-value-spherical-w1-versus-none-held-out-score-gain-ci-upper}]$
    \\
    Target-anchored W2 &
    \ppnum[3]{spatial-value-spherical-w2-rho} &
    \ppnum[1]{spatial-value-spherical-w2-loglik} &
    \ppnum[4]{spatial-value-spherical-w2-versus-none-qlr-pvalue} &
    $\ppnum[4]{spatial-value-spherical-w2-versus-none-held-out-score-gain}$
    $[\ppnum[4]{spatial-value-spherical-w2-versus-none-held-out-score-gain-ci-lower},
    \ppnum[4]{spatial-value-spherical-w2-versus-none-held-out-score-gain-ci-upper}]$
    \\
    \bottomrule
  \end{tabular}
\end{table}

For the primary field $W^\flat$, the factor-conditioned estimate is
$\hat\rho=\ppnum[3]{transport-w2-rho}$, with 95\% interval
$[\ppnum[3]{transport-w2-rho-ci-lower},
\ppnum[3]{transport-w2-rho-ci-upper}]$, and the corresponding
adjustment index is
$\hat\lambda=\ppnum[2]{transport-w2-lambda}$, with interval
$[\ppnum[2]{transport-w2-lambda-ci-lower},
\ppnum[2]{transport-w2-lambda-ci-upper}]$.
Factor conditioning matters for magnitude: the matched raw-return,
common-intercept estimate is $\ppnum[3]{transport-w2-unadjusted-rho}$, so the
factors absorb part, but only part, of the dependence that a raw-return model
assigns to the field.
The remaining dependence is stable across common-factor controls: the
estimate ranges from \ppnum[3]{transport-factor-ff3-w2-rho} under FF3 to
\ppnum[3]{transport-factor-ff5-mom-w2-rho} under FF5 plus momentum, and every
bootstrap interval excludes zero
(\ppAppendixRef{sec:app-information-criteria}).
That stability parallels the co-mention evidence of
\citet{ge_news-implied_2023}, whose firm coefficients and sample differ from
ours.

The improvement concerns how unexpected returns move together.
Under the unrestricted firm coefficients of
\Cref{prop:factor-beta-invariance}, reduced-form factor betas and fitted means
equal their firm-by-firm least-squares values; the estimated models reproduce
that identity to within $10^{-10}$.
The held-out gain therefore reflects residual covariance.
Allowing every firm its own innovation variance raises the estimates to
\ppnum[3]{spatial-value-diagonal-w1-rho} for W1 and
\ppnum[3]{spatial-value-diagonal-w2-rho} for W2, rejects $\rho=0$ for both
fields, and keeps both held-out intervals above zero
(\Cref{tab:spatial-value-diagonal}).

\subsection{Anatomy of the barycentric field}
\label{sec:results-anatomy}

\begin{figure}[htbp]
  \centering
  \import{images/}{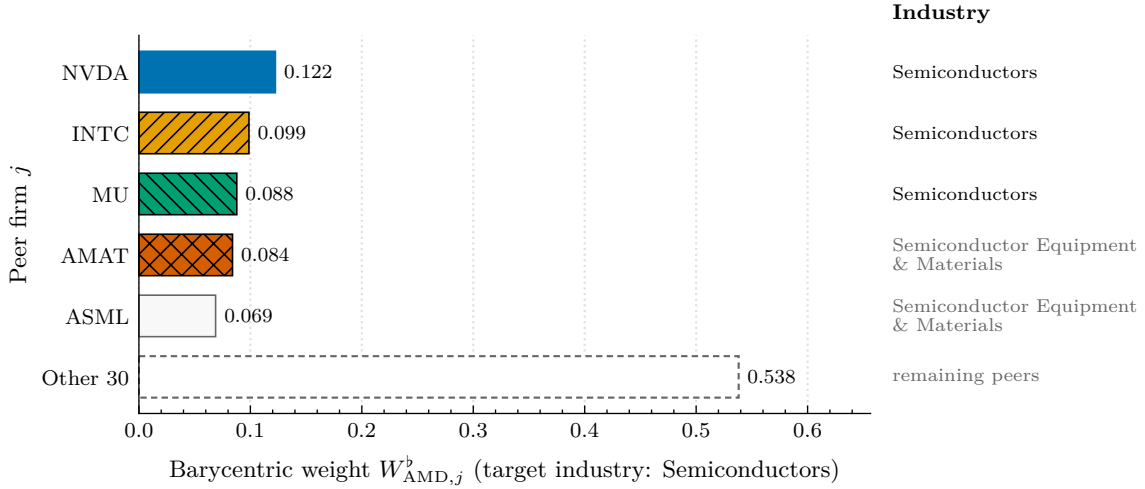}
  \caption{Anatomy of one barycentric peer basket. Bars show AMD's five largest
    weights $W^\flat_{ij}$ in the estimated field and the combined weight of
    all remaining peers; the row sums to one and excludes AMD. The right-hand
    column gives
  each peer's industry; peers in AMD's industry are set in black.}
  \label{fig:peer-basket}
  \figalttext{Horizontal bar chart for target AMD. Five named peer firms carry
    the largest reconstruction weights, each well below one half, and a final
  bar collects the weight spread over all remaining firms.}
\end{figure}

\Cref{fig:peer-basket} opens one row of the estimated field.
AMD's basket combines firms from complementary chip-design, equipment-supply,
and memory segments, led by NVDA, INTC, MU, AMAT, and ASML.
No single peer dominates: the five largest weights account for less than half
of the row, and the remaining mass is spread across many firms.
The row is a synthetic information-peer benchmark for AMD, constructed without
reference to industry codes or returns.

\begin{figure}[htbp]
  \centering
  \import{images/}{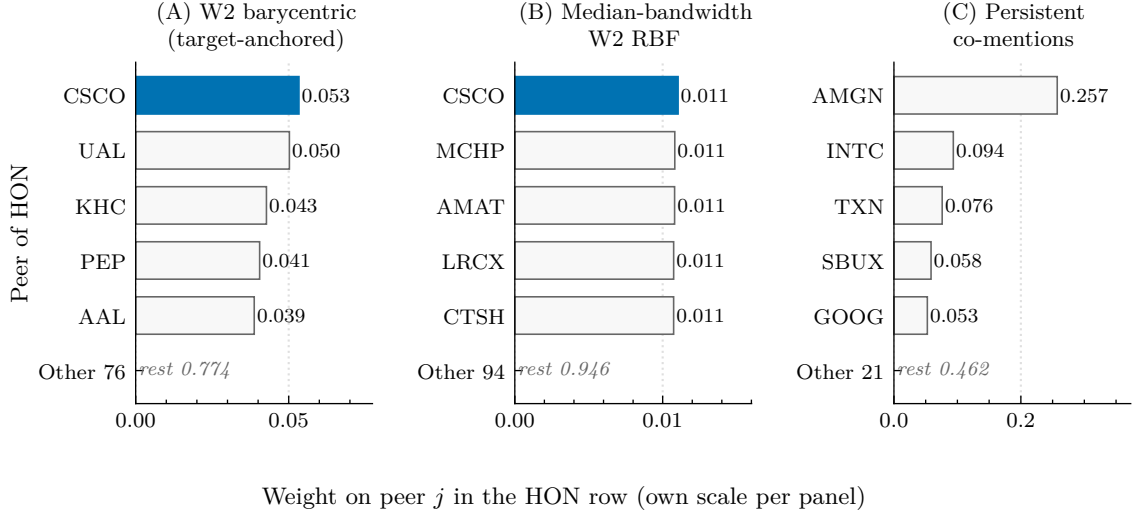}
  \caption{One target under three peer definitions. Panel-specific scales are
    used; compare composition, not bar length across panels. Each
    panel shows HON's
    five largest weights and the combined remaining weight under the
    barycentric field $W^\flat$ (left), the median-bandwidth pairwise
    RBF field built
    from the same Wasserstein distances (center), and the persistent news
    co-mention field (right). Shading marks
  a peer that appears in more than one panel.}
  \label{fig:peer-definitions}
  \figalttext{Three horizontal bar charts for target HON. The barycentric
    panel spreads weight over a mixed set of peers, including a firm that is
    far from HON in pairwise distance; the RBF panel is nearly flat; the
  co-mention panel concentrates on a few firms reported with HON in news.}
\end{figure}

Reconstruction relevance differs from pairwise proximity.
\Cref{fig:peer-definitions} places one target, HON, under three peer
definitions built from the same 2018--2022 information.
HON is selected mechanically as the target with the largest effective source
count, using only the frozen information field and no return outcomes; its
barycentric row is therefore the most diffuse in the sample, which makes the
contrast sharp. With the median bandwidth, the pairwise RBF row is nearly
uniform over all candidates, tilting only slightly toward the firms nearest
to HON in Wasserstein distance, and the co-mention row concentrates on a few
firms that appear alongside HON in news articles. The barycentric row selects
the combination that jointly reconstructs HON's footprint, and it assigns
UAL, only the 51st-closest firm by
pairwise distance, the second-largest weight
(\Cref{tab:p5-peer-interpretation}).
A firm far from the target on average can still contribute strongly to
reproducing one part of its information distribution, and the full field is
directed and heterogeneous across targets
(\ppAppendixRef{sec:app-field-anatomy}).

\subsection{Joint reconstruction and simpler peer definitions}
\label{sec:results-horse-race}

Joint reconstruction provides the largest held-out gain among the tested
peer-field specifications.
\Cref{tab:p5-economic-field-comparison} holds firms, dates, factor controls,
and the return likelihood fixed while changing the peer definition.
Among peer-field specifications, the two barycentric fields record the largest
held-out gains over the factor-only model,
\ppnum[4]{economic-w1-held-out-gain} for W1 and
\ppnum[4]{economic-w2-held-out-gain} for W2.

\begin{table}[H]
\centering
\scriptsize
\setlength{\tabcolsep}{3.0pt}
\begin{tabular}{lrrrrr}
\toprule
Dependence specification & $\hat\rho$ & QLL & Held-out score & Gain & 95\% CI \\
\midrule
Factor-only & 0.000 & 205748.3 & 2.468881 & 0.000000 & -- \\
Direct Ledoit--Wolf residual covariance & -- & -- & 2.558644 & 0.089763 & [-0.0028, 0.0903] \\
W1 barycentric & 0.684 & 209488.3 & 2.526065 & 0.057184 & [0.0497, 0.0656] \\
W2 barycentric & 0.687 & 209476.0 & 2.525957 & 0.057076 & [0.0496, 0.0653] \\
Median-bandwidth W2 RBF & 0.467 & 206227.6 & 2.479192 & 0.010311 & [0.0082, 0.0129] \\
Calibration-tuned W2 RBF & 0.572 & 206604.0 & 2.484870 & 0.015989 & [0.0135, 0.0191] \\
Centroid reconstruction & 0.470 & 209343.1 & 2.522691 & 0.053810 & [0.0460, 0.0625] \\
Equal active support & 0.619 & 207216.3 & 2.494105 & 0.025224 & [0.0218, 0.0294] \\
Persistent co-mentions & 0.466 & 209240.3 & 2.519320 & 0.050439 & [0.0428, 0.0586] \\
Industry/sector fallback & 0.229 & 207873.3 & 2.501014 & 0.032133 & [0.0259, 0.0390] \\
\bottomrule
\end{tabular}
\caption{Factor-conditioned peer-field and covariance benchmarks. All rows use the primary FF5-plus-momentum controls, the same firms and dates; fields use firm information dated through 2022 and are held fixed throughout the 2023--2026 evaluation, and covariance inputs are calibrated on 2018--2022 returns. The held-out score is the mean Gaussian quasi-log score per firm-day on 2023--2026; gain is relative to the factor-only fit. The direct Ledoit--Wolf row estimates a conventional factor-residual covariance from the calibration sample and therefore has no spatial coefficient or QLL. Intervals are paired stationary-block bootstrap intervals that refit every comparator on each calibration draw, matching the resampling architecture of \Cref{tab:spatial-value}.}
\label{tab:p5-economic-field-comparison}
\end{table}

The conventional residual-covariance benchmark has the higher point estimate
on the same held-out score: its gain is
\ppnum[4]{economic-ledoit-wolf-held-out-gain}, compared with
\ppnum[4]{economic-w2-held-out-gain} for $W^\flat$.
Its own gain interval spans zero once calibration uncertainty enters, while
the peer-field gain intervals remain positive.
Across the common draws the paired Ledoit--Wolf-minus-barycentric
difference is positive
in a share \ppnum[3]{economic-ledoit-wolf-minus-w2-positive-share} of
replications, with 95\% interval
$[\ppnum[4]{economic-ledoit-wolf-minus-w2-ci-lower},
\ppnum[4]{economic-ledoit-wolf-minus-w2-ci-upper}]$.
That interval includes zero, so the comparison does not establish which model
has the higher expected score.
Direct shrinkage estimates covariance itself; the barycentric construction
supplies an explicit peer structure whose financial content is evaluated by
the same score.

Pairwise proximity captures only a small part of that gain.
The median-bandwidth RBF field uses the same empirical laws and the same
pairwise Wasserstein distances as $W^\flat$ but converts each
distance separately
into a weight; its gain is \ppnum[4]{economic-rbf-median-held-out-gain}.
Choosing the RBF bandwidth on the 2018--2022 validation split raises the gain
only to \ppnum[4]{economic-rbf-tuned-held-out-gain}, and neither RBF interval
overlaps the barycentric intervals.
The large increase from proximity weighting to joint reconstruction locates
most of the field's return content in the joint representation of the target.

The fitted weights matter beyond the peer set they select.
Weighting the active peers of $W^\flat$ equally yields a gain of
\ppnum[4]{economic-equal-support-held-out-gain}, less than half the
barycentric gain, with a disjoint interval.
A centroid reconstruction, which collapses each firm's articles to their mean
embedding before solving the same leave-one-out simplex problem, attains
\ppnum[4]{economic-centroid-held-out-gain}.
Because the two models use the same evaluation dates, the paired
barycentric-minus-
centroid score difference is \ppnum[4]{economic-w2-minus-centroid-estimate}
with 95\% interval
$[\ppnum[4]{economic-w2-minus-centroid-ci-lower},
\ppnum[4]{economic-w2-minus-centroid-ci-upper}]$; $W^\flat$ is higher in a share
\ppnum[3]{economic-w2-minus-centroid-positive-share} of the draws.
The centroid therefore retains most of the field's score gain, while the full
distribution contributes a small but measurable increment beyond
target-specific reconstruction.
Persistent news co-mentions attain \ppnum[4]{economic-co-mentions-held-out-gain}
and the industry/sector fallback
\ppnum[4]{economic-industry-sector-held-out-gain}.
The barycentric point estimates exceed both, and the relation between the
language-model field and explicit news links is taken up directly in
\Cref{sec:field-overlap}.
The spatial coefficient itself does not rank the fields: median-bandwidth RBF
and co-mentions have nearly equal $\hat\rho$ but held-out gains that differ by
a factor of five.
In-sample quasi-log-likelihood in the raw-return, common-intercept
specification orders the barycentric, co-mention, equal-support, and RBF
fields in the same way (\Cref{tab:qmle-pooled}).

\subsection{Language-model peers and explicit news links}
\label{sec:field-overlap}

Information reconstruction and shared news coverage yield different peer
maps.
The barycentric and co-mention fields choose partly overlapping peer sets,
with different weights and related peer-return signals.
\Cref{tab:field-overlap} separates peer selection, cardinal weighting, and the
induced peer-return series, and evaluates overlap against matched-random
baskets because row sparsity sets a floor for peer overlap and common return
variation sets one for signal correlation.

\begin{table}[H]
  \centering
  \caption{Agreement between the frozen barycentric interaction field and the
    persistent co-mention field. The null draws
    \ppvalue{overlap-support-draws}
    fixed-seed baskets carrying $W^\flat$'s own row sparsities; $p$ is the
    share of null draws reaching the observed value. Rows with no mechanical
  floor carry no null.}
  \label{tab:field-overlap}
  \begin{tabular}{llrrr}
    \toprule
    Stage & Statistic & Observed & Null & $p$ \\
    \midrule
    Peer sets & Directed share of $W^\flat$ peers &
    \ppnum[3]{overlap-directed-share} &
    \ppnum[3]{overlap-support-null-share} &
    \ppnum[3]{overlap-support-null-pvalue} \\
    & Rank-matched share &
    \ppnum[3]{overlap-rank-matched-share} & --- & --- \\
    & Jaccard overlap &
    \ppnum[3]{overlap-jaccard-mean} & --- & --- \\
    \midrule
    Weights & Off-diagonal correlation &
    \ppnum[3]{overlap-weight-offdiagonal-pearson} & --- & --- \\
    & Cosine on the support union &
    \ppnum[3]{overlap-weight-union-cosine} & --- & --- \\
    & Rank correlation on common edges &
    \ppnum[3]{overlap-weight-common-spearman} & --- & --- \\
    \midrule
    Induced field & Correlation of $W r_t$ &
    \ppnum[3]{overlap-induced-pearson} &
    \ppnum[3]{overlap-induced-null-pearson} &
    \ppnum[3]{overlap-induced-null-pvalue} \\
    & Coefficient of determination &
    \ppnum[3]{overlap-induced-r-squared} & --- & --- \\
    \bottomrule
  \end{tabular}
\end{table}

The barycentric rows are the denser of the two, carrying
\ppnum[1]{overlap-w-flat-support-size} active peers on average against
\ppnum[1]{overlap-co-mentions-support-size} for the co-mention graph, and
\ppnum[3]{overlap-directed-share} of the barycentric peers are also
co-mention peers.
A random basket with the same row sparsities reaches
\ppnum[3]{overlap-support-null-share}, an excess of
\ppnum[1]{overlap-support-excess-pct} percentage points
($p=\ppnum[3]{overlap-support-null-pvalue}$): the two maps overlap more than
chance but remain distinct.
Their cardinal weights agree only partially, with off-diagonal correlation
\ppnum[3]{overlap-weight-offdiagonal-pearson} and a lower rank correlation of
\ppnum[3]{overlap-weight-common-spearman} over the
\ppnum[1]{overlap-weight-common-count} edges per row they share.
The induced peer-return series correlate more strongly, at
\ppnum[3]{overlap-induced-pearson}, but any two row-stochastic averages of a
co-moving cross-section are correlated: a random basket at $W^\flat$'s
sparsity already reaches \ppnum[3]{overlap-induced-null-pearson}.
The observed correlation exceeds that floor
($p=\ppnum[3]{overlap-induced-null-pvalue}$), and per-firm correlations range
from \ppnum[3]{overlap-induced-asset-min} to
\ppnum[3]{overlap-induced-asset-max}.
Common return variation therefore makes the fields look more alike as return
signals than as peer maps.

\subsection{Two channels in one adjustment problem}
\label{sec:two-field-results}

Combining the information field with explicit news links improves held-out
joint density beyond either field alone.
The joint model of \Cref{thm:two-field-closure} assigns a separate coefficient
to each peer definition.
\Cref{tab:two-field-pooled} reports the post-corpus factor-conditioned
estimates, with each channel's index given by \eqref{eq:channel-inversion}.

\begin{table}[H]
  \centering
  \caption{Pooled 2023--2026 joint barycentric and news-link estimates under
    the primary FF5-plus-momentum specification. The 95\% intervals are
    descriptive quantiles from 2,000 joint-date stationary-bootstrap refits.
    The working-null QLR comparisons use restricted factor-conditioned
  stationary-bootstrap calibrations.}
  \label{tab:two-field-pooled}
  \begin{tabular}{lrrrr}
    \toprule
    Channel & $\hat\rho$ & 95\% interval & $\hat\lambda$ & 95\% interval \\
    \midrule
    Barycentric ($W^\flat$) &
    \ppnum[3]{post-pooled-2023-2026-factor-two-field-rho-b} &
    $[\ppnum[3]{post-pooled-2023-2026-factor-two-field-rho-b-ci-lower},
    \ppnum[3]{post-pooled-2023-2026-factor-two-field-rho-b-ci-upper}]$ &
    \ppnum[2]{post-pooled-2023-2026-factor-two-field-lambda-b} &
    $[\ppnum[2]{post-pooled-2023-2026-factor-two-field-lambda-b-ci-lower},
    \ppnum[2]{post-pooled-2023-2026-factor-two-field-lambda-b-ci-upper}]$ \\
    News-link ($W^{\mathrm{news}}$) &
    \ppnum[3]{post-pooled-2023-2026-factor-two-field-rho-n} &
    $[\ppnum[3]{post-pooled-2023-2026-factor-two-field-rho-n-ci-lower},
    \ppnum[3]{post-pooled-2023-2026-factor-two-field-rho-n-ci-upper}]$ &
    \ppnum[2]{post-pooled-2023-2026-factor-two-field-lambda-n} &
    $[\ppnum[2]{post-pooled-2023-2026-factor-two-field-lambda-n-ci-lower},
    \ppnum[2]{post-pooled-2023-2026-factor-two-field-lambda-n-ci-upper}]$ \\
    Total &
    \ppnum[3]{post-pooled-2023-2026-factor-two-field-rho-total} &
    $[\ppnum[3]{post-pooled-2023-2026-factor-two-field-rho-total-ci-lower},
    \ppnum[3]{post-pooled-2023-2026-factor-two-field-rho-total-ci-upper}]$
    & & \\
    \midrule
    \bottomrule
  \end{tabular}
\end{table}

With parameters fitted on the calibration period, the joint model improves
the 2023--2026 score relative to the barycentric-only model by
\ppnum[4]{post-pooled-2023-2026-factor-two-field-held-out-joint-minus-barycentric-estimate}
with 95\% interval
$[\ppnum[4]{post-pooled-2023-2026-factor-two-field-held-out-joint-minus-barycentric-ci-lower},
\ppnum[4]{post-pooled-2023-2026-factor-two-field-held-out-joint-minus-barycentric-ci-upper}]$,
and relative to the news boundary by
\ppnum[4]{post-pooled-2023-2026-factor-two-field-held-out-joint-minus-news-estimate}
with interval
$[\ppnum[4]{post-pooled-2023-2026-factor-two-field-held-out-joint-minus-news-ci-lower},
\ppnum[4]{post-pooled-2023-2026-factor-two-field-held-out-joint-minus-news-ci-upper}]$.
Both intervals lie above zero, so each field adds out-of-period information
conditional on the other.
The restricted-null tests also reject each exclusion: adding the news-link
field to the barycentric boundary gives
$\mathrm{QLR}^{\mathrm{news}}=\ppnum[1]{post-pooled-2023-2026-factor-two-field-qlr-news-given-barycentric-statistic}$
($p=\ppnum[4]{post-pooled-2023-2026-factor-two-field-qlr-news-given-barycentric-pvalue}$),
and adding the barycentric field to the news boundary gives
$\mathrm{QLR}^\flat=\ppnum[1]{post-pooled-2023-2026-factor-two-field-qlr-barycentric-given-news-statistic}$
($p=\ppnum[4]{post-pooled-2023-2026-factor-two-field-qlr-barycentric-given-news-pvalue}$).
Across \ppvalue{post-pooled-2023-2026-factor-two-field-trading-days} common
dates, the joint model improves on the barycentric-only and news-only
boundaries by
\ppnum[1]{post-pooled-2023-2026-factor-two-field-loglik-gain-w-flat} and
\ppnum[1]{post-pooled-2023-2026-factor-two-field-loglik-gain-co-mentions}
quasi-log-likelihood units, respectively.
The barycentric channel receives the larger share: its index is
$\hat\lambda^\flat=\ppnum[2]{post-pooled-2023-2026-factor-two-field-lambda-b}$,
with
interval
$[\ppnum[2]{post-pooled-2023-2026-factor-two-field-lambda-b-ci-lower},
\ppnum[2]{post-pooled-2023-2026-factor-two-field-lambda-b-ci-upper}]$, and the
news-link index is
$\hat\lambda^{\mathrm{news}}=\ppnum[2]{post-pooled-2023-2026-factor-two-field-lambda-n}$,
with
interval
$[\ppnum[2]{post-pooled-2023-2026-factor-two-field-lambda-n-ci-lower},
\ppnum[2]{post-pooled-2023-2026-factor-two-field-lambda-n-ci-upper}]$.
Under the adjustment interpretation of \eqref{eq:two-field-objective}, the
larger barycentric index corresponds to a stronger penalty on misalignment
with the barycentric peer average than with co-mentioned firms.

The two channel coefficients share the total dependence assigned to peers.
The joint total $\hat\rho^\flat+\hat\rho^{\mathrm{news}}=
\ppnum[3]{post-pooled-2023-2026-factor-two-field-rho-total}$ has interval
$[\ppnum[3]{post-pooled-2023-2026-factor-two-field-rho-total-ci-lower},
\ppnum[3]{post-pooled-2023-2026-factor-two-field-rho-total-ci-upper}]$.
The channel coefficients have bootstrap correlation
\ppnum[3]{post-pooled-2023-2026-factor-two-field-channel-correlation},
so total peer
adjustment is estimated more sharply than its division between channels
under the factor-conditioned specification.
Language-model reconstruction and explicit co-coverage are thus related but
economically distinct definitions of a firm's information peers, and the
barycentric definition receives the larger share of adjustment.

The raw-return, common-intercept joint estimates and their uncertainty region
remain in \ppAppendixRef{sec:app-unadjusted} as a corroborating specification.

\subsection{Transport-order equivalence}
\label{sec:results-transport-order}

The financial evidence survives replacing quadratic transport costs with
linear costs.
\Cref{tab:transport-order} compares target-anchored W1 and W2 fields on the
same \ppvalue{transport-n-assets} firms and
\ppvalue{transport-evaluation-trading-days} dates; the two constructions share
the corpus, embeddings, candidate sets, and simplex restriction and differ only
in the power of the displacement cost.

\begin{table}[H]
  \centering
  \caption{Factor-conditioned transport-order sensitivity. Excess returns are
    conditioned on firm-specific intercepts and loadings on the Fama--French
    five factors and momentum. The 95\% intervals use 2,000 paired
    joint-date stationary-bootstrap refits. The held-out score fits all
    parameters on 2018--2022 returns and evaluates the mean Gaussian
    quasi-log score per firm-day on 2023--2026 returns; higher is better.
  The adjustment index is $\hat\lambda=\hat\rho/(1-\hat\rho)$.}
  \label{tab:transport-order}
  \begin{tabular}{lrrrr}
    \toprule
    Interaction field & $\hat\rho$ & 95\% interval & $\hat\lambda$
    & Held-out score \\
    \midrule
    Target-anchored W1 &
    \ppnum[3]{transport-w1-rho} &
    $[\ppnum[3]{transport-w1-rho-ci-lower},
    \ppnum[3]{transport-w1-rho-ci-upper}]$ &
    \ppnum[2]{transport-w1-lambda} &
    \ppnum[6]{transport-w1-predictive-score} \\
    Target-anchored W2 &
    \ppnum[3]{transport-w2-rho} &
    $[\ppnum[3]{transport-w2-rho-ci-lower},
    \ppnum[3]{transport-w2-rho-ci-upper}]$ &
    \ppnum[2]{transport-w2-lambda} &
    \ppnum[6]{transport-w2-predictive-score} \\
    \bottomrule
  \end{tabular}
\end{table}

W1 yields $\hat\rho=\ppnum[3]{transport-w1-rho}$, and the paired W2-minus-W1
difference is $\ppnum[6]{transport-delta-rho}$, with 95\% interval
$[\ppnum[6]{transport-delta-rho-ci-lower},
\ppnum[6]{transport-delta-rho-ci-upper}]$.
The difference is precisely estimated but economically negligible, and W2 has
the slightly lower post-corpus fit, with a quasi-log-likelihood difference of
$\ppnum[2]{transport-delta-loglik}$
$[\ppnum[2]{transport-delta-loglik-ci-lower},
\ppnum[2]{transport-delta-loglik-ci-upper}]$.
Out of period, the W2-minus-W1 score difference is
$\ppnum[6]{transport-delta-predictive-score}$
$[\ppnum[6]{transport-delta-predictive-score-ci-lower},
\ppnum[6]{transport-delta-predictive-score-ci-upper}]$.
Its 90\% interval, from \ppnum[6]{transport-equivalence-ci-lower} to
\ppnum[6]{transport-equivalence-ci-upper}, lies inside the
$\pm\ppnum[6]{transport-equivalence-margin}$ margin fixed in advance, so
the two fields are \ppvalue{transport-equivalence-status}, and the equivalence
holds across the factor ladder.

W1 and W2 are nearly the same object: their off-diagonal weights correlate at
\ppnum[3]{transport-weight-correlation},
their active supports have Jaccard overlap
\ppnum[3]{transport-active-support-jaccard}, and their induced peer-return
signals correlate at \ppnum[4]{transport-signal-correlation}, with a difference
whose standard deviation is \ppnum[3]{transport-relative-signal-difference-sd}
of the W2 signal's.
Factor residualization leaves that classification unchanged. Both separation
thresholds bind, so the comparison rests on the paired contrasts above; a joint
W1 and W2 specification, added after the standalone comparison, places its
optimum on the W1-only boundary with strongly negatively correlated channel
estimates (\Cref{tab:transport-order-joint}).

A dimension ladder over the normalized Qwen3-Embedding-8B prefix locates the
equivalence in the construction
(\ppAppendixRef{sec:p5-matryoshka},
\ppAppendixRef{sec:app-transport-order}).
Shorter prefixes change which articles the two orders match, but the changed
assignments do not survive reconstruction: operator and spatial-lag
correlations remain effectively one at every dimension, and W2 never improves
the held-out score.
Target reconstruction therefore maps different article-level alignments into
nearly identical financial operators.

\subsection{Representation robustness}
\label{sec:results-representation}

The representation checks examine dependence on the embedding model and
its reported training-information vintage.
Across Qwen3-Embedding widths from 4,096 to 64 coordinates, the smaller
Qwen3-Embedding-4B model, the BGE-large-v1.5 model family, and all eight
published ChronoBERT checkpoints with reported 2017--2024 training cutoffs,
the barycentric field attains a higher quasi-log-likelihood than its matching
RBF field in the raw-return, common-intercept specification, while both fields
retain positive spatial coefficients
(\Cref{tab:p5-representation-ablation}).
Magnitudes respond to aggressive compression: from 1,024 coordinates downward,
barycentric feedback and its margin over RBF decline.
The annual ChronoBERT staircase is non-monotone: many paired changes in
barycentric feedback and its gap relative to RBF reject zero after Holm
adjustment, while only a subset meets the post-specified joint equivalence
rule (\Cref{tab:p5-representation-contrasts}).
The first annual step rejects zero for both estimands and meets joint
equivalence; the 2017-to-2024 endpoint meets joint equivalence without
Holm-adjusted rejection for either estimand.
These cumulative-training comparisons do not identify a causal leakage effect.
The pre-period ChronoBERT 2017 check preserves a positive barycentric gain of
\ppnum[4]{point-in-time-w2-held-out-gain} and a barycentric-minus-RBF
difference of
\ppnum[4]{point-in-time-w2-minus-rbf-estimate}, with 95\% interval
$[\ppnum[4]{point-in-time-w2-minus-rbf-ci-lower},
\ppnum[4]{point-in-time-w2-minus-rbf-ci-upper}]$
(\ppAppendixRef{sec:p5-point-in-time}).
Annual re-estimation under the frozen field yields positive adjustment in each
of 2023, 2024, 2025, and the partial 2026 sample, for the barycentric field
alone and jointly with the news-link field
(\Cref{tab:qmle-annual,tab:two-field-annual}).

\section{Discussion}\label{sec:limitations}

The evidence supports a compositional interpretation of economic peers.
A firm receives weight because it helps a combination of firms represent the
target's information environment.
Its relevance therefore depends on the other firms available for
reconstruction, as well as on its own resemblance to the target.
This interpretation explains why a distant firm can receive substantial
weight and why reversing the target and candidate can change the relation.
The resulting basket is a synthetic information-peer benchmark.
Across firms, these baskets provide an empirical map of a latent information
field.
Its weights describe reconstruction relevance within the observed universe;
a direct commercial link between two firms is neither required nor inferred
from a high weight.

The distinction between joint representation and pairwise proximity also
explains the comparator results.
Pairwise distances describe how separately chosen firms differ.
Reconstruction uses the candidate firms together, allowing their information
footprints to represent different parts of the target.
Equal weighting loses much of this information even when the peer set is
unchanged.
Centroid reconstruction retains most of the financial content, showing that
the joint organization of peers matters more here than the detail of the
distributional representation.
Full article distributions refine that organization by retaining within-firm
heterogeneity.
This finding gives a concrete answer to when semantic measurement is useful:
the way representations are combined into an economic object materially
affects their information about return dependence.

The information field adds a dimension of dependence to conventional factor
analysis.
Its peer weights are measured from news, and subsequent returns evaluate
their financial relevance.
Under the unrestricted firm coefficients of
\Cref{prop:factor-beta-invariance}, standard factors continue to describe
exposure to common systematic risks, while the field organizes the covariance
of residual returns.
News co-mentions contribute additional information because shared coverage
and joint representability describe different aspects of firms' information
environments.
Their partly overlapping peer maps and positive conditional held-out gains
support using both when the research question calls for these distinct
relations.

The covariance benchmark helps delimit the field's practical role.
Direct residual shrinkage has a higher held-out point score, but its paired
difference from the barycentric model is uncertain once calibration
uncertainty is included.
The comparison establishes no ranking between the two approaches at the
reported confidence level.
The field supplies an explicit, text-derived peer structure that can be
inspected and interpreted alongside its financial performance.
Peer benchmarking and spillover research can use such a structure as an input,
with causal uses requiring the additional identification described in
\Cref{sec:estimand}.
Portfolio applications require their own decision-based validation; the
appendix GMV exercise does not establish a portfolio advantage
(\ppAppendixRef{sec:app-gmv}).

The application covers continuously and heavily covered large-cap firms in a
survivor-conditioned US cross-section.
The observed text, candidate universe, and embedding model all shape the map.
Broader markets, successive information vintages, and other corpora would
test whether the same measurement captures stable peer relations across
market segments and information regimes.
These extensions would establish the reach of the latent-field
interpretation beyond the present setting.

\section{Conclusion}\label{sec:conclusion}

Our results support a compositional view of a firm's economic peers.
A firm's relevant peer need not be the single firm most similar to it.
It can instead be a weighted combination of firms whose information
environments jointly represent different parts of its activities.
Language-model representations make those activities measurable from corporate
news, and target-anchored optimal transport converts them into firm-specific
peer baskets.
Across firms, those baskets form a directed information field.

The broader contribution is to make a familiar object in spatial finance, the
interaction structure linking firms, an object of measurement.
Spatial asset-pricing models require a matrix $W$ describing which firms
interact and with what weight.
Rather than specifying that matrix from geography, industry membership, or
contemporaneous returns, we measure it from an earlier information environment
and then ask whether subsequent return dependence follows the resulting
structure.
In this sense, language models serve as a measurement instrument rather than a
return-prediction device: they provide a representation from which an
economically interpretable financial object can be constructed and
independently evaluated.

The return evidence gives that measured object financial content.
The field constructed from 2018--2022 news improves held-out joint-density fit
over 2023--2026 after conditioning on the Fama--French five factors and
momentum.
Because unrestricted firm coefficients leave conventional factor betas
unchanged, the gain lies in residual covariance: information about which firms'
unexpected returns move together after common factor exposures are accounted
for.
The comparator evidence further identifies what matters in the measurement.
Target-specific joint reconstruction contains substantially more information
than pairwise proximity, fitted peer weights matter beyond peer selection, and
full article distributions add a smaller increment beyond centroid
reconstruction.
Explicit news co-mentions contribute complementary information, indicating
that joint representability and direct shared coverage capture distinct
dimensions of economic relatedness.

This distinction matters wherever economic information is richer than the
available history of return co-movement.
Firms accumulate observable information about products, customers, suppliers,
technologies, and risks that need not conform to a single industry
classification and can precede a long trading record.
Our evidence does not establish causal transmission or portfolio gains, nor
does it test newly listed firms directly.
It establishes the prior measurement result: a peer structure recovered from
earlier text contains information about subsequent financial dependence.

More generally, the value of AI for empirical finance need not lie only in
improving prediction.
It can lie in making previously latent economic objects measurable and placing
them back inside interpretable financial models.
Here, language models provide the representation, optimal transport organizes
it into economic peers, and spatial asset pricing supplies the financial
discipline with which that structure is evaluated.
The result is a measurable map of the information environment in which firms
are embedded, and a framework for asking how that environment is reflected in
capital markets.

\section*{Data availability statement}

Firm news records were obtained from Nasdaq's public ticker-indexed news
archive \citep{nasdaq_news_headlines_2026}.
Daily adjusted-price histories were obtained from Yahoo Finance
\citep{yahoo_finance_historical_2026} through the version-fixed
\texttt{yfinance} client \citep{aroussi_yfinance_2026}.
Article vectors were produced with \texttt{qwen/qwen3-embedding-8b} through
OpenRouter; the embedding model is documented by
\citet{qwen3_embedding_2025}.
The provider-hosted articles, model endpoint, and market data remain subject to
their providers' terms and are not redistributed.
A citable public release containing the versioned code, environment
specifications, derived interaction fields, estimation outputs,
computer-verified proofs and scripts needed to regenerate the manuscript's
tables and figures are publicly available to aid in reproducible science.

\section*{Funding}

The authors received no financial support for the research, authorship, or
publication of this article.

\section*{Disclosure statement}

The authors report no potential conflict of interest.

\section*{Declaration of generative AI use}

During the preparation of this work the authors used Codex~0.146.0 and
Claude Code~2.1.223 in order to assist with code refactoring and editorial
review.
After using these tools, the authors reviewed and edited the content as needed
and take full responsibility for the content of the publication.

\bibliographystyle{plainnat}
\bibliography{refs}

\appendix
\section{Cross-Sectional Wasserstein Dispersion and Spatial Attenuation}
\label{sec:dispersion}

The dispersion analysis is the only part of the paper that uses a free-center
problem.
Unlike the fixed-target reconstruction in \Cref{sec:interaction-operator}, the
unrestricted Wasserstein barycenter problem below selects a center distribution
$Q$ that summarizes heterogeneity across firms.

This appendix bounds how much characteristic-implied exposure heterogeneity
survives peer adjustment.
The answer is economically simple: characteristic
dispersion places a floor on stand-alone exposure dispersion, and
adjustment through an admissible peer field cannot erase more than a
coefficient-dependent share of that floor.
The carrier restrictions introduced below are maintained rather than
estimated.

Fix cross-sectional weights $q=(q_1,\ldots,q_N)$ in the unit simplex.
Assume that the laws considered below have finite second moments.
The simplex restriction makes the weights nonnegative with unit sum,
so they determine each firm's contribution without changing the scale
of the aggregate comparison.
Finite second moments make the quadratic transport costs below well defined.
Economically, $\mathcal D_q(C_1,\dots,C_N)$ is the least weighted
separation compatible with all observed firm-level characteristic laws.
It is therefore a conservative measure of cross-sectional information
heterogeneity rather than the dispersion generated by one selected matching.
Formally, for characteristic laws on $(\Omega,d_\Omega)$ and
$(X_1,\dots,X_N)\sim\gamma$, define
\[
  \mathcal D_q(C_1,\dots,C_N)
  =\inf_{\gamma\in\Pi}
  \mathbb E_\gamma\Big[\sum_{i<j}q_{i}q_{j}\,d_\Omega(X_i,X_j)^2\Big],
\]
with $\Pi$ the joint laws on $\Omega^N$ whose $i$th marginal is $C_i$.
The infimum asks how close the laws could be under their most
favorable common coupling.
Its units are squared embedding distance, and its value depends on
the chosen ground metric and weights.

The Hilbert-space restriction begins to matter for the center
representation: for any laws $R_1,\ldots,R_N$ on a Hilbert space, the
same functional satisfies
$\mathcal D_q(R_1,\ldots,R_N)=\inf_Q\sum_i q_i\,W_2^2(R_i,Q)$
\citep{agueh_carlier_barycenters_2011,gawronsky_distance_mpt_2026}.
Here $Q$ is free and summarizes cross-sectional dispersion.
Its minimizer is the unrestricted Wasserstein barycenter of the firm laws under
weights $q$.

For two firms, the aggregate functional reduces to the familiar
pairwise Wasserstein distance.

\begin{theorem}[Pairwise nesting]\label{thm:pairwise-nesting}
  For $N=2$,
  \begin{equation}\label{eq:pairwise-nesting}
    \mathcal D_{(q_1,q_2)}(C_1,C_2)=q_1q_2\,W_2^2(C_1,C_2).
  \end{equation}
\end{theorem}
For two marginals, $\Pi$ is exactly the set of couplings of $C_1,C_2$.
The objective is therefore $q_1q_2$ times the quadratic transport
cost, so taking the infimum gives Equation~\eqref{eq:pairwise-nesting}.

Thus the geometric object nests the squared Wasserstein term used to
derive the pairwise covariance envelope in \citet{gawronsky_continuous_2026}.
The equality establishes geometric nesting only; it does not import
that paper's covariance or coupling assumptions.

The Wasserstein dispersion measure is in embedding units, whereas
stand-alone exposures are in factor-loading units, so an explicit
transfer restriction is needed before the two can be compared.
Write $Z_i=\Gamma^{1/2}\xi_i$ and $P_i=\mathcal L(Z_i)$ for the
stand-alone exposure laws in covariance coordinates.
The transformation by $\Gamma^{1/2}$ places exposure differences in
the risk coordinates used by the dispersion certificate.
Let $t:\Omega\to\mathcal H$ be a common measurable carrier in those coordinates.
Commonness supplies one benchmark across firms, and measurability
makes the carrier image of each information draw a valid random exposure.
For some $L>0$, impose the noncollapse condition at the point where
characteristic distance must become exposure distance:
\[
  \norm{t(x)-t(y)}\geq L^{-1}d_\Omega(x,y)
  \quad\text{for every }x,y\in\Omega.
\]
This lower-distance restriction prevents the common carrier from
erasing separation between distinct information states.
No upper-distance bound enters the result.
For each firm, next allow synchronous slack $\tau_i\geq0$:
\[
  \norm{\Gamma^{1/2}T(X_i,U_i)-t(X_i)}\leq\tau_i
  \quad\text{almost surely}.
\]
The same draw $X_i$ appears on both sides because the bound must
couple firm $i$'s actual stand-alone exposure to the carrier image of
that information state.
The slack permits firm-specific transmission noise of at most
$\tau_i$ in risk-coordinate norm.
Together, noncollapse and synchronous slack state the entire
characteristic-to-exposure restriction used below.
They do not restrict $W$ or the peer-adjustment mechanism.

Writing $\tau_q=(\sum_i q_i\tau_i^2)^{1/2}$ for weighted
root-mean-square slack and $[a]_+=\max\{a,0\}$, we maintain the
following characteristic-to-exposure transfer inequality:
\begin{equation}\label{eq:imported-transfer}
  \sqrt{\mathcal D_q(P_1,\dots,P_N)}
  \geq
  \left[L^{-1}\sqrt{\mathcal D_q(C_1,\dots,C_N)}-\tau_q\right]_+.
\end{equation}
The root-mean-square aggregation matches the cross-sectional weights,
while the positive part sets the lower floor to zero when
transmission slack absorbs the carrier-adjusted separation.
This restriction is stated here in full;
\citet{gawronsky_distance_mpt_2026} study its portfolio implications.

Observable separation is informative whenever its carrier-adjusted
magnitude exceeds the root-mean-square transmission slack.
At zero slack \eqref{eq:imported-transfer} becomes $\mathcal
D_q(P_1,\ldots,P_N)\geq L^{-2}\mathcal D_q(C_1,\ldots,C_N)$.
The carrier, $L$, and $\tau_i$ are maintained inputs rather than
estimated objects, so \eqref{eq:imported-transfer} is a conditional
restriction and is not numerically calibrated here.

\subsection{Spatial attenuation under peer adjustment}

The transfer inequality stops at stand-alone exposure dispersion;
attenuation enters only to ask what the spatial closure does to that floor.
The reduced form in \Cref{thm:sar-reduced-form} maps stand-alone
exposure $\xi$ into peer-adjusted exposure $B$ through the normalized
spatial multiplier:

\begin{equation}\label{eq:sar-multiplier}
  S_\rho=(1-\rho)(I-\rho W)^{-1},
  \qquad B=S_\rho\,\xi .
\end{equation}

The restriction $0\leq\rho<1$ makes the Neumann expansion of $S_\rho$
a convex mixture of current and iterated peer averages.
Nonnegativity and unit row sums make each application of $W$ an
average, which is the property needed for Jensen's inequality below.
Because $W$ is directed, however, equal cross-sectional weights need
not be preserved by peer averaging.
Let $\pi$ instead be a strictly positive stationary distribution, so
$\pi^\top W=\pi^\top$.
Economically, $\pi_i$ measures firm $i$'s long-run influence under
repeated peer averaging.
Stationarity makes the aggregate mean invariant to $W$, while strict
positivity keeps every firm in the comparison and makes the weighted
quadratic norm nondegenerate.
For any exposure profile $z$, define

\begin{equation}\label{eq:cross-sectional-dispersion}
  V_\pi(z)=\sum_i\pi_i\norm{z_i-\bar z_\pi}^2,
  \qquad \bar z_\pi=\sum_i\pi_i z_i .
\end{equation}

Under these restrictions, peer adjustment cannot increase
cross-sectional dispersion.
It also cannot eliminate more than a coefficient-dependent share: at
least $\{(1-\rho)/(1+\rho)\}^2$ of stand-alone dispersion remains.
The finite, nonempty cross-section in the theorem keeps the weighted
mean and sums well defined.
The next result formalizes the bracket without requiring $W$ to be symmetric.

\begin{theorem}[Spatial attenuation]\label{thm:spatial-attenuation}
  Let $W$ be nonnegative and row stochastic, let $\pi$ be a strictly positive
  stationary distribution, and let $0\leq\rho<1$. For every finite nonempty
  cross-section $z$,
  \begin{equation}\label{eq:attenuation}
    \Big(\frac{1-\rho}{1+\rho}\Big)^{2}V_\pi(z)
    \;\leq\;
    V_\pi(S_\rho z)
    \;\leq\;
    V_\pi(z).
  \end{equation}
\end{theorem}
To establish the upper bound, let $f=z-\bar z_\pi\1$ and define
$\lVert f\rVert_\pi^2=V_\pi(z)$.
Stationarity makes $Wf$ centered whenever $f$ is, while nonnegative
unit rows permit Jensen's inequality:
\[
  \sum_i\pi_i\lVert(Wf)_i\rVert^2
  \leq\sum_{i,j}\pi_i W_{ij}\lVert f_j\rVert^2
  =\sum_j\pi_j\lVert f_j\rVert^2.
\]
The final equality uses stationarity, so $W$ contracts the centered
norm and hence so does every $W^k$.
Because $0\leq\rho<1$, the Neumann mixture
$S_\rho=(1-\rho)\sum_{k\geq0}\rho^k W^k$ is a convex combination of
these contractions.
This proves the upper bound.
For the lower bound, the centered profile $g=S_\rho f$ satisfies
$(1-\rho)f=(I-\rho W)g$.
The triangle inequality and the contraction $\lVert
Wg\rVert_\pi\leq\lVert g\rVert_\pi$ give
\[
  (1-\rho)\lVert f\rVert_\pi
  \leq\lVert g\rVert_\pi+\rho\lVert Wg\rVert_\pi
  \leq(1+\rho)\lVert g\rVert_\pi.
\]
Rearranging and squaring gives the lower bound in \eqref{eq:attenuation}.

At $\rho=0$, the two bounds coincide and peer adjustment leaves
dispersion unchanged.
For every admissible $\rho$, peer-adjusted exposures cannot be more
dispersed than stand-alone exposures, while the guaranteed retained
fraction $\{(1-\rho)/(1+\rho)\}^2$ declines with $\rho$.
Evaluating this fraction at the fitted working-model spatial-feedback
coefficient $\widehat\rho$ gives a descriptive plug-in lower
envelope, not the exact amount of attenuation or a calibration of
$t$, $L$, or $\tau_i$.
It has a structural interpretation as a retained-share bound only
under the return-projection conditions of
\Cref{sec:identification-estimation}.
The bracket needs neither symmetry nor an additional spectral-gap
restriction, but it need not be sharp for the fitted operator.

The transfer and attenuation bounds can now be composed because they
measure dispersion in the same risk coordinates.
Set $Y_i=\Gamma^{1/2}B_i$, the risk-coordinate form of peer-adjusted
exposure, so linearity gives $Y=S_\rho Z$.
Then set the dispersion weights $q=\pi$.
This is not an arbitrary aggregation choice: the stationary weights
both preserve the mean under the directed operator and determine each
firm's contribution to the characteristic dispersion floor.
\begin{corollary}[Spatial-dispersion certificate]
  \label{cor:spatial-dispersion-certificate}
  Under the carrier and slack conditions stated above and the hypotheses of
  \Cref{thm:spatial-attenuation},
  \begin{equation}\label{eq:spatial-dispersion-certificate}
    \mathbb E\,V_\pi(Y)
    \geq
    \Big(\frac{1-\rho}{1+\rho}\Big)^2
    \left[L^{-1}\sqrt{\mathcal D_\pi(C_1,\dots,C_N)}-\tau_\pi\right]_+^2.
  \end{equation}
\end{corollary}
The complete chain is visible in one display.
The identity $V_\pi(Z)=\sum_{i<j}\pi_i\pi_j\lVert Z_i-Z_j\rVert^2$
links weighted variance to the multimarginal objective, and the
actual joint law of $Z$ is one feasible coupling.
Therefore
\[
  \begin{aligned}
    \mathbb E\,V_\pi(Y)
    &\geq
    \Big(\frac{1-\rho}{1+\rho}\Big)^2\mathbb E\,V_\pi(Z) \\
    &\geq
    \Big(\frac{1-\rho}{1+\rho}\Big)^2
    \mathcal D_\pi(P_1,\ldots,P_N) \\
    &\geq
    \Big(\frac{1-\rho}{1+\rho}\Big)^2
    \left[
      L^{-1}\sqrt{\mathcal D_\pi(C_1,\ldots,C_N)}-\tau_\pi
    \right]_+^2.
  \end{aligned}
\]
The first step applies attenuation realization by realization, the
second uses feasibility of the actual stand-alone exposure coupling,
and the third squares the transfer inequality.
This ordering shows how observed information heterogeneity becomes a
stand-alone exposure floor and then a peer-adjusted exposure floor
under the common adjustment coefficient $\rho$.
Information separation, carrier distortion, transmission noise, and
peer adjustment each enter once with a known direction.

The certificate is expressed in the latent risk coordinates of
peer-adjusted exposure, whereas the application observes scalar returns.
\Cref{sec:identification-estimation} states the return projection that
preserves $W$ and $\rho$ and relates the structural coefficient to the
working-model index estimated by QMLE.

\section{Stability of the Barycentric Field}
\label{sec:asymptotics}

This appendix establishes when the constructed field is a stable
peer-averaging system and measures how far the fitted model lies from the
strong-interaction boundary.
We first state the maintained conditions and derive their implications for peer
averaging, stationary weighting, and the spatial multiplier.
We then report the fitted residual, Dobrushin, and finite-$\rho$ diagnostics
before characterizing the rank-one boundary.
These results certify the constructed field, although their matrix implications
apply to any row-stochastic interaction matrix that satisfies the stated
conditions.

\subsection{Maintained operator conditions}

The main argument requires a nonnegative row-stochastic interaction matrix
$W$, a strictly positive stationary probability vector $\pi$, and a coefficient
$0\leq\rho<1$, for which \Cref{prop:peer-adjustment} gives a unique equilibrium.
Row stochasticity defines the directed peer average, the stationary vector
supplies the weighting in \Cref{thm:spatial-attenuation}, and the stability
condition makes the spatial multiplier well defined.
Let $P=\1\pi^\top$.
All matrix norms here are maximum absolute row-sum norms.

\ppAppendixRef{sec:app-network-diagnostics} evaluates these conditions for the fitted
barycentric field $W^\flat$.

The boundary analysis also requires a rate at which repeated peer averaging
approaches $P$.
The following certificate states that condition.

\begin{definition}[Geometric Perron certificate]
  \label{def:perron-limit}
  A row-stochastic matrix $W$ has a geometric Perron certificate
  $(\pi,C,q)$ if $0\leq q<1$ and
  \[
    \lVert W^k-P\rVert \leq Cq^k
    \qquad\text{for every }k\geq0.
  \]
\end{definition}

\subsection{Implications for peer averaging and the multiplier}

Row stochasticity makes $Wx$ a peer average, while the stability condition
makes $(I-\rho W)^{-1}$, and hence the equilibrium mapping from stand-alone
exposures, well defined.
The geometric certificate adds the limiting results needed for stationary
weighting and the boundary analysis.

\begin{theorem}[Perron limit and stationary weights]
  \label{thm:perron-from-geometric}\label{thm:replication-centrality}
  \label{thm:stationarity}
  Let $W$ be row stochastic.
  (i) If $W$ has a geometric Perron certificate $(\pi,C,q)$, then $W^k\to P$
  as $k\to\infty$.
  If $W^k\to P$, then (ii) $W^k x\to(\pi^\top x)\1$ for every vector $x$, and
  (iii) $\pi^\top W=\pi^\top$.
\end{theorem}
Part (i) follows from $Cq^k\to0$, and (ii) from multiplying $W^k\to P$ by
$x$.
For (iii), $W^{k+1}=W^{k}W$ converges both to $PW$ and to $P$, so $PW=P$, which
is $\pi^\top W=\pi^\top$.
Repeated peer averaging therefore replaces any profile by its $\pi$-weighted
cross-sectional mean, and the weights of that mean are invariant under one
application of the operator.
Together, these results make $\pi_i$ firm $i$'s long-run influence and justify
the stationary centring in \Cref{thm:spatial-attenuation}.

\subsection{Fitted operator and finite-\texorpdfstring{$\rho$}{rho} diagnostics}
\label{sec:app-network-diagnostics}

The fitted field and coefficient are checked for whether they
satisfy these conditions and whether the boundary closely approximates the
empirical model.
The fitted $\ppvalue{n-tickers-priced}\times\ppvalue{n-tickers-priced}$
operator is row stochastic to a maximum residual of
\ppnum[8]{network-row-sum-residual}.
Because its entries are nonnegative and its rows sum to one, its maximum
absolute row-sum norm is one.
The pooled raw-return, common-intercept estimate
$\hat\rho=\ppnum[3]{post-pooled-2023-2026-w-flat-rho}$ therefore satisfies the
stability condition and makes the spatial multiplier well defined.
Power iteration also yields a strictly positive stationary distribution with
minimum mass \ppnum[4]{network-stationary-minimum} and stationarity residual
\ppnume[2]{network-stationary-residual}.
These fitted checks support, respectively, the peer average, the equilibrium
mapping from stand-alone exposures, and the stationary weighting used in the
main text.

The Dobrushin coefficient
$\delta(W)=\tfrac12\max_{i,j}\sum_m|W_{im}-W_{jm}|$ measures the largest
difference between any two firms' peer-weight profiles. A value below one means
that all row pairs retain some overlap and supplies the auditable envelope
$C=2$ and $q=\delta(W)$: every row of $W^k$ lies within
$2\delta(W)^k$ of the stationary distribution in $\ell_1$ distance.
For the fitted operator, the second power is strictly positive with minimum
entry \ppnum[6]{network-w2-minimum-weight}, and the Dobrushin coefficient is
\ppnum[3]{network-dobrushin}.
Thus the fitted operator has a geometric Perron certificate: its iterated peer
profiles converge even though the operator is directed.

The Perron limit describes the boundary $\rho\uparrow1$, whereas the empirical
application uses a finite working-model coefficient.
An exact decomposition separates the rank-one part of the multiplier from the
remaining cross-sectional variation at any admissible $\rho$.
Write $N=W-P$. Row stochasticity, stationarity, and unit mass imply
$WP=PW=P^2=P$. Whenever the displayed inverses exist and $\rho\neq1$, the
Leontief multiplier splits exactly as
\begin{equation}
  (I-\rho W)^{-1}
  =(1-\rho)^{-1}P+(I-\rho N)^{-1}(I-P).
  \label{eq:resolvent-split}
\end{equation}
The error made by replacing the normalized multiplier with its limiting
projection is therefore explicit rather than asymptotic:
\begin{equation}
  (1-\rho)(I-\rho W)^{-1}-P
  =(1-\rho)(I-\rho N)^{-1}(I-P).
  \label{eq:finite-rho-remainder}
\end{equation}

\begin{proposition}[Finite-$\rho$ resolvent bound]
  \label{prop:finite-rho-bound}
  If $\lVert\rho(W-P)\rVert<1$, then
  \[
    \left\lVert(1-\rho)(I-\rho W)^{-1}-P\right\rVert
    \leq
    |1-\rho|\{1-\lVert\rho(W-P)\rVert\}^{-1}\lVert I-P\rVert.
  \]
\end{proposition}
The proposition provides an auditable upper bound on the distance between the
finite-$\rho$ normalized multiplier and its limiting projection.
Taking norms in \eqref{eq:finite-rho-remainder} and applying the Neumann bound
\[
  \lVert(I-\rho N)^{-1}\rVert
  \leq\{1-\lVert\rho N\rVert\}^{-1},
\]
gives the displayed bound.

At that raw-return, common-intercept coefficient, the exact
normalized-resolvent error is
\ppnum[3]{network-w-flat-resolvent-error}, against a Dobrushin bound of
\ppnum[3]{network-w-flat-resolvent-bound}.
The Neumann envelope does not apply on this panel: it requires
$\lVert\rho N\rVert<1$, and the fitted coefficient gives
$\lVert\rho N\rVert=\ppnum[3]{network-w-flat-complement-norm}$, outside the
region of convergence.
Every fitted variant lies outside it on the
\ppnum[0]{n-tickers-priced}-firm panel, so the Dobrushin bound is the only
envelope reported here.
The exact error remains materially above zero, so the rank-one limit is not a
close approximation to the fitted normalized multiplier in this norm.
The Dobrushin bound is a valid but conservative envelope.
Thus the fitted coefficient lies inside the stable region, but the boundary
approximation is not the empirical working model.

The finite multiplier also gives a more direct exposure interpretation than
$\rho$ alone.  Define
\[
  G(\rho)=(I-\rho W)^{-1},
  \qquad
  H(\rho)=(1-\rho)G(\rho).
\]
The normalized multiplier $H$ is the object in the exposure equilibrium:
\begin{equation}
  B=H(\rho)\xi
  =(1-\rho)\sum_{k\geq 0}\rho^k W^k\xi .
  \label{eq:finite-rho-exposure-weights}
\end{equation}
For nonnegative row-stochastic $W$, every row of $H$ sums to one.  Hence its
diagonal entry is the model-implied ultimate own-exposure weight and
$1-H_{ii}$ is the corresponding total weight on other firms, including all
higher-order paths.  At the primary factor-conditioned coefficient on $W^\flat$, the
complete operator
reports mean own exposure between
\ppnum[4]{network-factor-w2-own-exposure-minimum} and
\ppnum[4]{network-factor-w2-own-exposure-maximum}, with mean
\ppnum[4]{network-factor-w2-own-exposure-mean}; the complementary mean other-firm
weight is \ppnum[4]{network-factor-w2-other-exposure-mean}.  These are
finite-network implications of the maintained closure, not estimates of return
variance shares or causal transmission.

The unnormalized resolvent and normalized exposure multiplier must be kept
distinct.  Because $W\1=\1$,
\[
  G(\rho)\1=(1-\rho)^{-1}\1,\qquad H(\rho)\1=\1.
\]
The certificate evaluates the corresponding means as
\ppnum[4]{network-factor-w2-g-common-mode-mean} and
\ppnum[4]{network-factor-w2-h-common-mode-mean}, with numerical identity errors
of respectively
\ppnume[2]{network-factor-w2-g-common-mode-error} and
\ppnume[2]{network-factor-w2-h-common-mode-error}.  Thus the first number is a
resolvent scale, not a claim that common factor exposure is amplified by that
amount; the normalized equilibrium preserves a common one-unit shift.

\subsection{The rank-one boundary}

As peer adjustment approaches its strong-interaction limit, cross-sectional
variation collapses onto one common direction.

\begin{theorem}[Rank-one boundary]
  \label{thm:resolvent-from-perron}\label{thm:rankone-collapse}
  Let $W$ be row stochastic with $W^k\to P$, and let
  $\Sigma_{\mathrm{SAR}}=(I-\rho W)^{-1}V(I-\rho W)^{-\top}$.
  As $\rho\uparrow1$, (i) $S_\rho=(1-\rho)(I-\rho W)^{-1}\to P$, and
  (ii) $(1-\rho)^2\Sigma_{\mathrm{SAR}}\to(\pi^\top V\pi)\1\1^\top$.
\end{theorem}
For (i), write $S_\rho=(1-\rho)\sum_{k\geq0}\rho^{k}W^{k}$ and
$P=(1-\rho)\sum_{k\geq0}\rho^{k}P$.
Given $\varepsilon>0$, choose $K$ with $\lVert W^k-P\rVert<\varepsilon$ for
$k\geq K$; since $\lVert W^k-P\rVert\leq2$, the head
$(1-\rho)\sum_{k<K}\rho^k\lVert W^k-P\rVert$ vanishes as $\rho\uparrow1$ and
the tail is at most $\varepsilon$.
For (ii), continuity of matrix multiplication gives
$S_\rho VS_\rho^\top\to PVP^\top=(\pi^\top V\pi)\1\1^\top$.

Part (ii) is the $\rho\uparrow1$ endpoint of the attenuation result: at the
boundary, spatial feedback removes cross-sectional dispersion entirely and the
covariance becomes rank one.
The finite-$\rho$ remainder shows why this endpoint is useful as a theoretical
boundary but not as an approximation to the fitted working model.

\section{Additional Empirical Evidence}
\label{sec:app-empirical}

\subsection{Sample composition and specification history}
\label{sec:app-sample}


\begin{table}[H]

\centering

\small
\begin{tabularx}{\linewidth}{Xrrrrr}
\toprule
Sector & $n$ & Articles & Volatility & $W_2$ within & $W_2$ cross \\
\midrule
Communication Services & 11 & 128.0 & 0.0239 & 1.042 & 1.060 \\
Consumer Cyclical & 18 & 128.0 & 0.0281 & 1.033 & 1.053 \\
Consumer Defensive & 7 & 128.0 & 0.0176 & 1.010 & 1.058 \\
Energy & 1 & 128.0 & 0.0386 & \textemdash & 1.045 \\
Financial Services & 1 & 128.0 & 0.0272 & \textemdash & 1.070 \\
Healthcare & 14 & 128.0 & 0.0233 & 1.025 & 1.065 \\
Industrials & 7 & 128.0 & 0.0240 & 1.030 & 1.057 \\
Technology & 38 & 128.0 & 0.0265 & 0.998 & 1.050 \\
Utilities & 3 & 128.0 & 0.0159 & 0.882 & 1.058 \\
\midrule
All sectors & 100 & 128.0 & 0.0251 & 1.008 & 1.056 \\
\bottomrule
\end{tabularx}

\caption{Per-sector composition of the news frame and its balanced-cloud sample: firm count, mean article count per firm, mean daily return volatility, and mean within- versus cross-sector rooted $W_2$ distance under the frozen Qwen3-Embedding-8B geometry. Firm counts, volatility, and distance summaries use the same estimation sample. The Articles column reports the constant balanced-cloud size after per-firm truncation, not raw frame coverage, and therefore carries no information about a firm's true news volume. Volatility is daily and unannualized. A dash marks a statistic undefined for that sector, such as within-sector distance for a singleton.}

\label{tab:p5-sample-descriptives}

\end{table}

\Cref{tab:p5-sample-descriptives} records the news frame's industry
composition and its return and text dispersion.
Sector sizes are heavily unbalanced, with Technology holding roughly a third of
the frame and two sectors holding fewer than three firms, so peer sets built
from the text geometry inherit that imbalance.
Mean within-sector $W_2$ distance is below the corresponding cross-sector
distance in every sector holding more than one firm: the frozen geometry
contains industry structure before any return is used.

The reported specification expands the priced universe from an earlier 52-firm
specification to \ppvalue{n-tickers-priced} firms and ends the return window
on 6 April 2026, the final common date before HOLX's take-private.
\Cref{tab:p5-universe-horizon-bridge} bridges the three observed
specifications.
At 52 firms, extending the common-date window from 6 April to 15 July 2026
moves $\hat\rho$ from
$\ppnum[3]{bridge-reference-52-short-rho}$ to
$\ppnum[3]{bridge-reference-52-long-rho}$, a difference of
$\ppnum[3]{bridge-horizon-rho-difference}$.
At the short horizon, expanding the universe from 52 to 100 firms moves
$\hat\rho$ from $\ppnum[3]{bridge-reference-52-short-rho}$ to
$\ppnum[3]{bridge-current-100-short-rho}$, a difference of
$\ppnum[3]{bridge-universe-rho-difference}$.
Both are arithmetic differences reported without contrast-level inference.
The universe comparison changes the frozen operator as well as the roster, and
its complete-case intersection has one fewer date.
The 52 shared return histories are identical through the common cutoff, so
data revision within those histories does not explain the comparison.

\begin{table}[H]
\centering
\small
\begin{tabular}{lrrrl}
\toprule
Specification & Firms & End date & Trading days & $\hat\rho$ [95\% CI] \\
\midrule
Historical reference & 52 & 15 Jul 2026 & 885 & 0.776 [0.743, 0.807] \\
Horizon-matched & 52 & 6 Apr 2026 & 816 & 0.771 [0.735, 0.806] \\
Current universe & 100 & 6 Apr 2026 & 815 & 0.843 [0.815, 0.869] \\
\bottomrule
\end{tabular}
\caption{Sequential universe-and-horizon bridge for the barycentric $W^\flat$ single-field SAR QMLE in the raw-return, common-intercept specification (\ppAppendixRef{sec:app-unadjusted}); it is not the primary factor-conditioned estimate. Every row starts on 3 January 2023. The historical row uses the recovered 52-firm operator and the long return window. The horizon-matched row holds that operator and universe fixed while shortening the return endpoint; its interval is newly estimated with the joint-date stationary bootstrap. The current-universe row expands to the 100-firm operator at the same endpoint, so it changes both the roster and the operator; the first and third intervals come from the reported source results. Complete-case trading-day counts can differ across universes.}
\label{tab:p5-universe-horizon-bridge}
\end{table}

\subsection{The complete barycentric field}
\label{sec:app-field-anatomy}

\begin{figure}[htbp]
  \centering
  \import{images/}{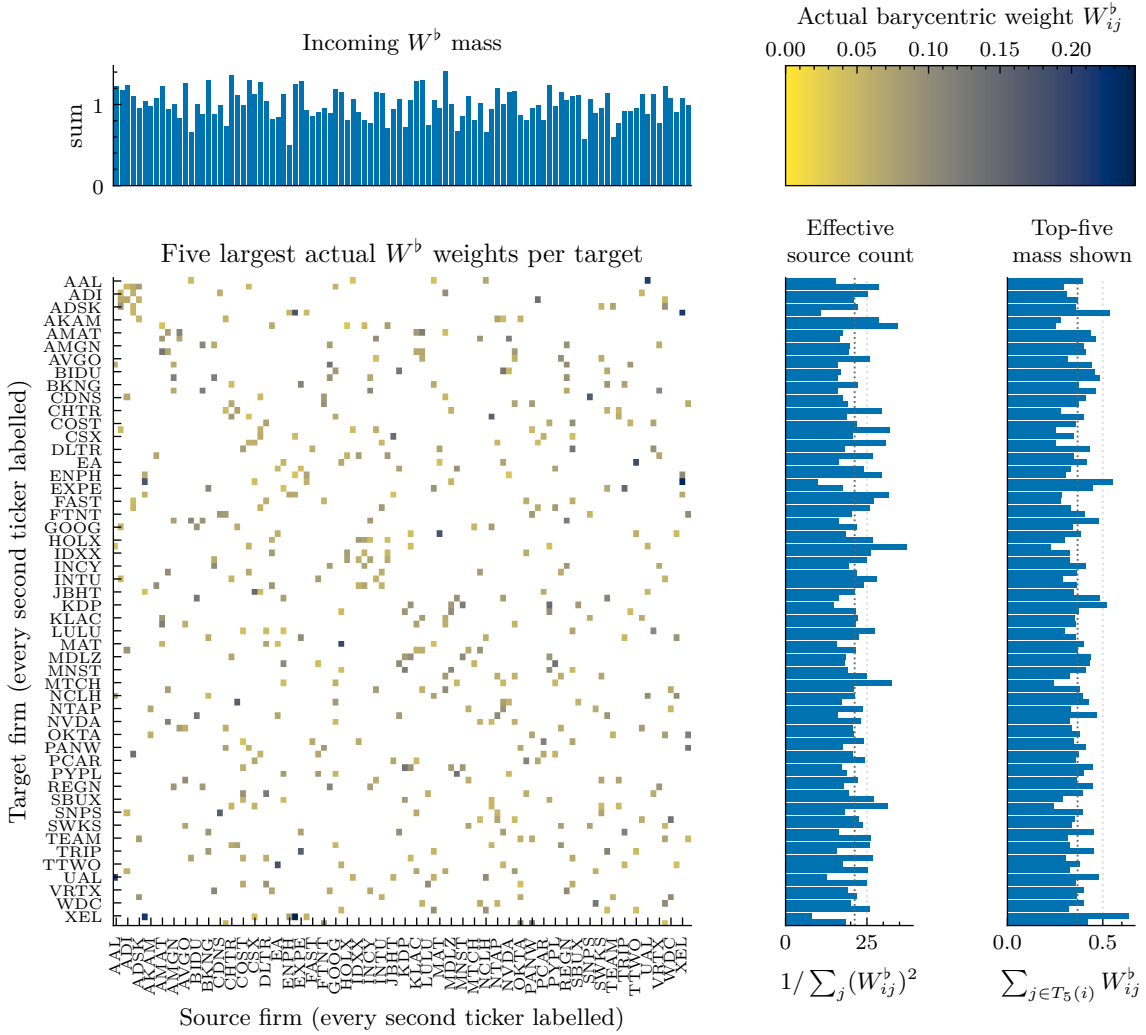}
  \caption{Illustrative barycentric interaction field from the target-anchored
    Wasserstein barycentric reconstruction. Each row is the actual
    leave-one-out solution $W^\flat_{i\cdot}$ to \eqref{eq:w2-barycentre} over
    the frozen 2018--2022 article clouds: weights are nonnegative, sum to one,
    and assign zero self-weight. The heat map shows each target's five largest
    \emph{actual} coefficients without renormalizing them; the remaining
    positive coefficients still belong to $W^\flat$ but are not colored. The
  upper and right-hand diagnostics are calculated from the complete field.}
  \label{fig:wflat-barycentric-mixtures}
  \figalttext{Rows index target firms and columns index source firms. Colored
    cells mark each target's five largest actual barycentric weights; their
    scattered, row-specific locations show that the field is directed and
    heterogeneous. Marginal bar charts show heterogeneous incoming mass,
    effective source counts, and the fraction of each full row contained in the
  displayed top five. Uncolored positive weights remain in the operator.}
\end{figure}

\Cref{fig:wflat-barycentric-mixtures} displays the complete field.
Each row is a target firm and each column a candidate source; color marks the
five largest actual coefficients in each row without renormalizing them.
The upper panel reports each source's total incoming mass across target rows,
and the right-hand panels report each target's effective source count and the
share of its row captured by the five displayed cells.

Both concentration within rows and total incoming mass across rows vary across
firms, and the colored locations are scattered and row specific: the field is
directed and heterogeneous rather than a block-diagonal industry map.
\Cref{tab:p5-peer-interpretation} lists actual reconstruction weights $W^\flat_{ij}$,
pairwise-distance ranks, sector and industry matches, and persistent
co-mention status for AMD and for the targets with minimum, median, and maximum
effective source count.
A large weight records a peer's usefulness within the target's reconstructing
combination, which need not coincide with its pairwise rank.

\begin{table}[H]
\centering
\scriptsize
\setlength{\tabcolsep}{2.5pt}
\begin{tabular}{ll l r r c c c}
\toprule
Target & Selection role & Peer & $W^\flat_{ij}$ & W2 rank & Same sector & Same industry & Co-mention \\
\midrule
AMD & declared AMD & NVDA & 0.1225 & 1 & Yes & Yes & Yes \\
AMD & declared AMD & INTC & 0.0987 & 4 & Yes & Yes & Yes \\
AMD & declared AMD & MU & 0.0878 & 3 & Yes & Yes & Yes \\
AMD & declared AMD & AMAT & 0.0840 & 2 & Yes & No & Yes \\
AMD & declared AMD & ASML & 0.0688 & 6 & Yes & No & Yes \\
\addlinespace[1.5pt]
HON & maximum effective source count & CSCO & 0.0534 & 1 & No & No & Yes \\
HON & maximum effective source count & UAL & 0.0502 & 51 & Yes & No & Yes \\
HON & maximum effective source count & KHC & 0.0427 & 24 & No & No & No \\
HON & maximum effective source count & PEP & 0.0405 & 7 & No & No & Yes \\
HON & maximum effective source count & AAL & 0.0387 & 60 & Yes & No & Yes \\
\addlinespace[1.5pt]
JBHT & median effective source count & CSX & 0.1473 & 1 & Yes & No & Yes \\
JBHT & median effective source count & PCAR & 0.0629 & 3 & Yes & No & No \\
JBHT & median effective source count & UAL & 0.0524 & 43 & Yes & No & Yes \\
JBHT & median effective source count & CTSH & 0.0458 & 2 & No & No & No \\
JBHT & median effective source count & AEP & 0.0387 & 5 & No & No & No \\
\addlinespace[1.5pt]
XEL & minimum effective source count & EXC & 0.2443 & 1 & Yes & Yes & Yes \\
XEL & minimum effective source count & AEP & 0.2173 & 2 & Yes & Yes & Yes \\
XEL & minimum effective source count & ENPH & 0.0797 & 3 & No & No & Yes \\
XEL & minimum effective source count & FANG & 0.0541 & 5 & No & No & No \\
XEL & minimum effective source count & CSX & 0.0409 & 4 & No & No & No \\
\bottomrule
\end{tabular}
\caption{Descriptive W2 peer interpretation panel. Targets are AMD and deterministic minimum, median, and maximum effective source-count firms, with lexical tie-breaking. Each row lists one of the target's five largest actual reconstruction weights; the distance rank is the target's pairwise W2 rank. Sector, industry, and co-mention columns report metadata associated with each candidate.}
\label{tab:p5-peer-interpretation}
\end{table}

\subsection{The common-factor ladder and its information criteria}
\label{sec:app-information-criteria}

\Cref{tab:factor-set-sensitivity} repeats the primary specification under
each common-factor control set on the same dates.
The spatial coefficient is stable across the ladder and every bootstrap
interval excludes zero; adding momentum lowers $\hat\rho$ by about one
percentage point within either factor family.

\begin{table}[H]
  \centering
  \caption{Common-factor sensitivity for the barycentric field $W^\flat$.
    Each row profiles firm-specific intercepts and loadings on the indicated
    factors. The empirical 95\% intervals use 2,000 paired joint-date
    stationary-bootstrap refits. The held-out score fits all parameters on
    2018--2022 and evaluates the mean Gaussian quasi-log score per firm-day on
    2023--2026 returns; higher is better. FF5 plus momentum is the prespecified
  primary model.}
  \label{tab:factor-set-sensitivity}
  \begin{tabular}{lrrrr}
    \toprule
    Factor controls & $K$ & $\hat\rho$ & 95\% interval & Held-out score \\
    \midrule
    FF3 &
    \ppvalue{transport-factor-ff3-n-factors} &
    \ppnum[3]{transport-factor-ff3-w2-rho} &
    $[\ppnum[3]{transport-factor-ff3-w2-rho-ci-lower},
    \ppnum[3]{transport-factor-ff3-w2-rho-ci-upper}]$ &
    \ppnum[6]{transport-factor-ff3-w2-predictive-score} \\
    FF3 plus momentum &
    \ppvalue{transport-factor-ff3-mom-n-factors} &
    \ppnum[3]{transport-factor-ff3-mom-w2-rho} &
    $[\ppnum[3]{transport-factor-ff3-mom-w2-rho-ci-lower},
    \ppnum[3]{transport-factor-ff3-mom-w2-rho-ci-upper}]$ &
    \ppnum[6]{transport-factor-ff3-mom-w2-predictive-score} \\
    FF5 &
    \ppvalue{transport-factor-ff5-n-factors} &
    \ppnum[3]{transport-factor-ff5-w2-rho} &
    $[\ppnum[3]{transport-factor-ff5-w2-rho-ci-lower},
    \ppnum[3]{transport-factor-ff5-w2-rho-ci-upper}]$ &
    \ppnum[6]{transport-factor-ff5-w2-predictive-score} \\
    FF5 plus momentum (primary) &
    \ppvalue{transport-factor-ff5-mom-n-factors} &
    \ppnum[3]{transport-factor-ff5-mom-w2-rho} &
    $[\ppnum[3]{transport-factor-ff5-mom-w2-rho-ci-lower},
    \ppnum[3]{transport-factor-ff5-mom-w2-rho-ci-upper}]$ &
    \ppnum[6]{transport-factor-ff5-mom-w2-predictive-score} \\
    \bottomrule
  \end{tabular}
\end{table}

The held-out score is highest under FF3, and working information criteria rank
the ladder differently according to their penalty; FF5 plus momentum remains
primary because it conditions on the broadest prespecified set of observed
factors.

\begin{table}[H]
  \centering
  \caption{Working information criteria for the barycentric factor ladder.
    The first panel reports the maximized post-corpus quasi-log-likelihood and
    AIC. The second reports HQC and BIC using either multivariate dates, $T$,
    or firm-days, $NT$, as the effective sample size. Lower information
  criteria are better.}
  \label{tab:factor-information-criteria}
  \begin{tabular}{lrrr}
    \toprule
    Factor controls & Parameters & QLL & AIC \\
    \midrule
    FF3 &
    \ppvalue{spatial-value-factor-ff3-parameter-count} &
    \ppnum[1]{spatial-value-factor-ff3-loglik} &
    \ppnum[1]{spatial-value-factor-ff3-aic} \\
    FF3 plus momentum &
    \ppvalue{spatial-value-factor-ff3-mom-parameter-count} &
    \ppnum[1]{spatial-value-factor-ff3-mom-loglik} &
    \ppnum[1]{spatial-value-factor-ff3-mom-aic} \\
    FF5 &
    \ppvalue{spatial-value-factor-ff5-parameter-count} &
    \ppnum[1]{spatial-value-factor-ff5-loglik} &
    \ppnum[1]{spatial-value-factor-ff5-aic} \\
    FF5 plus momentum &
    \ppvalue{spatial-value-factor-ff5-mom-parameter-count} &
    \ppnum[1]{spatial-value-factor-ff5-mom-loglik} &
    \ppnum[1]{spatial-value-factor-ff5-mom-aic} \\
    \bottomrule
  \end{tabular}

  \begin{tabular}{lrrrr}
    \toprule
    Factor controls & $\mathrm{HQC}_T$ & $\mathrm{HQC}_{NT}$ &
    $\mathrm{BIC}_T$ & $\mathrm{BIC}_{NT}$ \\
    \midrule
    FF3 &
    \ppnum[1]{spatial-value-factor-ff3-hqc-dates} &
    \ppnum[1]{spatial-value-factor-ff3-hqc-firm-days} &
    \ppnum[1]{spatial-value-factor-ff3-bic-dates} &
    \ppnum[1]{spatial-value-factor-ff3-bic-firm-days} \\
    FF3 plus momentum &
    \ppnum[1]{spatial-value-factor-ff3-mom-hqc-dates} &
    \ppnum[1]{spatial-value-factor-ff3-mom-hqc-firm-days} &
    \ppnum[1]{spatial-value-factor-ff3-mom-bic-dates} &
    \ppnum[1]{spatial-value-factor-ff3-mom-bic-firm-days} \\
    FF5 &
    \ppnum[1]{spatial-value-factor-ff5-hqc-dates} &
    \ppnum[1]{spatial-value-factor-ff5-hqc-firm-days} &
    \ppnum[1]{spatial-value-factor-ff5-bic-dates} &
    \ppnum[1]{spatial-value-factor-ff5-bic-firm-days} \\
    FF5 plus momentum &
    \ppnum[1]{spatial-value-factor-ff5-mom-hqc-dates} &
    \ppnum[1]{spatial-value-factor-ff5-mom-hqc-firm-days} &
    \ppnum[1]{spatial-value-factor-ff5-mom-bic-dates} &
    \ppnum[1]{spatial-value-factor-ff5-mom-bic-firm-days} \\
    \bottomrule
  \end{tabular}
\end{table}

Because each added factor introduces one loading for every firm, the working
information criteria impose materially different penalties from the held-out
score comparison.
Working AIC and both HQC conventions are lowest for FF5 plus momentum, whereas
both BIC conventions are lowest for FF3 plus momentum; the ranking depends on
the penalty and the effective sample size.
The criteria are
$\mathrm{AIC}=-2\ell+2k$,
$\mathrm{HQC}_T=-2\ell+2k\log\log T$,
$\mathrm{HQC}_{NT}=-2\ell+2k\log\log(NT)$,
$\mathrm{BIC}_T=-2\ell+k\log T$, and
$\mathrm{BIC}_{NT}=-2\ell+k\log(NT)$, using dates and firm-days,
respectively, as the effective sample size; lower values are preferred.
For $K$ factors, $k=N(K+1)+1$ under the spherical factor-only model and
$N(K+1)+2$ under a spatial alternative; the diagonal-variance models have
$N(K+1)+N$ and $N(K+1)+N+1$ parameters.
Momentum additions are nested within each factor family, whereas the official
FF3 and FF5 files differ in their SMB construction.

\subsection{Firm-specific innovation variances}
\label{sec:app-diagonal}

\begin{table}[H]
  \centering
  \caption{Spatial-value robustness with firm-specific innovation variances.
    QLL gains, restricted-null QLR tests, and held-out score gains compare each
  field with the matched diagonal-variance factor-only model.}
  \label{tab:spatial-value-diagonal}
  \begin{tabular}{lrrrr}
    \toprule
    Field & $\hat\rho$ & QLL gain & QLR $p$ & Held-out gain \\
    \midrule
    Target-anchored W1 &
    \ppnum[3]{spatial-value-diagonal-w1-rho} &
    \ppnum[1]{spatial-value-diagonal-w1-versus-none-loglik-gain} &
    \ppnum[4]{spatial-value-diagonal-w1-versus-none-qlr-pvalue} &
    $\ppnum[4]{spatial-value-diagonal-w1-versus-none-held-out-score-gain}$
    $[\ppnum[4]{spatial-value-diagonal-w1-versus-none-held-out-score-gain-ci-lower},
    \ppnum[4]{spatial-value-diagonal-w1-versus-none-held-out-score-gain-ci-upper}]$
    \\
    Target-anchored W2 &
    \ppnum[3]{spatial-value-diagonal-w2-rho} &
    \ppnum[1]{spatial-value-diagonal-w2-versus-none-loglik-gain} &
    \ppnum[4]{spatial-value-diagonal-w2-versus-none-qlr-pvalue} &
    $\ppnum[4]{spatial-value-diagonal-w2-versus-none-held-out-score-gain}$
    $[\ppnum[4]{spatial-value-diagonal-w2-versus-none-held-out-score-gain-ci-lower},
    \ppnum[4]{spatial-value-diagonal-w2-versus-none-held-out-score-gain-ci-upper}]$
    \\
    \bottomrule
  \end{tabular}
\end{table}

With firm-specific variances, each spatial model also has lower working AIC,
HQC, and BIC than its matched factor-only benchmark, and the reduced-form
coefficient identity holds to within its $10^{-10}$ absolute and relative
tolerance.

\subsection{Covariance-sensitive portfolio diagnostic}
\label{sec:app-gmv}

Proposition~\ref{prop:factor-beta-invariance} implies that the economic
difference across fields lies in the covariance assigned to factor-residual
returns. For each fitted field, the stored total covariance is
\begin{equation}
  \widehat{\Sigma}_{r,m}
  =\widehat{B}\widehat{\Sigma}_{f}\widehat{B}^{\prime}
  + A_m^{-1}\widehat{\Omega}_m A_m^{-\prime},
  \qquad A_m=I-\hat\rho_m W_m,
  \label{eq:app-gmv-covariance}
\end{equation}
where $m$ indexes the factor-only, W1, and W2 specifications and
$\widehat{\Omega}_m$ is the constant-correlation Ledoit--Wolf estimate of the
factor-residualized spatial innovations. The factor-only row sets
$\hat\rho_m=0$ and uses the same residual estimator. Thus the covariance inputs
combine common-factor covariance with the same estimator-defined local
factor-residual covariance rule across candidates.
We translate these covariance estimates into a fixed-weight allocation
diagnostic. The weights are formed on the factor-complete 2018--2022 panel,
frozen at the formation vintage, and evaluated on 2023--2026 excess returns
under a fully invested long-only constraint with a 15\% single-name cap.

\begin{table}[H]
  \centering
  \scriptsize
  \setlength{\tabcolsep}{3.0pt}
  \begin{tabular}{lrrrrr}
    \toprule
    Covariance input & Realized variance & Ann. vol. & Rel. to factor only & Eff. $N$ & Max. wt. \\
    \midrule
Factor only + LW residual covariance & 0.5441 & 11.71\% & 100.0\% & 9.76 & 15.0\% \\
Target-anchored W1 & 0.5459 & 11.73\% & 100.3\% & 9.82 & 15.0\% \\
Target-anchored W2 & 0.5455 & 11.72\% & 100.3\% & 9.83 & 15.0\% \\
Sample GMV & 0.5428 & 11.70\% & 99.8\% & 9.57 & 15.0\% \\
Direct Ledoit--Wolf GMV & 0.5395 & 11.66\% & 99.2\% & 9.96 & 15.0\% \\
Equal weight & 1.1606 & 17.10\% & 213.3\% & 100.00 & 1.0\% \\
    \bottomrule
  \end{tabular}
  \caption{Covariance-sensitive portfolio diagnostic. Candidate weights are formed from factor-complete 2018--2022 excess returns and frozen at the 2022 formation vintage. They are long-only, fully invested, and subject to a 15\% single-name cap; realized risk is evaluated on constant-weight excess-return mixtures over 2023--2026. The factor-only, W1, and W2 rows use diagonal innovation variances with common constant-correlation residual shrinkage. Realized variance is scaled by $10^4$; lower values are better.}
  \label{tab:p5-gmv-validation}
\end{table}

Under this fixed-weight GMV convention, W1 and W2 do not reduce realized
variance relative to the factor-only covariance benchmark; direct
Ledoit--Wolf shrinkage is slightly lower. The exhibit is therefore a diagnostic
of the residual-covariance interpretation, while the paper's primary evidence
remains the held-out joint-density comparison.

\subsection{Transport order at the article level and jointly}
\label{sec:app-transport-order}

The transport-order equivalence holds across the factor ladder.
The W2-minus-W1 coefficient difference ranges from
\ppnum[6]{transport-factor-ff3-mom-delta-rho} to
\ppnum[6]{transport-factor-ff5-delta-rho}, and the held-out score difference
from \ppnum[6]{transport-factor-ff5-delta-predictive-score} to
\ppnum[6]{transport-factor-ff3-delta-predictive-score} per firm-day.
After projection on the intercept and factors, the induced peer-return signals
correlate at
\ppnum[6]{transport-factor-residualized-signal-correlation} with a normalized
difference of
\ppnum[3]{transport-factor-residualized-relative-signal-difference-sd}, so the
factor-residualized quantities leave the weak-separation classification
unchanged.

At full dimension the two transport orders send
\ppnum[0]{transport-switch-d4096-switched-atoms} target atoms to different peer
atoms.
Scoring both assignments on the same chord distance, the mean W2-minus-W1
difference is \ppnum[4]{transport-switch-d4096-max-difference} in the longest
matched distance and \ppnum[4]{transport-switch-d4096-p95-difference} at its
95th percentile, against
$+\ppnum[4]{transport-switch-d4096-switched-difference}$ in mean matched
distance on the reassigned atoms.
Quadratic transport thus buys a shorter displacement tail with slightly longer
typical matches, which is the economic content of squaring the ground cost.
W2 gives a strictly shorter longest match in
\ppnum[3]{transport-switch-d4096-max-reduction-share} of firm pairs, at a mean
difference of \ppnum[3]{transport-switch-d4096-max-reduction-given-lower}, and
a strictly longer one in
\ppnum[3]{transport-switch-d4096-max-increase-share}, at
$+\ppnum[3]{transport-switch-d4096-max-increase-given-higher}$; the remaining
pairs share the same longest match.
The pattern strengthens monotonically down the dimension ladder, reaching
\ppnum[4]{transport-switch-d64-max-difference} in the longest match and a
shorter-tail share of \ppnum[3]{transport-switch-d64-max-reduction-share} at 64
dimensions.

After observing the standalone comparison, we estimated both fields jointly,
entering each supplied matrix unchanged:

\begin{equation}\label{eq:factor-two-field-transport}
  (I-\rho^\flat_1 W^\flat_1-\rho^\flat W^\flat)r_t^e
  =\alpha+A f_t+\varepsilon_t,
  \qquad
  \rho^\flat_1,\rho^\flat\geq 0,
  \quad \rho^\flat_1+\rho^\flat<1.
\end{equation}

At every candidate $(\rho^\flat_1,\rho^\flat)$ the specification re-profiles the same
firm intercepts, factor loadings, and innovation variance as
\eqref{eq:factor-spatial-lag}, and the restrictions $\rho^\flat=0$ and $\rho^\flat_1=0$
recover the single-field models.
The incremental statistic for W2 given W1 is
\[
  \operatorname{QLR}_{2\mid 1}
  =2\left\{
    \ell^Q(\hat\rho^\flat_1,\hat\rho^\flat)
    -\ell^Q(\hat\rho^{\flat,0}_1,0)
  \right\},
\]
with the reverse statistic defined analogously.
Because each null places a coefficient on the boundary, reference distributions
come from 2,000 restricted-null stationary-block innovation resamples generated
through the restricted spatial inverse, with the factor nuisance space projected
anew before both models are refit.

\begin{table}[H]
  \centering
  \caption{Factor-conditioned joint W1 and W2 specification. Both
    supplied fields
    enter unchanged under nonnegative coefficients whose sum is below one.
    Intervals are descriptive 95\% bootstrap quantiles, and the
    coefficient-zero share is the fraction of draws in which that coefficient is
    estimated at zero; both use 2,000 joint-date stationary-bootstrap refits.
  The specification was added after the standalone transport-order comparison.}
  \label{tab:transport-order-joint}
  \begin{tabular}{lrrr}
    \toprule
    Quantity & Estimate & 95\% interval & Coefficient-zero share \\
    \midrule
    W1 coefficient $\rho^\flat_1$ &
    \ppnum[3]{transport-joint-w1-rho} &
    $[\ppnum[3]{transport-joint-rho-w1-ci-lower},
    \ppnum[3]{transport-joint-rho-w1-ci-upper}]$ &
    \ppnum[3]{transport-joint-w2-only-boundary-share} \\
    W2 coefficient $\rho^\flat$ &
    \ppnum[3]{transport-joint-w2-rho} &
    $[\ppnum[3]{transport-joint-rho-w2-ci-lower},
    \ppnum[3]{transport-joint-rho-w2-ci-upper}]$ &
    \ppnum[3]{transport-joint-w1-only-boundary-share} \\
    Total feedback $\rho^\flat_1+\rho^\flat$ &
    \ppnum[3]{transport-joint-rho-total} &
    $[\ppnum[3]{transport-joint-rho-total-ci-lower},
    \ppnum[3]{transport-joint-rho-total-ci-upper}]$ & --- \\
    \bottomrule
  \end{tabular}
\end{table}

The joint estimates are $\hat\rho^\flat_1=
\ppnum[3]{transport-joint-w1-rho}$ and $\hat\rho^\flat=
\ppnum[3]{transport-joint-w2-rho}$, so allowing W2 adds
\ppnum[2]{transport-joint-loglik-gain-w1} quasi-log-likelihood units beyond
W1 alone.
The bootstrap places \ppnum[3]{transport-joint-w1-only-boundary-share} of draws
on the W1-only boundary, where $\rho^\flat=0$, and the two channel estimates
correlate at
\ppnum[3]{transport-joint-channel-correlation}: total feedback is precisely
positive, while its allocation between two nearly collinear fields is weakly
determined.
For W2 conditional on W1,
$\operatorname{QLR}_{2\mid 1}=
\ppnum[3]{transport-joint-qlr-w2-given-w1-statistic}$ with working-null
$p=\ppnum[4]{transport-joint-qlr-w2-given-w1-pvalue}$; for W1 conditional on
W2, $\operatorname{QLR}_{1\mid 2}=
\ppnum[3]{transport-joint-qlr-w1-given-w2-statistic}$ with
$p=\ppnum[4]{transport-joint-qlr-w1-given-w2-pvalue}$.
The in-sample nested fit therefore favors W1, while the held-out scores remain
predictively equivalent.

\subsection{Raw-return, common-intercept estimates}
\label{sec:app-unadjusted}

\begin{table}[H]
  \centering
  \caption{Pooled 2023--2026 single-field estimates in the
    raw-return, common-intercept
    specification. The models use
    raw returns and a common intercept. The 95\% intervals use 2,000
    joint-date stationary-bootstrap refits, and
    $\hat\rho=\hat\lambda/(1+\hat\lambda)$. The final column reports the
  maximized conditional quasi-log-likelihood on the common return sample.}
  \label{tab:qmle-pooled}
  \begin{tabular}{lrrrr}
    \toprule
    Interaction field & $\hat\lambda$ & 95\% interval & Implied $\hat\rho$
    & $\ell^Q$ \\
    \midrule
    Barycentric field ($W^\flat$) &
    \ppnum[2]{post-pooled-2023-2026-w-flat-lambda} &
    $[\ppnum[2]{post-pooled-2023-2026-w-flat-lambda-ci-lower},
    \ppnum[2]{post-pooled-2023-2026-w-flat-lambda-ci-upper}]$ &
    \ppnum[3]{post-pooled-2023-2026-w-flat-rho} &
    \ppnum[1]{post-pooled-2023-2026-w-flat-loglik} \\
    RBF--Wasserstein diffusion &
    \ppnum[2]{post-pooled-2023-2026-w-h-lambda} &
    $[\ppnum[2]{post-pooled-2023-2026-w-h-lambda-ci-lower},
    \ppnum[2]{post-pooled-2023-2026-w-h-lambda-ci-upper}]$ &
    \ppnum[3]{post-pooled-2023-2026-w-h-rho} &
    \ppnum[1]{post-pooled-2023-2026-w-h-loglik} \\
    Persistent news co-mentions &
    \ppnum[2]{post-pooled-2023-2026-w-co-mentions-lambda} &
    $[\ppnum[2]{post-pooled-2023-2026-w-co-mentions-lambda-ci-lower},
    \ppnum[2]{post-pooled-2023-2026-w-co-mentions-lambda-ci-upper}]$ &
    \ppnum[3]{post-pooled-2023-2026-w-co-mentions-rho} &
    \ppnum[1]{post-pooled-2023-2026-w-co-mentions-loglik} \\
    \midrule
    Equal active support &
    \ppnum[2]{post-pooled-2023-2026-equal-support-lambda} &
    $[\ppnum[2]{post-pooled-2023-2026-equal-support-lambda-ci-lower},
    \ppnum[2]{post-pooled-2023-2026-equal-support-lambda-ci-upper}]$ &
    \ppnum[3]{post-pooled-2023-2026-equal-support-rho} &
    \ppnum[1]{post-pooled-2023-2026-equal-support-loglik} \\
    \bottomrule
  \end{tabular}
\end{table}

In the raw-return, common-intercept specification, the barycentric field has
pooled adjustment index
$\hat\lambda=\ppnum[2]{post-pooled-2023-2026-w-flat-lambda}$, with interval
$[\ppnum[2]{post-pooled-2023-2026-w-flat-lambda-ci-lower},
\ppnum[2]{post-pooled-2023-2026-w-flat-lambda-ci-upper}]$, and feedback
coefficient $\hat\rho=\ppnum[3]{post-pooled-2023-2026-w-flat-rho}$.
When the identification condition in \Cref{sec:estimand} holds, this index
corresponds to an adjustment criterion that places about
\ppnum[2]{post-pooled-2023-2026-w-flat-lambda} times as much weight on
misalignment with the barycentric peer average as on departure from
stand-alone exposure; the interval determines the sign more sharply than the
magnitude.
Equal weighting of the same active support and pairwise RBF proximity both
yield positive feedback and lower quasi-log-likelihood than the barycentric
field, and the news-link field attains higher likelihood than the RBF field
with a smaller adjustment index.

The raw-return joint fit provides a corroborating channel comparison. Its
pooled barycentric and news-link coefficients are estimated jointly with a
common intercept, while the factor-conditioned estimates in
\Cref{tab:two-field-pooled} are the primary specification.

\begin{table}[H]
  \centering
  \caption{Raw-return, common-intercept joint barycentric and news-link model.
  The 95\% intervals use 2,000 joint-date stationary-bootstrap refits.}
  \label{tab:two-field-raw-appendix}
  \begin{tabular}{lrrr}
    \toprule
    Channel & $\hat\rho$ & 95\% interval & $\hat\lambda$ \\
    \midrule
    Barycentric ($W^\flat$) &
    \ppnum[3]{post-pooled-2023-2026-two-field-rho-b} &
    $[\ppnum[3]{post-pooled-2023-2026-two-field-rho-b-ci-lower},
    \ppnum[3]{post-pooled-2023-2026-two-field-rho-b-ci-upper}]$ &
    \ppnum[2]{post-pooled-2023-2026-two-field-lambda-b} \\
    News-link ($W^{\mathrm{news}}$) &
    \ppnum[3]{post-pooled-2023-2026-two-field-rho-n} &
    $[\ppnum[3]{post-pooled-2023-2026-two-field-rho-n-ci-lower},
    \ppnum[3]{post-pooled-2023-2026-two-field-rho-n-ci-upper}]$ &
    \ppnum[2]{post-pooled-2023-2026-two-field-lambda-n} \\
    Total &
    \ppnum[3]{post-pooled-2023-2026-two-field-rho-total} &
    $[\ppnum[3]{post-pooled-2023-2026-two-field-rho-total-ci-lower},
    \ppnum[3]{post-pooled-2023-2026-two-field-rho-total-ci-upper}]$ & \\
    \bottomrule
  \end{tabular}
\end{table}

\begin{figure}[htbp]
  \centering
  \import{images/}{two_field_region\ppfiguresuffix.pgf}
  \caption{Joint two-field likelihood and bootstrap geometry in the raw-return,
    common-intercept specification of \Cref{tab:two-field-raw-appendix}. The horizontal
    axis is barycentric-channel feedback $\rho^\flat$, and the vertical axis is
    news-link feedback $\rho^{\mathrm{news}}$. The left panel places the joint estimate
    inside the admissible triangle, with both single-field fits on the axes and
    the stability boundary $\rho^\flat+\rho^{\mathrm{news}}=1$ marked. The
    right panel is the same
    profiled surface zoomed to the bootstrap cloud, with a descriptive
    two-parameter quasi-log-likelihood contour. The cloud's diagonal
    tilt records
    the trade-off between channels, while its dispersion in both directions
  shows the sampling variation in their allocation.}
  \label{fig:two-field-region}
  \figalttext{Two contour panels plot barycentric-channel feedback horizontally
    and news-link feedback vertically. The left panel shows an interior joint
    estimate within the triangular stability region and single-field estimates
    on the axes. The right panel magnifies an elongated downward-sloping
    bootstrap cloud around the joint estimate, indicating a trade-off in
  allocating feedback between channels.}
\end{figure}

\Cref{fig:two-field-region} shows the profiled joint surface behind
\Cref{tab:two-field-raw-appendix}.
The bootstrap cloud tilts along the diagonal, reflecting the trade-off between
two channels whose peer-return signals correlate at
\ppnum[3]{overlap-induced-pearson}, yet it varies in both directions rather
than collapsing onto a one-dimensional ridge.

\begin{table}[H]
  \centering
  \caption{Annual unadjusted model-implied adjustment under the frozen
    barycentric interaction field. The 2026 period ends on
  \ppvalue{post-evaluation-date-end}.}
  \label{tab:qmle-annual}
  \begin{tabular}{lrrr}
    \toprule
    Evaluation year & $\hat\lambda$ & 95\% interval & Implied $\hat\rho$ \\
    \midrule
    2023 & \ppnum[2]{post-eval2023-w-flat-lambda} &
    $[\ppnum[2]{post-eval2023-w-flat-lambda-ci-lower},
    \ppnum[2]{post-eval2023-w-flat-lambda-ci-upper}]$ &
    \ppnum[3]{post-eval2023-w-flat-rho} \\
    2024 & \ppnum[2]{post-eval2024-w-flat-lambda} &
    $[\ppnum[2]{post-eval2024-w-flat-lambda-ci-lower},
    \ppnum[2]{post-eval2024-w-flat-lambda-ci-upper}]$ &
    \ppnum[3]{post-eval2024-w-flat-rho} \\
    2025 & \ppnum[2]{post-eval2025-w-flat-lambda} &
    $[\ppnum[2]{post-eval2025-w-flat-lambda-ci-lower},
    \ppnum[2]{post-eval2025-w-flat-lambda-ci-upper}]$ &
    \ppnum[3]{post-eval2025-w-flat-rho} \\
    2026 & \ppnum[2]{post-eval2026-w-flat-lambda} &
    $[\ppnum[2]{post-eval2026-w-flat-lambda-ci-lower},
    \ppnum[2]{post-eval2026-w-flat-lambda-ci-upper}]$ &
    \ppnum[3]{post-eval2026-w-flat-rho} \\
    \bottomrule
  \end{tabular}
\end{table}

With the 2018--2022 field held fixed, re-estimating only the common coefficient
and nuisance parameters within each calendar year yields adjustment indices
between \ppnum[2]{post-eval2024-w-flat-lambda} and
\ppnum[2]{post-eval2026-w-flat-lambda}, and all four intervals lie above zero.
Interval widths vary, especially in 2025, and the 2026 estimate uses
\ppvalue{post-eval2026-w-flat-trading-days} dates.

\section{Supporting Interaction-Field Results and Implementation}
\label{sec:supporting-results}

This appendix documents the sample roster, the resampling and inference
architecture, the transport-dimension ladder, the representation sensitivity,
and the annual joint-field estimates.

The firms used to construct the primary barycentric interaction field are
listed in \Cref{tab:sample-universe}.


\begin{table}[htbp]
\centering
\small
\begin{tabularx}{\linewidth}{@{}lr>{\raggedright\arraybackslash}X@{}}
\toprule
Sector & Firms & Ticker symbols \\
\midrule
Communication Services & 11 & BIDU, CHTR, CMCSA, EA, GOOG, MTCH, NFLX, NTES, SIRI, TMUS, TTWO \\
Consumer Cyclical & 18 & BKNG, EBAY, EXPE, HAS, JD, LULU, MAR, MAT, MELI, NCLH, ORLY, ROST, SBUX, TCOM, TRIP, TSCO, ULTA, WYNN \\
Consumer Defensive & 7 & COST, DLTR, KDP, KHC, MDLZ, MNST, PEP \\
Energy & 1 & FANG \\
Financial Services & 1 & PYPL \\
Healthcare & 14 & ALGN, AMGN, AZN, BIIB, BMRN, DXCM, GILD, HOLX, IDXX, ILMN, INCY, ISRG, REGN, VRTX \\
Industrials & 7 & AAL, CSX, FAST, HON, JBHT, PCAR, UAL \\
Technology & 38 & ADBE, ADI, ADP, ADSK, AKAM, AMAT, AMD, ASML, AVGO, CDNS, CHKP, CSCO, CTSH, ENPH, FISV, FTNT, INTC, INTU, KLAC, LRCX, MCHP, MRVL, MU, NTAP, NVDA, NXPI, OKTA, PANW, PAYX, QCOM, SNPS, STX, SWKS, TEAM, TXN, WDAY, WDC, ZS \\
Utilities & 3 & AEP, EXC, XEL \\
\bottomrule
\end{tabularx}
\caption{Canonical empirical roster: ticker symbols of the firms in the canonical priced $W_2$ geometry, grouped by the sector label in Nasdaq summary metadata. Firm names accompany the symbols in the replication archive's universe metadata.}
\label{tab:sample-universe}
\end{table}

\subsection{Resampling and Inference Implementation}
\label{sec:app-inference}\label{sec:p5-two-field-implementation}

Every interval and every held-out comparison in the paper draws on one
resampling architecture, so a gain reported in more than one table carries one
interval.
Each of the 2,000 draws uses three stationary-block date schedules with
expected block length 21, taken from a single seed: one for the 2018--2022
calibration panel on which the scored parameters are fitted, one for the
post-corpus panel on which in-period coefficients are refitted, and one for the
post-corpus dates over which the fitted model is scored.
Every model, every comparator field, and both transport orders share the same
draws, so score differences are paired, and each draw preserves the
cross-section observed on its resampled dates.
The procedure refits the factor-only restriction, the direct Ledoit--Wolf
residual covariance, and every field coefficient on each calibration draw; the
calibration-tuned RBF bandwidth is held at its full-calibration selection, so
that operator is frozen while its coefficient is refitted.
Annual folds add one to the base seed in period order, and within a period the
joint fit and both single-field boundaries share one bootstrap bank.
The draws condition on the corpus, embeddings, and constructed fields, so they
propagate return-sampling and nuisance-parameter uncertainty but not
uncertainty from article sampling, from the representation, or from tied
optimal assignments.
All reported intervals are bootstrap quantiles; numerical-Hessian standard
errors are retained only as diagnostics, and boundary-constrained joint
intervals are descriptive.

The joint search profiles the mixture weight and total feedback on a
\ppvalue{two-field-theta-points} by \ppvalue{two-field-rho-points} grid and
refines the maximum continuously within its grid cell; every bootstrap draw
receives the same refinement, so coefficients and intervals do not inherit the
grid spacing.
The Jacobian term is evaluated from the eigenvalues of each field mixture on
the grid and by direct LU factorization during refinement; tests verify that
the two routes agree to floating-point tolerance.

The restricted-null QLR calibration is used only for the nested restrictions
$\rho=0$, $\rho^\flat=0$, and $\rho^{\mathrm{news}}=0$, never for the
non-nested comparison
between fields.
For each null, the procedure recovers and centers the residuals of the
restricted fit, standardizes each asset path, and draws an independent
stationary-block path for every asset and null replicate, which preserves
temporal blocks while removing contemporaneous residual alignment under the
working null.
After restoring the fitted common or firm-specific variances and factor means,
it maps the innovations back through the restricted spatial inverse and refits
both models with the same refinement; non-converged draws are rejected.
The $p$-value is $(1+\#\{QLR^*\geq QLR\})/(B+1)$ with $B=2{,}000$ null draws,
reported at its floor of $1/2{,}001$ when no draw exceeds the observed
statistic.
The simulation accommodates the nonnegative-channel boundary, so no chi-square
reference is used; this working null does not provide size control under
common stochastic volatility.

The transport-order equivalence margin, fixed before the final estimation, is
1\% of the W2-versus-factor-only mean calibration-score gain: the fields are
predictively equivalent when the 90\% paired interval for the W2-minus-W1
score lies inside it.
The weak-separation rule, fixed after a pilot computation and before the same
final estimation, classifies the fields as nearly collinear when their induced
peer-return signals correlate at $0.98$ or more and the standard deviation of
their difference is at most $0.10$ of the W2 signal's.

\subsection{Matryoshka Dimension Ladder}
\label{sec:p5-matryoshka}

The ladder repeats both transport orders at $D\in\{4096,1024,256,64\}$ within
Qwen3-Embedding 8B \citep{kusupati_matryoshka_2022,qwen3_embedding_2025}.
For each article embedding $z$, the representation retains the leading $D$
coordinates and then normalizes that prefix,
\[
  z^{(D)}=\frac{(z_1,\ldots,z_D)}{\lVert(z_1,\ldots,z_D)\rVert_2}.
\]
The ground cost remains unscaled Euclidean chord distance at every dimension.
Unit normalization bounds chord distance in $[0,2]$ and squared chord cost in
$[0,4]$, and for two unit prefixes $d^2=2-2\langle z_i^{(D)},z_j^{(D)}\rangle$,
so the W2 assignment cost is twice their cosine dissimilarity.
A common scalar rescaling multiplies the W1 and W2 assignment objectives by
positive constants, so it alters neither optimal assignment nor the minimizing
reconstruction weights.

The mechanism under study is concentration rather than scale.
For any reference distance $d_0$, the exact identity
\[
  d^2=2d_0d-d_0^2+(d-d_0)^2
\]
shows that squared cost is affine in cost up to the final nonnegative
remainder.
When the observed distances occupy a narrow neighborhood of $d_0$, that
remainder is small and minimum-sum $d$ and $d^2$ assignments can coincide.
The ladder tests this local approximation empirically.
At each dimension it enumerates every article-to-article cost in every
unordered cross-firm balanced assignment problem, reports the population
coefficient of variation of chord distance, and traces W1 and W2 agreement
through the share of identical atom assignments, the correlation of
off-diagonal operator weights, the correlation of induced peer-return signals,
and the W2-minus-W1 held-out score under the primary specification.
This ordering separates concentration in the embedding geometry from
convergence introduced by target reconstruction or by common variation in
returns.

\begin{table}[H]
\centering
\scriptsize
\setlength{\tabcolsep}{3.2pt}
\begin{tabular}{rrrrrr}
\toprule
Dim. & Chord-distance CV & Assignment overlap & Weight corr. & Spatial-lag corr. & $\Delta$ held-out score \\
\midrule
4,096 & 5.75\% & 90.1\% & 0.9979 & 0.9998 & -0.000109 \\
1,024 & 5.97\% & 89.5\% & 0.9978 & 0.9998 & -0.000339 \\
256 & 7.16\% & 87.4\% & 0.9977 & 0.9998 & -0.000354 \\
64 & 10.33\% & 82.4\% & 0.9975 & 0.9997 & -0.000443 \\
\bottomrule
\end{tabular}
\caption{Matryoshka dimension ladder for target-anchored W1 and W2 transport. At each dimension, leading coordinates are truncated and then row-wise L2-normalized; the ground cost is unscaled Euclidean chord distance. Chord-distance CV uses all 81,100,800 article-pair costs across 4,950 unordered firm pairs. Assignment overlap is the share of the 633,600 matched target atoms receiving the same peer atom under minimum-sum $d$ and $d^2$ assignments. Weight correlation uses off-diagonal target-operator entries; spatial-lag correlation uses the common 2023--2026 excess-return panel. The final column is W2 minus W1 mean Gaussian quasi-log score per firm-day after fitting the FF5-plus-momentum model on 2018--2022; positive values favour W2.}
\label{tab:transport-dimension-ladder}
\end{table}

\subsection{Embedding-Representation Sensitivity}
\label{sec:p5-representation-ablation}

The representation exercises ask whether the field comparison depends on one
text representation.
The sensitivity roster separates four design margins: output width, model
capacity, model family, and information vintage.
An encoder maps each article into a vector, whose dimension determines the
coordinates entering both the pairwise $W_2$ distances and the target-anchored
reconstruction.

The output-width comparison uses Qwen3-Embedding 8B at full width and at 1,024,
256, and 64 coordinates.
Qwen's Matryoshka training is designed to keep leading coordinate prefixes
informative, so the shorter vectors test how much compression the interaction
geometry tolerates.
The common 1,024-coordinate Qwen3-Embedding 8B and 4B rows then vary model
capacity while holding output width fixed.
Their full-width comparison changes capacity and output width together, so it
does not isolate capacity.
The 1,024-coordinate BGE-large-v1.5 row changes model family and pretraining
relative to Qwen3-Embedding 8B, making it an external sensitivity check rather
than an identified training-mechanism effect
\citep{qwen3_embedding_2025,xiao_cpack_2023}.

The primary Qwen encoder postdates the 2018--2022 article window, so the
information-vintage comparison asks whether the field comparison depends on an
encoder's information cutoff.
It uses the published ChronoBERT checkpoint for each year-end from 2017 through
2024 \citep{he_chronobert_2025}.
These ModernBERT-based masked-language encoders produce 768-coordinate
representations by averaging final-layer token states over all non-padding
positions, including special tokens.
We hold inference fixed at 8,000 characters and 1,024 tokens per article,
using bfloat16 autocast with float32 model parameters and token-mean pooling.
We apply row-L2 normalization once before constructing distances.
The same 128 articles per firm enter every checkpoint; neither layers nor
checkpoints are selected using returns.
Later checkpoints continue training their predecessors, so their contrasts
combine later information with cumulative training exposure, not matched
compute or an identified causal leakage effect.
The 2017 cutoff precedes article construction; 2018--2022 falls within it,
and 2023--2024 postdates construction but overlaps the full 2023--2026
return-evaluation period.
No arm is a post-evaluation negative control.
Since publication occurred later, the exercise concerns reported training
cutoffs, not historical deployment availability or point-in-time validity
of all other study inputs.

Every row in \Cref{tab:p5-representation-ablation} uses the same article inputs
through 2022, the same 2023--2026 return dates, and one common joint-date
bootstrap schedule.
Pairing each contrast within the same resample removes common bootstrap
variation, while the Holm correction accounts for testing several
representations in each family.
The table reports direct differences in $\rho$ for the barycentric field and
changes in its $\rho$ gap relative to RBF.
For representation $r$, write
$g_r=\rho^\flat_r-\rho^h_r$.
The two candidate-minus-reference estimands are
$\rho^\flat_c-\rho^\flat_r$ and $g_c-g_r$; negative values indicate,
respectively, weaker fitted feedback for the barycentric field and a smaller
advantage over the matching RBF operator.
The persistent co-mention graph is embedding-free and is therefore reported
once.
The equal-active-support comparator replaces the fitted coordinate magnitudes
with uniform weights on the same selected support, so it asks whether the
barycentric coordinate magnitudes improve conditional spatial fit beyond peer
selection.

\subsection{Point-in-Time Encoder Check}
\label{sec:p5-point-in-time}

The primary Qwen encoder was released after the article-construction window.
This sensitivity therefore repeats the primary FF5-plus-momentum held-out
score comparison with ChronoBERT 2017, whose reported training cutoff is
31 December 2017, before both article construction and calibration
\citep{he_chronobert_2025}.
The barycentric and median-bandwidth RBF fields are rebuilt from the pre-period
representation, held fixed during evaluation, and compared with the same
factor-only restriction.

\begin{table}[H]
\centering
\scriptsize
\setlength{\tabcolsep}{3.2pt}
\begin{tabular}{lrrrr}
\toprule
Interaction field & $\hat\rho$ & Held-out score & Gain & 95\% CI \\
\midrule
Factor-only & 0.000 & 2.468881 & 0.000000 & -- \\
ChronoBERT 2017 W2 barycentric & 0.561 & 2.497084 & 0.028203 & [0.0249, 0.0318] \\
ChronoBERT 2017 median-bandwidth W2 RBF & 0.452 & 2.478885 & 0.010004 & [0.0082, 0.0120] \\
\midrule
W2 minus median RBF & -- & -- & 0.018199 & [0.0152, 0.0214] \\
\bottomrule
\end{tabular}
\caption{Point-in-time representation sensitivity under the primary FF5-plus-momentum specification. ChronoBERT 2017 has a reported training-information cutoff of 2017-12-31, pre-period relative to the 2018--2022 article and calibration windows. This is not a claim of historical checkpoint availability. Fields are held fixed throughout the 2023--2026 evaluation. The score is the mean Gaussian quasi-log score per firm-day; intervals use paired stationary-block bootstrap draws. The RBF field uses the median off-diagonal squared W2 distance as its bandwidth.}
\label{tab:p5-point-in-time-encoder}
\end{table}

The pre-period representation preserves a positive barycentric gain
and a positive
paired barycentric-minus-RBF difference under the primary score design.

The contrast table states one joint equivalence rule for the vintage rows and
reports its two benchmark-calibrated bounds.
A pair meets the rule only when the marginal 95\% paired intervals for both
estimands lie strictly inside their corresponding bounds.
Holm adjusts the $p$-values separately for the 28 vintage comparisons in each
metric; the intervals are not simultaneous.
These post-specified thresholds organize a descriptive sensitivity exercise;
they are not prespecified confirmatory equivalence margins.

\begingroup
\scriptsize
\setlength{\tabcolsep}{3.3pt}
\begin{longtable}{@{}p{0.35\linewidth}llrrr@{}}
\caption{Pooled 2023--2026 representation sensitivity of the single-field spatial horse race across seven base representations and eight published annual ChronoBERT vintages. Each encoder row compares the barycentric interaction field $W^\flat$, obtained by target-anchored Wasserstein-barycentric reconstruction, with the median-bandwidth RBF transform of its matching squared-W$_2$ matrix. Intervals use the same 2,000 joint-date stationary bootstrap on the common 815-date schedule. $\hat\lambda=\hat\rho/(1-\hat\rho)$ is the implied adjustment index; log-likelihood is descriptive. Persistent news co-mentions are embedding-free; equal active support is shown once for the primary barycentric interaction field. These are paired conditional-fit sensitivity comparisons across the registered representation cells. Published cumulative-training checkpoints are not a matched-compute causal leakage test.}
\label{tab:p5-representation-ablation} \\
\toprule
Encoder & Dim. & Interaction matrix & $\hat\rho$ [95\% CI] & $\hat\lambda$ & Log-likelihood \\
\midrule
\endfirsthead
\multicolumn{6}{c}{\tablename\ \thetable\ (continued)} \\
\toprule
Encoder & Dim. & Interaction matrix & $\hat\rho$ [95\% CI] & $\hat\lambda$ & Log-likelihood \\
\midrule
\endhead
\midrule
\endfoot
\bottomrule
\endlastfoot
Qwen3-8B full (primary) & 4096 & Barycentric $W^\flat$ & 0.843 [0.815, 0.869] & 5.36 & 205146.3 \\
 &  & RBF--Wasserstein $W^h$ & 0.812 [0.780, 0.842] & 4.33 & 200099.7 \\
\addlinespace[1.5pt]
Qwen3-4B full & 2560 & Barycentric $W^\flat$ & 0.844 [0.816, 0.870] & 5.42 & 204835.4 \\
 &  & RBF--Wasserstein $W^h$ & 0.812 [0.779, 0.842] & 4.32 & 200068.5 \\
\addlinespace[1.5pt]
Qwen3-8B@1024 & 1024 & Barycentric $W^\flat$ & 0.844 [0.815, 0.870] & 5.40 & 205070.3 \\
 &  & RBF--Wasserstein $W^h$ & 0.812 [0.780, 0.842] & 4.33 & 200104.7 \\
\addlinespace[1.5pt]
Qwen3-4B@1024 & 1024 & Barycentric $W^\flat$ & 0.842 [0.813, 0.868] & 5.33 & 204859.1 \\
 &  & RBF--Wasserstein $W^h$ & 0.812 [0.780, 0.842] & 4.33 & 200094.3 \\
\addlinespace[1.5pt]
BGE-large-v1.5 full & 1024 & Barycentric $W^\flat$ & 0.840 [0.811, 0.866] & 5.24 & 204072.1 \\
 &  & RBF--Wasserstein $W^h$ & 0.811 [0.778, 0.841] & 4.29 & 200042.4 \\
\addlinespace[1.5pt]
Qwen3-8B@256 & 256 & Barycentric $W^\flat$ & 0.839 [0.810, 0.866] & 5.22 & 204870.9 \\
 &  & RBF--Wasserstein $W^h$ & 0.814 [0.781, 0.844] & 4.37 & 200127.5 \\
\addlinespace[1.5pt]
Qwen3-8B@64 & 64 & Barycentric $W^\flat$ & 0.821 [0.787, 0.852] & 4.59 & 204193.7 \\
 &  & RBF--Wasserstein $W^h$ & 0.816 [0.784, 0.845] & 4.43 & 200215.3 \\
\addlinespace[1.5pt]
ChronoBERT 2017 & 768 & Barycentric $W^\flat$ & 0.784 [0.741, 0.822] & 3.62 & 201973.8 \\
 &  & RBF--Wasserstein $W^h$ & 0.811 [0.777, 0.841] & 4.28 & 200041.7 \\
\addlinespace[1.5pt]
ChronoBERT 2018 & 768 & Barycentric $W^\flat$ & 0.781 [0.738, 0.820] & 3.56 & 201824.1 \\
 &  & RBF--Wasserstein $W^h$ & 0.810 [0.777, 0.841] & 4.28 & 200031.9 \\
\addlinespace[1.5pt]
ChronoBERT 2019 & 768 & Barycentric $W^\flat$ & 0.781 [0.738, 0.821] & 3.57 & 201786.3 \\
 &  & RBF--Wasserstein $W^h$ & 0.810 [0.776, 0.841] & 4.27 & 200026.7 \\
\addlinespace[1.5pt]
ChronoBERT 2020 & 768 & Barycentric $W^\flat$ & 0.781 [0.738, 0.820] & 3.57 & 201811.2 \\
 &  & RBF--Wasserstein $W^h$ & 0.810 [0.776, 0.841] & 4.27 & 200027.5 \\
\addlinespace[1.5pt]
ChronoBERT 2021 & 768 & Barycentric $W^\flat$ & 0.776 [0.732, 0.816] & 3.47 & 201664.2 \\
 &  & RBF--Wasserstein $W^h$ & 0.810 [0.776, 0.840] & 4.26 & 200027.8 \\
\addlinespace[1.5pt]
ChronoBERT 2022 & 768 & Barycentric $W^\flat$ & 0.782 [0.739, 0.820] & 3.58 & 201883.2 \\
 &  & RBF--Wasserstein $W^h$ & 0.810 [0.777, 0.841] & 4.27 & 200040.1 \\
\addlinespace[1.5pt]
ChronoBERT 2023 & 768 & Barycentric $W^\flat$ & 0.786 [0.744, 0.824] & 3.67 & 202016.8 \\
 &  & RBF--Wasserstein $W^h$ & 0.810 [0.777, 0.840] & 4.26 & 200043.2 \\
\addlinespace[1.5pt]
ChronoBERT 2024 & 768 & Barycentric $W^\flat$ & 0.783 [0.741, 0.822] & 3.62 & 201936.1 \\
 &  & RBF--Wasserstein $W^h$ & 0.810 [0.777, 0.840] & 4.27 & 200036.6 \\
\addlinespace[1.5pt]
\midrule
Embedding-free & -- & Persistent news $W^{\mathrm{news}}$ & 0.667 [0.618, 0.716] & 2.01 & 203493.8 \\
Primary support & -- & Equal active support & 0.841 [0.814, 0.865] & 5.27 & 202124.6 \\
\end{longtable}
\endgroup
\begingroup
\scriptsize
\setlength{\tabcolsep}{3.0pt}
\begin{longtable}{@{}p{0.35\linewidth}rrrrrr@{}}
\caption{Paired common-schedule QMLE contrasts. Panel A estimates $\rho^\flat_c-\rho^\flat_r$; Panel B estimates $(\rho^\flat_c-\rho^h_c)-(\rho^\flat_r-\rho^h_r)$. $p$ denotes two-sided add-one sign-tail probabilities; $p_H$ denotes Holm adjustment separately within each metric's seven base and 28 vintage contrasts. Estimates and intervals retain five decimal places so endpoints near zero remain visible. Vintage equivalence requires both intervals to lie strictly inside their respective BGE-large-minus-Qwen3-8B@1024 benchmark-calibrated, post-specified bounds; Eq. is this joint two-metric decision. Intervals are marginal, not simultaneous. The contrasts compare conditional fit on a common return schedule.}
\label{tab:p5-representation-contrasts} \\
\toprule
Contrast ($c-r$) & Estimate & 95\% CI & $p$ & $p_H$ & Bound & Eq. \\
\midrule
\endfirsthead
\multicolumn{7}{c}{\tablename\ \thetable\ (continued)} \\
\toprule
Contrast ($c-r$) & Estimate & 95\% CI & $p$ & $p_H$ & Bound & Eq. \\
\midrule
\endhead
\midrule
\endfoot
\bottomrule
\endlastfoot
\multicolumn{7}{l}{\textit{Panel A: direct barycentric-$\rho$ contrasts}} \\
Qwen3-4B full $-$ Qwen3-8B full & 0.00134 & [0.00058, 0.00219] & 0.001 & 0.007 & -- & -- \\
Qwen3-4B@1024 $-$ Qwen3-8B@1024 & -0.00179 & [-0.00291, -0.00077] & 0.004 & 0.008 & -- & -- \\
Qwen3-8B@1024 $-$ Qwen3-8B full & 0.00090 & [0.00001, 0.00164] & 0.049 & 0.049 & -- & -- \\
Qwen3-8B@256 $-$ Qwen3-8B@1024 & -0.00449 & [-0.00600, -0.00324] & 0.001 & 0.007 & -- & -- \\
Qwen3-8B@64 $-$ Qwen3-8B@256 & -0.01805 & [-0.02296, -0.01353] & 0.001 & 0.007 & -- & -- \\
BGE-large-v1.5 full $-$ Qwen3-8B@1024 & -0.00392 & [-0.00572, -0.00245] & 0.001 & 0.007 & -- & -- \\
BGE-large-v1.5 full $-$ Qwen3-4B@1024 & -0.00213 & [-0.00359, -0.00080] & 0.002 & 0.007 & -- & -- \\
\addlinespace[1.5pt]
ChronoBERT 2018 $-$ ChronoBERT 2017 & -0.00297 & [-0.00406, -0.00215] & 0.001 & 0.028 & 0.00572 & Yes \\
ChronoBERT 2019 $-$ ChronoBERT 2017 & -0.00240 & [-0.00367, -0.00147] & 0.001 & 0.028 & 0.00572 & Yes \\
ChronoBERT 2019 $-$ ChronoBERT 2018 & 0.00056 & [-0.00057, 0.00142] & 0.257 & 1.000 & 0.00572 & Yes \\
ChronoBERT 2020 $-$ ChronoBERT 2017 & -0.00267 & [-0.00371, -0.00171] & 0.001 & 0.028 & 0.00572 & Yes \\
ChronoBERT 2020 $-$ ChronoBERT 2018 & 0.00030 & [-0.00052, 0.00135] & 0.527 & 1.000 & 0.00572 & Yes \\
ChronoBERT 2020 $-$ ChronoBERT 2019 & -0.00026 & [-0.00134, 0.00109] & 0.693 & 1.000 & 0.00572 & Yes \\
ChronoBERT 2021 $-$ ChronoBERT 2017 & -0.00740 & [-0.00931, -0.00605] & 0.001 & 0.028 & 0.00572 & No \\
ChronoBERT 2021 $-$ ChronoBERT 2018 & -0.00444 & [-0.00577, -0.00340] & 0.001 & 0.028 & 0.00572 & No \\
ChronoBERT 2021 $-$ ChronoBERT 2019 & -0.00500 & [-0.00646, -0.00381] & 0.001 & 0.028 & 0.00572 & No \\
ChronoBERT 2021 $-$ ChronoBERT 2020 & -0.00474 & [-0.00631, -0.00352] & 0.001 & 0.028 & 0.00572 & No \\
ChronoBERT 2022 $-$ ChronoBERT 2017 & -0.00200 & [-0.00281, -0.00128] & 0.001 & 0.028 & 0.00572 & Yes \\
ChronoBERT 2022 $-$ ChronoBERT 2018 & 0.00096 & [-0.00001, 0.00214] & 0.052 & 0.364 & 0.00572 & Yes \\
ChronoBERT 2022 $-$ ChronoBERT 2019 & 0.00040 & [-0.00055, 0.00170] & 0.454 & 1.000 & 0.00572 & Yes \\
ChronoBERT 2022 $-$ ChronoBERT 2020 & 0.00066 & [-0.00044, 0.00174] & 0.242 & 1.000 & 0.00572 & Yes \\
ChronoBERT 2022 $-$ ChronoBERT 2021 & 0.00540 & [0.00408, 0.00713] & 0.001 & 0.028 & 0.00572 & No \\
ChronoBERT 2023 $-$ ChronoBERT 2017 & 0.00217 & [0.00099, 0.00367] & 0.001 & 0.028 & 0.00572 & No \\
ChronoBERT 2023 $-$ ChronoBERT 2018 & 0.00514 & [0.00359, 0.00720] & 0.001 & 0.028 & 0.00572 & No \\
ChronoBERT 2023 $-$ ChronoBERT 2019 & 0.00458 & [0.00297, 0.00673] & 0.001 & 0.028 & 0.00572 & No \\
ChronoBERT 2023 $-$ ChronoBERT 2020 & 0.00484 & [0.00345, 0.00665] & 0.001 & 0.028 & 0.00572 & No \\
ChronoBERT 2023 $-$ ChronoBERT 2021 & 0.00958 & [0.00734, 0.01257] & 0.001 & 0.028 & 0.00572 & No \\
ChronoBERT 2023 $-$ ChronoBERT 2022 & 0.00418 & [0.00286, 0.00593] & 0.001 & 0.028 & 0.00572 & No \\
ChronoBERT 2024 $-$ ChronoBERT 2017 & -0.00042 & [-0.00125, 0.00056] & 0.369 & 1.000 & 0.00572 & Yes \\
ChronoBERT 2024 $-$ ChronoBERT 2018 & 0.00255 & [0.00146, 0.00410] & 0.001 & 0.028 & 0.00572 & No \\
ChronoBERT 2024 $-$ ChronoBERT 2019 & 0.00199 & [0.00093, 0.00350] & 0.001 & 0.028 & 0.00572 & Yes \\
ChronoBERT 2024 $-$ ChronoBERT 2020 & 0.00225 & [0.00116, 0.00349] & 0.001 & 0.028 & 0.00572 & Yes \\
ChronoBERT 2024 $-$ ChronoBERT 2021 & 0.00699 & [0.00531, 0.00930] & 0.001 & 0.028 & 0.00572 & No \\
ChronoBERT 2024 $-$ ChronoBERT 2022 & 0.00159 & [0.00054, 0.00284] & 0.005 & 0.040 & 0.00572 & Yes \\
ChronoBERT 2024 $-$ ChronoBERT 2023 & -0.00259 & [-0.00361, -0.00177] & 0.001 & 0.028 & 0.00572 & Yes \\
\midrule
\multicolumn{7}{l}{\textit{Panel B: barycentric-minus-RBF gap contrasts}} \\
Qwen3-4B full $-$ Qwen3-8B full & 0.00158 & [0.00076, 0.00255] & 0.001 & 0.007 & -- & -- \\
Qwen3-4B@1024 $-$ Qwen3-8B@1024 & -0.00173 & [-0.00283, -0.00075] & 0.004 & 0.012 & -- & -- \\
Qwen3-8B@1024 $-$ Qwen3-8B full & 0.00083 & [-0.00007, 0.00159] & 0.081 & 0.162 & -- & -- \\
Qwen3-8B@256 $-$ Qwen3-8B@1024 & -0.00593 & [-0.00762, -0.00451] & 0.001 & 0.007 & -- & -- \\
Qwen3-8B@64 $-$ Qwen3-8B@256 & -0.02002 & [-0.02565, -0.01496] & 0.001 & 0.007 & -- & -- \\
BGE-large-v1.5 full $-$ Qwen3-8B@1024 & -0.00244 & [-0.00392, -0.00091] & 0.001 & 0.007 & -- & -- \\
BGE-large-v1.5 full $-$ Qwen3-4B@1024 & -0.00071 & [-0.00192, 0.00073] & 0.276 & 0.276 & -- & -- \\
\addlinespace[1.5pt]
ChronoBERT 2018 $-$ ChronoBERT 2017 & -0.00268 & [-0.00375, -0.00190] & 0.001 & 0.028 & 0.00392 & Yes \\
ChronoBERT 2019 $-$ ChronoBERT 2017 & -0.00174 & [-0.00296, -0.00081] & 0.001 & 0.028 & 0.00392 & Yes \\
ChronoBERT 2019 $-$ ChronoBERT 2018 & 0.00095 & [-0.00015, 0.00181] & 0.083 & 0.498 & 0.00392 & Yes \\
ChronoBERT 2020 $-$ ChronoBERT 2017 & -0.00203 & [-0.00301, -0.00110] & 0.001 & 0.028 & 0.00392 & Yes \\
ChronoBERT 2020 $-$ ChronoBERT 2018 & 0.00065 & [-0.00015, 0.00169] & 0.132 & 0.660 & 0.00392 & Yes \\
ChronoBERT 2020 $-$ ChronoBERT 2019 & -0.00029 & [-0.00137, 0.00105] & 0.650 & 1.000 & 0.00392 & Yes \\
ChronoBERT 2021 $-$ ChronoBERT 2017 & -0.00648 & [-0.00829, -0.00521] & 0.001 & 0.028 & 0.00392 & No \\
ChronoBERT 2021 $-$ ChronoBERT 2018 & -0.00380 & [-0.00509, -0.00280] & 0.001 & 0.028 & 0.00392 & No \\
ChronoBERT 2021 $-$ ChronoBERT 2019 & -0.00475 & [-0.00621, -0.00357] & 0.001 & 0.028 & 0.00392 & No \\
ChronoBERT 2021 $-$ ChronoBERT 2020 & -0.00445 & [-0.00603, -0.00322] & 0.001 & 0.028 & 0.00392 & No \\
ChronoBERT 2022 $-$ ChronoBERT 2017 & -0.00144 & [-0.00229, -0.00068] & 0.001 & 0.028 & 0.00392 & Yes \\
ChronoBERT 2022 $-$ ChronoBERT 2018 & 0.00124 & [0.00026, 0.00242] & 0.015 & 0.105 & 0.00392 & Yes \\
ChronoBERT 2022 $-$ ChronoBERT 2019 & 0.00030 & [-0.00066, 0.00160] & 0.613 & 1.000 & 0.00392 & Yes \\
ChronoBERT 2022 $-$ ChronoBERT 2020 & 0.00059 & [-0.00053, 0.00167] & 0.312 & 1.000 & 0.00392 & Yes \\
ChronoBERT 2022 $-$ ChronoBERT 2021 & 0.00504 & [0.00378, 0.00669] & 0.001 & 0.028 & 0.00392 & No \\
ChronoBERT 2023 $-$ ChronoBERT 2017 & 0.00293 & [0.00175, 0.00436] & 0.001 & 0.028 & 0.00392 & No \\
ChronoBERT 2023 $-$ ChronoBERT 2018 & 0.00561 & [0.00413, 0.00759] & 0.001 & 0.028 & 0.00392 & No \\
ChronoBERT 2023 $-$ ChronoBERT 2019 & 0.00466 & [0.00314, 0.00668] & 0.001 & 0.028 & 0.00392 & No \\
ChronoBERT 2023 $-$ ChronoBERT 2020 & 0.00496 & [0.00363, 0.00665] & 0.001 & 0.028 & 0.00392 & No \\
ChronoBERT 2023 $-$ ChronoBERT 2021 & 0.00941 & [0.00727, 0.01219] & 0.001 & 0.028 & 0.00392 & No \\
ChronoBERT 2023 $-$ ChronoBERT 2022 & 0.00437 & [0.00308, 0.00609] & 0.001 & 0.028 & 0.00392 & No \\
ChronoBERT 2024 $-$ ChronoBERT 2017 & 0.00026 & [-0.00058, 0.00122] & 0.548 & 1.000 & 0.00392 & Yes \\
ChronoBERT 2024 $-$ ChronoBERT 2018 & 0.00295 & [0.00189, 0.00446] & 0.001 & 0.028 & 0.00392 & No \\
ChronoBERT 2024 $-$ ChronoBERT 2019 & 0.00200 & [0.00102, 0.00342] & 0.001 & 0.028 & 0.00392 & Yes \\
ChronoBERT 2024 $-$ ChronoBERT 2020 & 0.00229 & [0.00123, 0.00346] & 0.001 & 0.028 & 0.00392 & Yes \\
ChronoBERT 2024 $-$ ChronoBERT 2021 & 0.00675 & [0.00515, 0.00894] & 0.001 & 0.028 & 0.00392 & No \\
ChronoBERT 2024 $-$ ChronoBERT 2022 & 0.00170 & [0.00062, 0.00293] & 0.004 & 0.032 & 0.00392 & Yes \\
ChronoBERT 2024 $-$ ChronoBERT 2023 & -0.00266 & [-0.00365, -0.00186] & 0.001 & 0.028 & 0.00392 & Yes \\
\end{longtable}
\endgroup

For moderate Qwen3-8B truncation from full width to 1,024 coordinates, the
direct-$\rho$ contrast narrowly rejects zero after Holm adjustment, whereas the
barycentric-minus-RBF gap interval contains zero.
The subsequent 1,024-to-256 and 256-to-64 steps yield lower fitted feedback for
the barycentric field and smaller gaps relative to RBF after Holm adjustment.
The full-width 4B cell raises both estimands, whereas the fixed-width 4B
cell lowers both after Holm adjustment; the opposing native- and fixed-width
comparisons therefore cannot be read as one pure capacity effect.
BGE-large lowers both estimands relative to Qwen3-8B@1,024, but the comparison
does not isolate architecture.

The ChronoBERT staircase is non-monotone in both estimands.
Many annual pairs reject zero after Holm adjustment, but only a subset meets
the joint equivalence rule.
The first annual step lowers both estimands after adjustment: its direct-$\rho$
and gap intervals lie within their respective equivalence bounds.
The 2017-to-2024 endpoint meets both bounds without Holm-adjusted rejection.
Thus bounded differences for selected pairs do not establish staircase-wide
invariance, an exact null, or a causal effect of later information.
Log-likelihood remains descriptive throughout.

\subsection{Annual Joint-Field Persistence}
\label{sec:p5-two-field-annual}

\Cref{tab:two-field-annual} re-estimates both coefficients within each
non-overlapping annual return sample while holding the two 2018--2022 matrices
fixed, using the raw-return, common-intercept specification of
\ppAppendixRef{sec:app-unadjusted}.

\begin{table}[htbp]
  \centering
  \caption{Annual joint estimates under the frozen barycentric and news-link
    fields in the raw-return, common-intercept specification. The 2026 period
  ends on \ppvalue{post-evaluation-date-end}.}
  \label{tab:two-field-annual}
  \begin{tabular}{lrrrr}
    \toprule
    Evaluation year & $\hat\rho^\flat$ & $\hat\rho^{\mathrm{news}}$ &
    $\hat\lambda^\flat$
    & $\hat\lambda^{\mathrm{news}}$ \\
    \midrule
    2023 & \ppnum[3]{post-eval2023-two-field-rho-b} &
    \ppnum[3]{post-eval2023-two-field-rho-n} &
    \ppnum[2]{post-eval2023-two-field-lambda-b} &
    \ppnum[2]{post-eval2023-two-field-lambda-n} \\
    2024 & \ppnum[3]{post-eval2024-two-field-rho-b} &
    \ppnum[3]{post-eval2024-two-field-rho-n} &
    \ppnum[2]{post-eval2024-two-field-lambda-b} &
    \ppnum[2]{post-eval2024-two-field-lambda-n} \\
    2025 & \ppnum[3]{post-eval2025-two-field-rho-b} &
    \ppnum[3]{post-eval2025-two-field-rho-n} &
    \ppnum[2]{post-eval2025-two-field-lambda-b} &
    \ppnum[2]{post-eval2025-two-field-lambda-n} \\
    2026 & \ppnum[3]{post-eval2026-two-field-rho-b} &
    \ppnum[3]{post-eval2026-two-field-rho-n} &
    \ppnum[2]{post-eval2026-two-field-lambda-b} &
    \ppnum[2]{post-eval2026-two-field-lambda-n} \\
    \bottomrule
  \end{tabular}
\end{table}

Both channel coefficients remain positive in every year, and the barycentric
estimate is larger throughout.
The gap narrows in the incomplete 2026 period, which uses only
\ppvalue{post-eval2026-two-field-trading-days} dates and is not directly
comparable with a complete-year estimate.

\section{Formal Verification Status}\label{sec:lean-status}

Selected algebraic closure and scalar-projection results are machine-checked
in Lean 4 with mathlib under the hypotheses recorded below.
The factor-beta invariance proposition is not among these
machine-checked statements.
These checks concern the mathematical statements; they do not certify the
text-derived interaction weights or the statistical interpretation of the
estimates, which depends on the empirical design in \Cref{sec:empirical-design}.
The scripts and theorem inventory are included in the replication archive.

\ppLeanStatusList

\end{document}